\documentclass{ar-1col}
\usepackage{url}
\usepackage{natbib}
\usepackage{verbatim}
\usepackage{mathrsfs}			% to use \mathscr

\usepackage{comprehensive_abbrevs}
\usepackage{aas_macros}

\def\lesssim{\mathbin{\lower 3pt\hbox
{$\rlap{\raise 5pt\hbox{$\char'074$}}\mathchar"7218$}}}   %< or of order
\def\gtrsim{\mathbin{\lower 3pt\hbox
{$\rlap{\raise 5pt\hbox{$\char'076$}}\mathchar"7218$}}}

\jname{Xxxx. Xxx. Xxx. Xxx.}
\jvol{AA}
\jyear{2019}
\doi{10.1146/(TBD) ... ... Preprint prepared by the authors.}

\begin{document}

% Page header
\markboth{Cordes \& Chatterjee}{Fast Radio Bursts}

% Title
\title{Fast Radio Bursts: \\ An Extragalactic Enigma }

%Authors, affiliations address.
\author{James M. Cordes$^1$ and Shami Chatterjee$^1$
\affil{$^1$Department of Astronomy and Cornell Center for Astrophysics and Planetary Science, Cornell University, Ithaca, NY 14853, USA\\
email: jmc33@cornell.edu, sc99@cornell.edu}
}
%Abstract
%\begin{abstract}
%Abstract text, approximately 150 words.
%\end{abstract}

%Abstract text, approximately 150 words.
\begin{abstract}
We summarize our understanding of  millisecond radio bursts
from an  extragalactic population of sources.  FRBs occur at an extraordinary rate,
thousands per day over the entire sky with radiation energy densities at the source about ten billion times larger than those from Galactic pulsars.  We survey FRB phenomenology, source models and host galaxies, coherent radiation models, and the role of plasma propagation effects in burst detection.  The FRB field is guaranteed to be exciting: new telescopes will expand the sample from the current $\sim 80$ unique burst sources (and a few secure localizations and redshifts) to thousands, with  burst localizations that enable host-galaxy redshifts emerging directly from  interferometric surveys.
\\
$\bullet$ FRBs are now established as an extragalactic phenomenon. \\
$\bullet$ Only a few sources are known to repeat.  Despite the failure to redetect other FRBs, they are not inconsistent with all being repeaters. \\
$\bullet$ FRB sources may be new, exotic kinds of objects or known types in extreme circumstances.
Many inventive models exist, ranging from alien spacecraft to cosmic strings but those concerning compact objects and supermassive black holes have gained  the most attention.
A rapidly rotating magnetar is a promising explanation for FRB~121102 along with the persistent source associated with it, but alternative source models are not ruled out for it or other FRBs.  \\
$\bullet$ FRBs  are powerful  tracers of circumsource environments, `missing baryons'  in the  IGM, and dark matter.    \\
$\bullet$ The relative contributions of host galaxies and the IGM to propagation effects have yet to be disentangled, so dispersion measure distances have large uncertainties.

\end{abstract}

%Keywords, etc.
\maketitle

%Table of Contents
\tableofcontents

\section{INTRODUCTION}
\label{sec:intro}

%\bigskip

{\it Fast radio bursts} (FRBs) are  millisecond-duration pulses that  originate from as-yet unidentified extragalactic sources.   They are similar in some respects to  pulses from Galactic radio pulsars, but  the  flux density is of order ten billion times larger and their spectra are radically different from most pulsar spectra and most other radio sources.  To date\footnote{Up to 2019 February 1. Literature review covered up to December 1, 2018 except for early results from CHIME.}  bursts from over 80 distinct sources have been reported in the literature since the discovery of the first FRB \cite[][]{lbm+07}. Of these,  multiple  bursts have been detected from only a few FRB sources and only a few have secure localizations. FRB~121102, the first to be localized, is in a star-forming region in a dwarf galaxy with a luminosity distance of about one Gpc.

The nature of FRBs and their sources  are thus first and foremost a {\it bona fide} mystery about which we have several important clues that will likely lead soon to an understanding of the phenomenon.     In addition, FRBs  are also superb tools for probing the diverse media with dramatically different conditions along their lines of sight, including the immediate source environment, their   host galaxies,  and the cosmic web.

\begin{figure}[h]
\centerline{\includegraphics[width=2in]{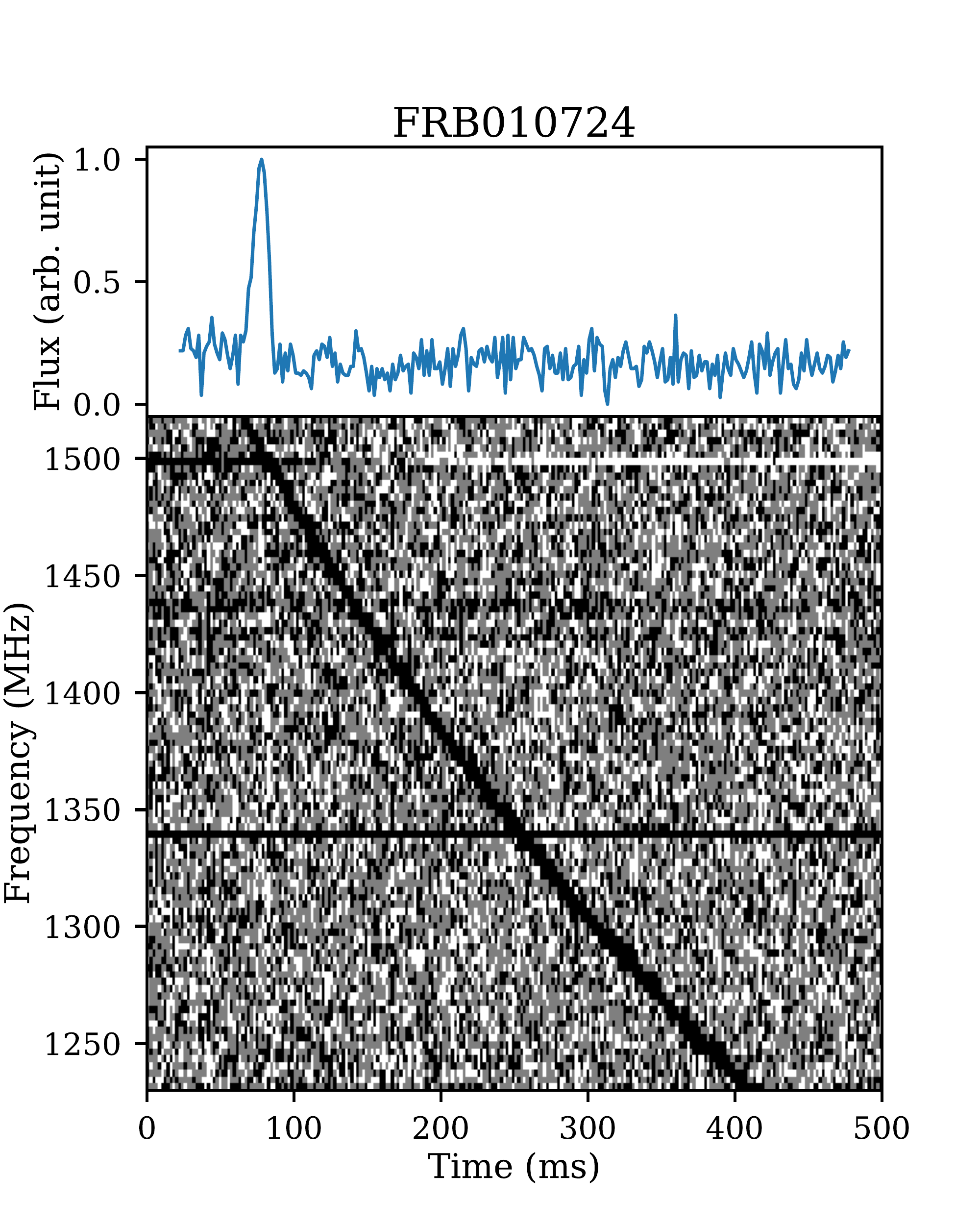}
\hspace{-3.4in}
\includegraphics[width=2in]{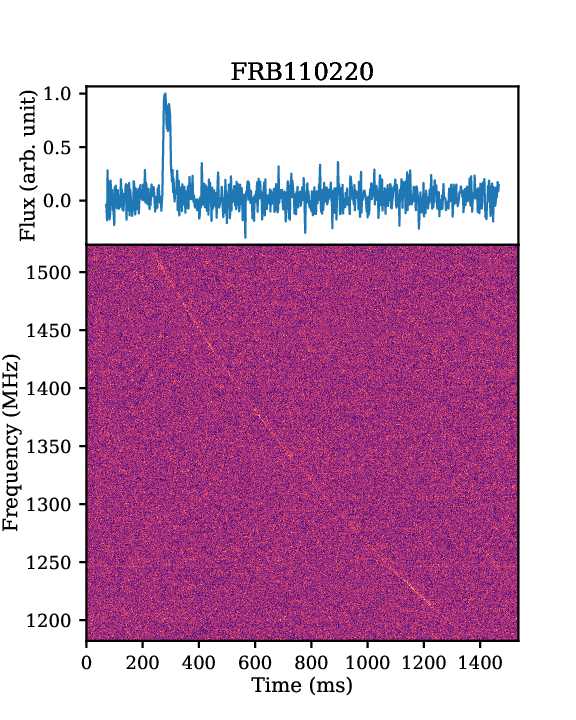}
\hspace{-3.4in}
\includegraphics[width=1.8in]{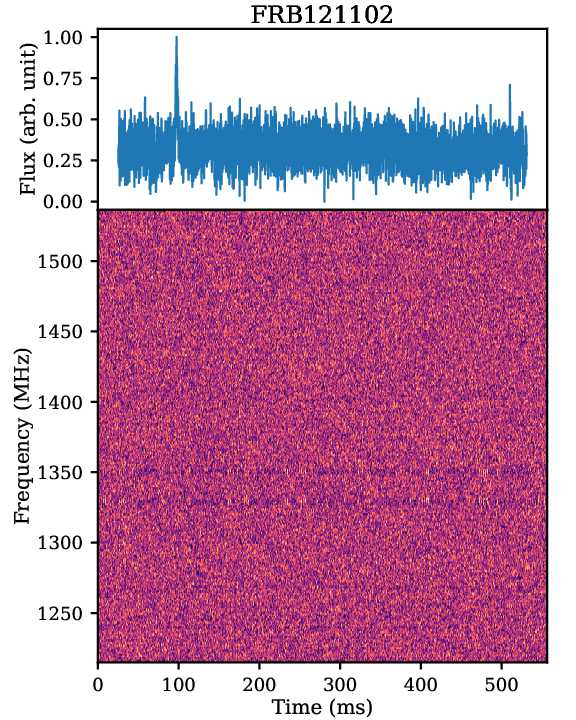}
}
\caption{Fast Radio Burst dynamic spectra. In each case, the lower panel shows the sweep of the burst across the time--frequency plane, and the upper panel shows the total pulse intensity after removing the best-fit quadratic dispersion sweep and frequency-averaging across the band. Time and frequency resolutions vary, depending on the instrument.
Left: FRB~010724, the first-reported fast radio burst \citep{lbm+07}, with DM $=375~\DMunits$.
Middle: FRB~110220, detected at Parkes \citep{tsb+13} with DM $=944.4~\DMunits$, leading to the realization that FRBs were most likely astrophysical in nature.
Right: The original detection of FRB~121102 at Arecibo \citep{sch+14}, the first reported non-Parkes FRB, with DM $=557.4~\DMunits$.
\label{fig:bursts1}
}
\end{figure}

\begin{figure}[h]
\centerline{\includegraphics[width=1.9in]{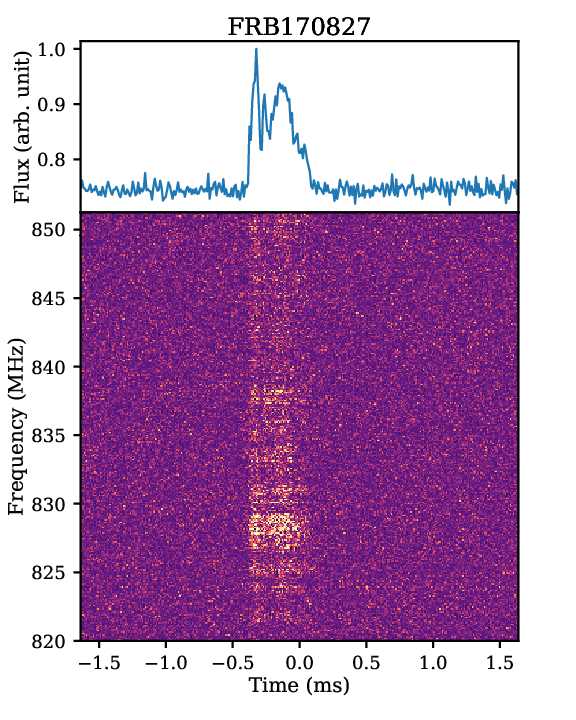}
\hspace{-3.4in}
\includegraphics[width=2in]{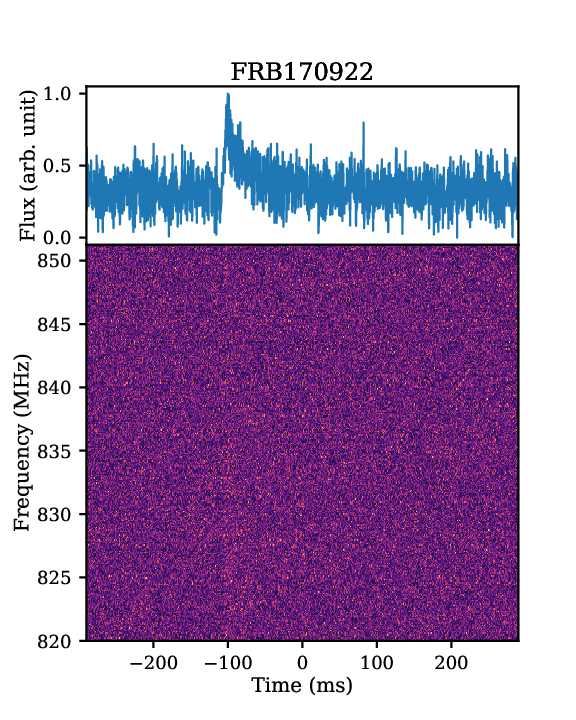}
\hspace{-3.4in}
\includegraphics[width=2in]{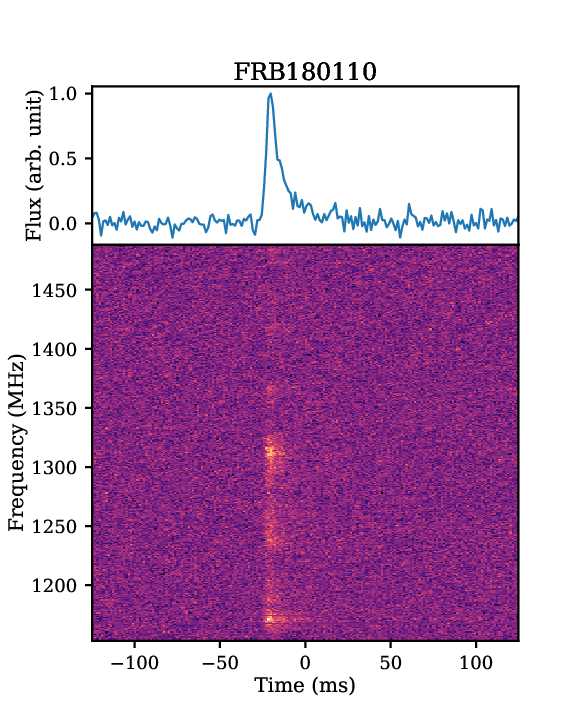}
}
\caption{Fast Radio Burst dynamic spectra. In each case, the lower panel shows the burst on the time--frequency plane after removing the best-fit quadratic pulse dispersion sweep, and the upper panel shows the dedispersed pulse total intensity after frequency-averaging across the band. Time and frequency resolutions vary, depending on the instrument.
Left: FRB~170827, detected at UTMOST \citep{2018ATel11675....1F} with DM $=899~\DMunits$. Voltage capture was triggered after real-time detection, and coherent de-dispersion reveals fine structure in the burst.
Middle: FRB~170922, detected at UTMOST
\citep[][]{2017ATel10867....1F}
 with DM $=1111~\DMunits$ and very significant pulse scattering.
Right: FRB~180110, a bright burst detected with ASKAP in fly's-eye mode \citep{smb+18} with DM $=716~\DMunits$.
\label{fig:bursts2}
}
\end{figure}

\begin{figure}[h]
\centerline{\includegraphics[width=2in]{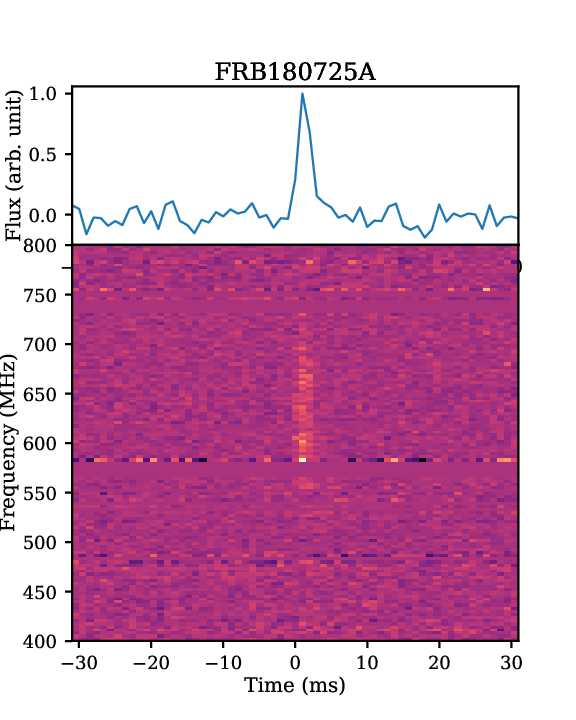}
\hspace{-3.4in}
\includegraphics[width=2in]{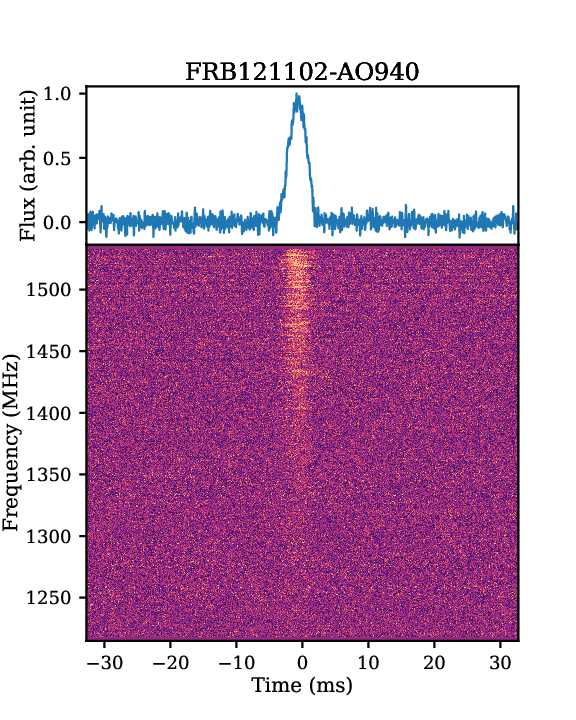}
\hspace{-3.4in}
\includegraphics[width=2in]{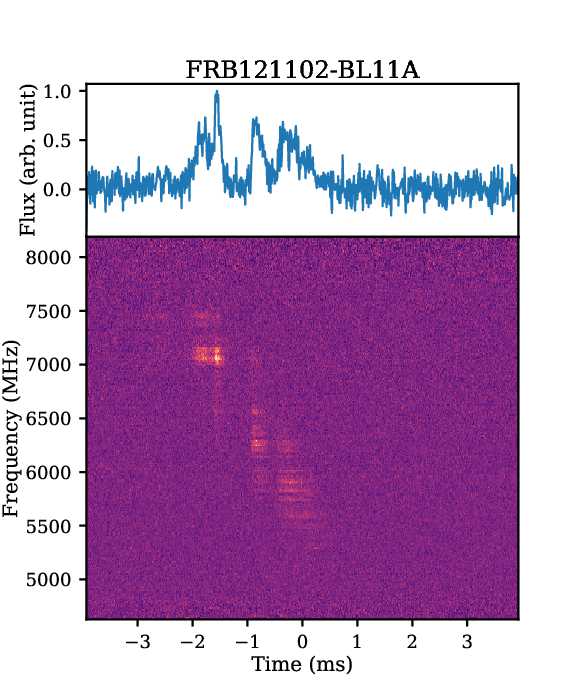}
}
\caption{Fast Radio Burst dynamic spectra. As in Figure~\ref{fig:bursts2}, the lower panel shows the burst on the time--frequency plane after removing the best-fit quadratic pulse dispersion sweep, and the upper panel shows the dedispersed pulse total intensity after frequency-averaging across the band. Time and frequency resolutions vary, depending on the instrument.
Left: FRB~180725A, the first FRB detected at CHIME and at frequencies down to 550~MHz \citep{2019Natur.566..230C}, with DM $=716.6~\DMunits$.
Middle: One of the first and brightest re-detections of FRB~121102 at Arecibo \citep{ssh+16a} at 1.2--1.6~GHz.
Right: Another detection of FRB~121102 at the Green Bank Telescope \citep{gsp+18}, but at 4--8~GHz. Precise localization enabled high frequency observations, and the known DM allowed coherent dedispersion, revealing extensive pulse structure at these higher frequencies.
\label{fig:bursts3}
}
\end{figure}

Short duration pulses  have been known in radio astronomy since the discovery of pulsars in 1967.  Recognition that their radio-frequency ($\nu$) dependent arrival times followed the characteristic $\DM\, \nu^{-2}$  scaling law expected for a tenuous, cold plasma was central to establishing an early distance scale for the first pulsars and the same approach has been taken for   FRBs.  Here $\DM = \int_0^d ds\, \nelec$ is the {\it dispersion measure}, the integral of the electron density to a source at distance $d$.    Values for Galactic pulsars range from $\sim 1$~to~$1700~\DMunits$, where the units follow from distances expressed in parsecs and electron densities in cm$^{-3}$.
Figure~\ref{fig:bursts1} shows FRB dynamic spectra where the dispersion delays have been retained, whereas Figures~\ref{fig:bursts2} and \ref{fig:bursts3} show them with the delay removed.

%\vspace{3.in}

%\vspace{3.in}

The first FRB \citep[][]{lbm+07} showed dispersive arrival times  combined with  broadening  by   multipath  scattering from small-scale fluctuations in electron density.  The
 too-large-to-be-Galactic value of DM led to the conjecture that the source was extragalactic.
 FRBs viewed at high Galactic latitudes $b$ receive Galactic contributions of about $30~\DMunits / \sin \vert b \vert$ compared to $\DM \sim 375~\DMunits$ for the Lorimer burst (FRB~010724).
  However,  establishing FRBs as an astrophysical phenomenon took another six years with the identification of further examples \citep[][]{tsb+13}, albeit from the same Parkes telescope in Australia.   The discovery of a burst  using the Arecibo radio telescope \citep[][]{sch+14} gave further credence to the phenomenon.
  %At present, a few dozen distinct burst sources have been identified and there are about double that number to %be published soon.

 The slow acceptance of FRBs as an extragalactic phenomenon  is a consequence  of a rather long history of false positives, the fact that some  Galactic objects, including pulsars, show a high degree of intermittency, and the idea that unmodeled HII regions  \citep[][]{2014ApJ...797...70K} or stars \citep[][]{mls+15a} in the Milky Way might be responsible for  the large measured values of DM.     And, of course,  radio-frequency interference (RFI) from artificial terrestrial sources  can mimic dispersed pulses \citep[][]{pkb+15}.

 Some Galactic pulsars are sufficiently intermittent that they were discovered as single-pulse emitters and only subsequently determined to be periodic with properties otherwise identical to pulsars.    They were consequently described as {\it rotating radio transients}
 \citep[RRATs,][]{mll+06}.  A handful of RRATs, like most FRBs,  have defied redetection.  However, their DM values are consistent with residence in the Milky Way, so --- as the jargon currently stands --- they are not FRBs.   Eventually, a Galactic FRB may be identified as a very bright event that saturates radio receivers in telescopes and in telephones.

In the early days of pulsar astronomy, attempts were made to detect dispersed radio pulses from high-energy objects, such as the  X-ray sources Sco~X-1 and Cyg~X-1  \citep[][]{1972ApJ...172L..17T}.   None were found and it is now understood  that strong coherent radio emission does not  occur in accreting X-ray sources. Later,
\citet[][]{1980ApJ...236L.109L} used a fast (sub-ms) spectrometer at Arecibo to search for bursts from the blazar M87. In this case, dispersed pulses were claimed but  not confirmed by others  and therefore were dismissed as  RFI or artifacts of the hardware.    In retrospect, given the {\it apparent}  non-repeatibility of most FRBs, one wonders  whether the M87 pulses were in fact real and await confirmation!

Other attempts were made at low frequencies, including work using the Molonglo telescope by \citet[][]{alv89} at 0.84~GHz that detected many pulsars in $\sim 4000$~hr along with considerable RFI and unclassified events in the $\sim$ ms range that were not obviously due to RFI.   On average per 12~hr observation
about one such unclassified event occurred that
 was ``not ruled out as being of celestrial origin'' \citep[][page 173]{alv89}.  The threshold for this survey was $\sim 15$-50~Jy~ms for widths from 1 to 10~ms.   A two-station survey at 0.27~GHz \citep[][]{1974ApJ...187L..57H}  was motivated by  predictions that supernovae would emit narrow $\lesssim 1$~s pulses.   The system had $\sim 1$~sr field of view and relied on pulse dispersion to discriminate RFI from events of celestial origin.   Unfortunately none were found at levels above
$\sim 10^4$~Jy for widths from 20~ms to 1~s.  The notion that evaporating blackholes would emit narrow radio bursts \citep[][]{1977Natur.266..333R}  motivated several detection attempts \citep[e.g.][]{1978Natur.276..590O}
and has been referred to as a possible source class for FRBs.  However,  the original idea involved physics that is no longer thought applicable (M. Rees, private communication).

An early example of multimessenger astronomy involved searches for radio pulses that coincided with gravitational wave (GW)  bursts from the Galactic center \citep[][and references therein]{1973Natur.242..105H, 1974Natur.247..444E} claimed by
\citet[][]{1970PhRvL..25..180W}.   Dispersion delays between GW and radio bursts were considered in the analyses but the time resolutions were long  (1 and 10~s, respectively).   No GW-radio coincidences were found but astrophysical radio events were claimed to contribute to the overall results obtained by
\citet[][]{1973Natur.242..105H}.

The role of dispersion delays between any prompt radio emission from gamma-ray bursts \citep[][]{pal93} was built into response time goals and detection criteria for low-frequency observations triggered by GRB detections
\citep[e.g.][]{1975ApJ...196L..11B, 1996MNRAS.281..977D}.    Similar considerations about dispersion effects enter into multiwavelength/messenger studies of FRBs.

This incomplete but representative historical summary illustrates that the techniques used today for FRB studies have their precedents in long-ago experiments, albeit now with much greater sensitivity and insights into the properties of transient radio emission.

The timing of this  review is at an inflection point when FRBs  have been  well  established as an extragalactic phenomenon but there is not yet a deep understanding of the underlying astrophysics nor their population statistics.   This will change with the rise of high duty cycle, wide-field surveys that are just beginning to dramatically  increase the discovery rate.   The review necessarily excludes insights that will emerge from these surveys and their multiwavelength followup. However, topics in fundamental astrophysics, methods, and interpretation that we discuss will hopefully have sustained relevance.

In this review we summarize what is currently known about the FRB phenomenon, the source physics that may underly them, and their potential as tools for extragalactic astrophysics and extreme physics.  The following questions motivate the content and organization of this review:
Do all FRB sources repeat?
What is the FRB distance scale?
 Do all FRBs originate from  the same type of object?
What can FRBs be used for?
And finally, where will the FRB field be in the long term (10 yr)?

\section{SUMMARY OF THE FRB PHENOMENON}
\label{sec:synopsis}

\newcommand{\Nv}{N_{\rm v}}
\newcommand{\Nr}{N_{\rm r}}
\newcommand{\Nrone}{N_{\rm r_1}}
\newcommand{\nsfour}{{\ns}_4}
\newcommand{\etai}{\eta_{1_i}}

FRBs are found  using data   that are essentially the same as those used in   pulsar surveys, namely high time resolution spectra ($\sim 100~\mu s$)  with $\sim 1000$ frequency channels across a total bandwidth of hundreds of MHz.  The key difference is that pulsar surveys seek  periodic signals using Fourier methods, which become insensitive to periods not small compared to data spans and of course are completely insensitive to single pulses.  Although individual `giant' pulses from pulsars have long been a known phenomenon, from the mid-1970s on, researchers were largely focused   on finding relativistic binary pulsars for tests of General Relativity and millisecond pulsars for use in pulsar timing arrays to detect nanohertz gravitational waves (with exceptions of course).  Giant pulses from the Crab pulsar and a few other objects were well studied during this period (and to the present) but were considered a niche subject with ties to high-energy emission.

 Attitudes changed during the 1990s because of interests in finding radio counterparts to gamma-ray bursts (GRBs) and recognition that the discovery phase space for fast transients was essentially unexplored territory.   Discoveries of RRATs and FRBs followed directly from the decision to search for single pulses in pulsar survey data.   In other words,  FRBs were discovered because of attitude adjustment, not from  technological innovation. However, followup observations of FRBs (especially localizations) have required innovations.

\subsection{Numbers and Rates}

To date FRBs have been detected from over 80 distinct sources  in a variety of surveys (Table~\ref{tab:surveys_large}) since the original event from 2001 was reported in 2007\footnote{As of 2019 Feb 1; The {\tt FRBCAT} catalog is
at {\url{\http://www.frbcat.org}} \citep[][]{pbj+16}}.
Until recently, most FRBs were discovered predominantly at $\sim 1.5$~GHz, initially with the Parkes telescope followed by  the first non-Parkes FRB using  the  Arecibo telescope, and a few detections at $\sim 0.8$~MHz with  the Green Bank Telescope and UTMOST  telescopes (Table~\ref{tab:surveys_low}).  In the last year, the discovery rate has accelerated with the advent of widefield surveys using the ASKAP telescope in a ``fly's eye'' mode at $\sim 1.3$ GHz \citep[][]{smb+18}  and the CHIME cylinder array \citep[][]{2018ApJ...863...48T} in the 0.4 -- 0.8 GHz band.   The CHIME detections \citep{2019Natur.566..230C,2019Natur.566..235C} are the first to be found below 0.8~GHz and contrast with the non-detections of FRBs with the GBT at  0.35 GHz \citep[][]{2017ApJ...844..140C} and LOFAR at 0.15~GHz \cite[][]{kca+15}. Burst detections are made on the basis of matched filtering (see sidebar).

\begin{textbox}[h]
\section{Matched Filtering}
\label{sb:mf}
Burst detection is based on the principle of matched filtering.
% for which the optimal filter is a combination of the signal shape and the power spectrum of additive noise.
A general model
$I({\Xarray}, {\thetavec}) = a A({\Xarray}, \thetavec) + N({\Xarray})$
comprises a signal $A$,
a design matrix of  variables $\Xarray$ with dependence on
 a vector of parameters $\thetavec$,
a scale factor $a$,
and noise $N$.
If the noise is white,  the matched filter is the signal shape,
$A({\Xarray}, \thetavec)$.
In practice, some aspects of the signal are known (e.g. dispersion delays) while burst shapes are not, requiring
searches over a template bank of burst shapes. FRBs show  stochastic structure that includes spectral confinement less than observing bandwidths and temporal substructure, knowledge of which can provide the
basis for   detection algorithms with better sensitivity.

The detection test statistic is the cross correlation function (CCF) of the template and events are found by requiring it to exceed a threshold.   A Bayesian approach calculates the posterior PDF using priors and the likelihood function for the parameters $a,\thetavec$.

Dispersed bursts have the form
$I(t,\nu) = a A(\nu, t-\tdm(\nu)-t_0) + N(\nu, t)$,
where $\tdm$ is the dispersion delay and $A(\nu, t)$ is the shape in time and frequency.
The parameters for this model are $a, t_0, \DM$ and  width $W$.   The signal-to-noise ratio of the CCF maximum is
$A_{\rm max} \sqrt{W} / \sigma_N$ or, equivalently,  $\F/\sqrt{W}\sigma_N$ where $\F$ is the fluence (area of $A$) and $\sigma_N$ is the RMS noise.  In some cases, the true DM differs slightly  from those found by maximizing the CCF.

%The dispersion delay for a cold, tenuous, unmagnetized plasma is
%$$
%\tdm(\nu)
    %= \frac{e^2}{2\pi \melec c \nu^2} \DM
    %= 4.15~{\rm ms} \frac{\DM}{\nu^2}.
%$$

\end{textbox}

% Table of large FRB surveys
\begin{table}[h]
\tabcolsep7.5pt
\renewcommand{\arraystretch}{1.1} % General space between rows (1 standard)
\caption{Large Scale Surveys at 1.4 GHz  that Constrain FRB Population Estimates}
\label{tab:surveys_large}
\begin{center}
\begin{tabular}{|lccccl|}
%\begin{tabular}{lccccl}
\hline
%&&&&& \\
{\bf Telescope/Survey} & ${\F}_{\rm min}$  & $\Omegas T$      & $\Nfrb$   & $\Gamma_{\rm frb}$(note a)    & References \\
                 & ${\rm (Jy\ ms)}$  &  (deg$^2$ h)    &           & (sky$^{-1}$ day$^{-1}$) & \\
\hline
% telescope      Fmin   OmegaT      Nfrb        Rate                                References
%\vspace{0.05in}      % counterintuitively, this vspace affects the spacing of the next two lines \\
%%%%%%%%%%%%%%%%%%%
{\bf Parkes / all}${\rm ^b}$    & $2$   & $4400$    & 19    & $1.7^{+1.5}_{-0.9}\times 10^3$    & Bhandari et al. (2018) \\
%\\
\quad Parkes / HTRU(h)  & $2$   & 1549      & 9     & $2.5^{+3.2}_{-1.6}\times 10^3$    & Thornton et al. (2013), \\
                        &       &           &       &                                   &Champion et al. (2016) \\
%\\
%\vspace{0.05in}
\quad Parkes / HTRU(m)  & $2$   &  694      & 0     & $\lesssim 1.4\times 10^3$
	& \citet[][]{2014ApJ...789L..26P} \\
% & Petroff et al. 2014 \\
%\\
%Q\vspace{0.05in}
\quad Parkes / SUPERB   & $2$   & 1621      & 5     & $1.7^{+1.5}_{-0.9}\times 10^3$    & Bhandari et al. (2018) \\
		&&&&&								\citet[][]{2018MNRAS.473..116K} \\
\hline
%%%%%%%%%%%%%%%%%%%
{\bf Arecibo/PALFA}${\rm ^c}$ &&&&&           \\
\quad Outer Galaxy      &       &           &       &                                   & Spitler et al. (2014) \\
\quad\quad Main beam    & $0.065$ & 6.2     & 1     & $1.6^{+6}_{-1.5}\times 10^5$      & (FRB~121102)\\
\quad\quad Sidelobes    & $0.350$ & 29.7    & 1     & $3.1^{+12}_{-3.1}\times 10^4$     & (FRB~121102)\\
\quad Outer+inner  Galaxy &       &         &       &                                   & Scholz et al. (2016) \\
\quad\quad Main beam    & $0.057$ & 19.5    & 1     & $5.1^{+17.8}_{-4.8}\times 10^4$   & (FRB~121102) \\
\quad\quad Sidelobes    & $0.300$ & 93      & 1     & $1.1^{+3.7}_{-1.0}\times 10^4$     & (FRB~121102)\\
\quad Outer+inner Galaxy &       &           &       &                                   & Patel et al. (2018) \\
\quad\quad Main beam    & $0.044$ & 12.7    & 1     & $7.8^{+25.6}_{-7.6}\times 10^4$   & (FRB~141113) \\
\quad\quad Sidelobes    & $0.239$ & 60      & 1     & $1.6^{+7.5}_{-1.6}\times 10^4$     & (FRB~141113)\\
%%%%%%%%%%%%%%%%%%%
\hline
{\bf ASKAP/Fly's Eye}         & 29.2  & $5.1\times 10^5$ & 20 & $37\pm 8$                     & Shannon et al. (2018) \\
\hline
\end{tabular}
\end{center}
\begin{tabnote}
$^{\rm a}$ The mean FRB rate is $4\pi \times (180/\pi)^2 \times 24 \times N_{\rm frb} / \Omegas T$ but the rates given take into account fluence completeness (Keane \& Petroff 2015).
\hfil
\break
$^{\rm b}$ This line includes all Parkes observations reported in Bhandari et al.~(2018, their Table 5), which includes their FRB detections in addition to the 14 from the HTRU and SUPERB surveys.
\break
$^{\rm c}$ Arecibo values are for the subsurveys yielding FRB~121102 (Spitler et al.  2014 and Scholz et al. 2016)
or FRB~141113 (Patel et al. 2018).   The analyses consider detection in the main lobes of the 7-beam ALFA receiver or
in the sidelobes, which a larger solid angle at lower sensitivity.  The Spitler et al. analysis considers only
the subsurvey of the outer Galaxy, while the other analyses consider the inner and outer Galaxy subsurveys together.
% for detection in either
\end{tabnote}
\end{table}

When sky coverage and selection effects are taken into account,
the small  number of bursts detected from distinct sources
translates into an astoundingly large all-sky rate
$\Ratefrb(\rm > 1~Jy~ms) \sim 10^3 - 10^4$~sky$^{-1}$~day$^{-1}$
above a 1~Jy~ms fluence threshold\footnote{
1~Jy = $10^{-23}$~erg~cm$^{-2}$~s$^{-1}$~Hz$^{-1}$
= $10^{-26}$~watts~m$^{-2}$~Hz$^{-1}$.}
\citep[][]{tsb+13, sch+14, ssh+16b, cpk+16, kp15, 2016MNRAS.461..984O,  2017AJ....154..117L}.
Although different surveys yield rates  that vary by about an order of magnitude,  allowance for  survey thresholds,  sky coverage, and small number statistics yields general consistency.    The salient point  is that the FRB rate is
{\it large}, about $10^3$ times greater than the GRB rate for FRB fluences larger than 1~Jy~ms.

% Table of low frequency surveys
\begin{table}[h]
\tabcolsep7.5pt
\renewcommand{\arraystretch}{1.1} % General space between rows (1 standard)
\caption{Low Frequency Surveys }
\label{tab:surveys_low}
\begin{center}
\begin{tabular}{|lccccl|}
%\begin{tabular}{lccccl}
\hline
% &&&&& \\
{\bf Telescope/Survey} & $\Fmin$  & $\Omegas T$      & $\Nfrb$   & $\Gamma_{\rm frb}$    & References \\
                 & ${\rm (Jy\ ms)}$  &  (deg$^2$ h)    &           & (sky$^{-1}$ day$^{-1}$) & \\
\hline
% telescope      Fmin   OmegaT      Nfrb        Rate                                References
{\bf LOFAR/ARTEMIS}$^{\rm a}$   & $139 $   & $3.4\times 10^4$      & 0    & $< 29 $    & Karastergiou et al.\ (2015) \\
\hfill  0.145 GHz &&&&& \\
\hline
{\bf GBT/GBNCC}$^{\rm b}$ & $1.4 $        & 580  & 0    & $< 3.6\times10^3 $    & Chawla et al. (2017) \\
\hfill 0.35 GHz &&&&& \\
\hline
{\bf UTMOST}                & 11  & $3.8\times 10^4$ & 3 & $78^{+12.4}_{-0.57}$  &
\citet{2016MNRAS.458..718C} \\
\hfill  0.84 GHz &&&&& \\
\hline
{\bf CHIME}         & TBD & TBD & 13 &  NA
	&  CHIME/FRB Collab \\
\hfill  0.4 - 0.8 GHz &&&&& (2019a,b) \\
\hline
\end{tabular}
\end{center}
\begin{tabnote}
$^{\rm a}$ Reported $\Fmin = 62~{\rm Jy}  \times 5~{\rm ms}$ has been scaled to $W=1$~ms using $\Fmin \propto \sqrt{W}$.
\hfil\break
$^{\rm b}$ Reported $\Fmin = 0.63~{\rm Jy}  \times 5~{\rm ms}$ has been scaled to $W=1$~ms.
\end{tabnote}
\end{table}

The Galactic latitude dependence of burst detection rates is of high interest because it would
 implicate propagation effects, especially interstellar scintillation,  from  the Milky Way's ISM in the detectability of FRBs and estimated population sizes.  However, the
 empirical evidence for latitude dependence  is murky \citep[][]{cpo16}.
%\citep[][]{2016MNRAS.458L..89C}.
 Early analyses suggested a deficit of mid-latitude FRBs that might be associated with the latitude dependence of interstellar scintillation
\citep[][]{2014ApJ...789L..26P, mj15}, which can more favorably enhance high-latitude detections.  More recent analyses  corroborate or assume a latitude dependence
\citep[][]{2017AJ....154..117L, 2018MNRAS.474.1900M} while \citet[][]{2018MNRAS.475.1427B} argue against it.
This debate is based largely on studies of less than 20 objects from heterogeneous surveys.
FRBs detected in directions through the Galactic plane do not seem to imply a low-latitude deficit.
The repeating FRB~121102 \citep[][]{sch+14}  and a new candidate FRB~141113
\citep[][]{pab+18} were both found in the deep  Arecibo  pulsar survey (PALFA) covering a small field at low latitudes
$\vert b \vert < 1^{\circ}$ in the Galactic anticenter direction.
Detection of the latter FRB  implies a large rate if it was found in the main lobe of
the telescope beam,
$
\Ratefrb(>0.044~{\rm Jy}) = 7.8^{+27.2}_{-7.4} \times \ 10^4 ~{\rm sky^{-1}~ day^{-1}},
$
or a factor of $\sim 5$ smaller rate if instead the burst was found in a sidelobe.
Interstellar scintillation and other selection effects are discussed in Section~\ref{sec:ao}.

Recently,  a shallow  ``fly's eye'' survey with very wide angular coverage using the ASKAP telescope yielded  20 large-amplitude bursts at 1.3 GHz, implying a rate
$\Ratefrb(> 26(W/1.26~{\rm Jy \ ms})^{-1/2}) = 37 \pm 8$~sky$^{-1}$~day$^{-1}$ \citep[][]{smb+18}.
The survey's narrow range of Galactic latitudes,
$\vert b \vert = 50^{\circ}\pm 5^{\circ}$, minimized any  latitude dependence as a factor in survey results.
 Comparison with deeper surveys and application of a ${V / V_{\rm m}}$ test  both indicate a steep
fluence dependence of the rate, $\Ratefrb \propto F^{-2}$.   This contrasts with other studies that indicate
shallower dependences, $\Ratefrb \propto F^{-0.6}$
%(Vendantham et al.)
\citep[][]{vrhs16}
based on a heterogeneous set of bursts,
but is consistent with the analysis of \citet[][]{2018MNRAS.481.2320L}.
As with the latitude dependence, knowledge of the rate's dependence on fluence  is currently  limited by small samples of bursts whose positions within the telescope beam at the time of discovery are not known, leading to significant uncertainties on fluences.   Surveys with interferometric arrays that also localize bursts
\citep[][]{lbb+15}
%(e.g. Law et al.)
will resolve this issue.  We note that a previous fly's eye survey with the Allen Telescope Array using smaller antennas (5-m vs.\ 12-m diameter) and smaller aggregate on-sky time yielded no FRB detections \citep[][]{2012ApJ...744..109S}.

\subsection{Follow-Up Observations: Trials and Tribulations}

The directions of all FRBs  have been searched for repeat bursts and several have been investigated in comprehensive multiwavelength observations.   Followup observations from radio  to $\gamma$-ray energies include those made as soon as possible after a radio burst detection using Astronomical Telegrams and an alert system based on VOEvents now under development \citep[][]{2017arXiv171008155P}.  Panchromatic observations have yielded no burst detections and, apart from FRB~121102, no persistent counterparts
\citep[][]{pwt+14, pbb+15, ssh+16b, 2018AJ....155..227B, 2018MNRAS.475.1427B}.

Several FRBs have shown repeat bursts at radio frequencies from $\sim$0.4~to~8~GHz.  FRB~121102, discussed in detail in Section~\ref{sec:repeater}, was found to repeat \citep[][]{ssh+16a} about 2.5~y after its initial detection \citep[][]{sch+14}, but after only 10.3~h of total on-source time.

Some FRB lines of sight have been reobserved for more than $10^3$~h without any redetections
\citep[e.g.][]{pjk+15, smb+18},
leading  some authors to argue that most  FRBs differ in physical nature from FRB~121102.  If so, this would  sustain
  the prospect  that most FRBs are from   one-off catastrophic events rather than from objects with persistent  activity.   However, if most or all FRBs ultimately repeat, the time to repeat may vary significantly between sources, particularly when amplitude distributions, scintillations and lensing, and detection thresholds are taken into account.
To assess repeatability,  the number of statistical trials  in a large survey that yields multiple FRBs needs to be considered, and this depends on the (unknown) size of the source population in the sampled volume.
The number of reobservations needed for repeat detections may be very large, especially in shallow surveys.

Consider a survey that yields  $\Nd$ bursts from $\Nd$ distinct sources.   Each of the $M$ active sources
in the surveyed volume repeats with an average rate $\rateone$.   However, no repeats are detected even though each sky position is visited $\Nv \gg 1$ times for a time $T$ per visit.
The total number of detected events is $\Nd \sim \rateone  T \Nv M$.  Because
at most one event per source is seen in  $\Nv$ visits, we have  $\rateone  T\Nv \ll 1$.    The number
of reobservations $\Nrone$ needed on average to redetect a single source is given by $\rateone T \Nrone = 1$.
Using the survey yield, $\Nrone \sim \Nv M / \Nd$.    But to  have detected a single repeat  from {\it any one} of
the $\Nd$ sources requires $\rateone T \Nrone \Nd \sim 1$, which gives
the number of repeats needed (per source) $\Nrone \sim \Nv M / \Nd^2$.

For the ASKAP survey \citep[][]{smb+18}, $\Nd = 20$, $T \sim 0.93$~h and $\Nv \sim$~17 to 1308, corresponding to 16 to 1200~h of reobservations.
Using the median $\Nv = 570$, the required number of reobservations  to see
a single repeat is $\Nrone \sim 1.4M$.    A plausible fiducial population size sampled in the ASKAP survey
is $M = 10^4 M_4$.  \citet[][]{2017ApJ...843...84N}, for example,  estimate a population number density $\ns \sim 10^4 \nsfour$~Gpc$^{-3}$ and the ASKAP survey may have sampled  a volume of $\sim 1$~Gpc$^3$.  This implies that a much  larger number  of reobservations $\sim 10^4M_4$ instead of the median 570 or the maximum $\sim 1308$ reported by \citet[][]{smb+18}  is needed to  expect any one of the ASKAP FRBs to have repeated in ASKAP observations.

Of course higher sensitivity telescopes can significantly reduce the time-to-redetection.
From the survey, the  implied  burst rate per source is
$\rateone \sim \Nd / T \Nv M \sim 3.8\times 10^{-6} M_4 $~h$^{-1}$,
or about $0.033M_4$ bursts per year, a very small rate.
For a differential burst amplitude distribution $\propto S^{-\beta}$ for sources distributed uniformly in Euclidean space ($\beta=5/2$), scaling from the ASKAP survey to the Parkes surveys  and assuming detection  thresholds are  bracketed by the distribution's cutoffs ($S_1 \ll S_{\rm A, P} \ll S_2$),  we obtain a predicted rate for  Parkes observations $\rateone({\rm P}) = \rateone({\rm A}) (S_{\rm A} / S_{\rm P})^{\beta-1} \sim 0.2$ - $1.7\times 10^{-3}$~h$^{-1}$ (for the nominal threshold or the `fluence complete' threshold, respectively),
compared to a rate using the Arecibo telescope $\rateone({\rm AO}) \sim 0.065$~h$^{-1}$.   These rates imply roughly 30~y, 600 - 4800~h, and 15~h of reobservation between detections for the ASKAP, Parkes, and Arecibo surveys respectively.
Parkes (let alone ASKAP) reobservations have not reached the required time-to-redetection values, whereas the first repeat burst from FRB~121102 was found after 10.3~h of on-source time spread over $\sim 2.5$~y of elapsed time in Arecibo followup observations.    {\it The rates and repeat times estimated here are therefore consistent with sources distributed uniformly in  Euclidean space that all produce multiple bursts. }
The possibility that all FRBs repeat removes a major argument for the conjecture that  there are multiple populations of FRBs \citep[][]{2018ApJ...854L..12P}.

Some caveats on these estimates are needed. First, calculated yields assume all observations are statistically independent. This is not the case if burst rates or amplitudes are variable with correlation times longer than a typical observation time $T$.   Episodic detections are expected if the correlation time is between $T$ and the total span of observations on any FRB source. This is the case for FRB~121102 but it is not yet known if rate variations are intrinsic or due to propagation effects; this is discussed further in later sections. If $\etai$ is the intrinsic, Poisson burst rate per source and a large modulation lasting $W_{\rm g}$ occurring at intervals $T_{\rm g}$ is required to produce detectable bursts, the propensity for FRBs to occur singly (except for FRB~121102) implies $\rateone W_{\rm g} < 1$ and the {\it apparent burst rate} is $\rateone = \etai W_{\rm g} / T_{\rm g}$.  For the repeater,  $\rateone$ during episodes lasting $\sim$~days is much larger than the apparent rate, signifying that some kind of modulation is active that yields a variable  mean burst rate (which may or may not correspond to
Poisson statistics).

% discussed further in \S~\ref{sec:repeater}.

\newcommand{\td}{t_{\rm d}}

\subsection{Dispersion and Scattering of FRBs}

The  arrival times of FRBs are inversely proportional to the line-of-sight integral of the group velocity.
For a magnetized plasma the  leading terms in the frequency-dependent part of the arrival time are
\citep[e.g.][]{1968Sci...160..760T, 2014MNRAS.441L..26T, sc19}
%(e.g. Tanenbaum et al. 1968; Tuntsov 2014;  Suresh \& Cordes 2018)
\be
t(\nu) =
	4.15~{\rm ms} \left(\frac{\DM}{\nu^2}\right)
	 \pm 28.6~{\rm ps} \left(\frac{\RM}{\nu^3}\right)
	 +  0.251~{\rm ps} \left(\frac{\EM}{\nu^4}\right),
	 {\rm ~for~\nu~in~GHz}
\label{eq:tnu}
\ee
where  terms are included   up to second order in $(\ompe/\omega)^2$ and linear in $\omcye/\omega$ (where $\ompe$ and $\omcye$ are the electron plasma and cyclotron frequencies, respectively).
Each term has an associated line-of-sight  integral measure. First is the   dispersion measure \DM\ (defined previously) with standard units of $\DMunits$.  The second term includes the Faraday rotation measure $\RM = 0.81 \int ds \, \nelec \Bpar$  with standard units of $\RMunits$ when the electron density $\nelec$ is in cm$^{-3}$ and the parallel (to the line of sight) magnetic field is in microgauss units.  The third involves the emission measure, $\EM = \int ds \, \nelec^2$, with standard units of $\EMunits$. The two signs of the second term correspond to the two hands of circular polarization.

Early analyses of  pulsars
\citep[][]{1968Sci...160..760T}
  and FRBs tested arrival times against the  dispersion law  $t(\nu) \propto \nu^{-\beta}$ and found $\beta = 2$ to within one percent or better
 \citep[][]{tsb+13, sch+14, ssh+16b, cpk+16, kp15, 2017AJ....154..117L}.
  % (Thornton et al. 2013; Spitler et al. 2014; Champion et al. 2016; others? Scholz et al. 2016?),
The resulting upper limits on the $\nu^{-3}$ and $\nu^{-4}$ terms and the absence of  free-free absorption (associated with EM) ruled   out the association of FRBs  with very dense plasmas
\citep[][]{lg14, 2014MNRAS.441L..26T, 2014MNRAS.443L..11D, 2014arXiv1409.5766K}, such as stellar envelopes
\citep[][]{lsm14}.   However, future observations of FRBs with large RMs may show  distorted burst shapes at low frequencies
$\nu \ll 1$~GHz  due to the  birefringent  delays for the two hands of circular polarization.

\begin{figure}[h]
\centerline{
\includegraphics[width=3.0in]{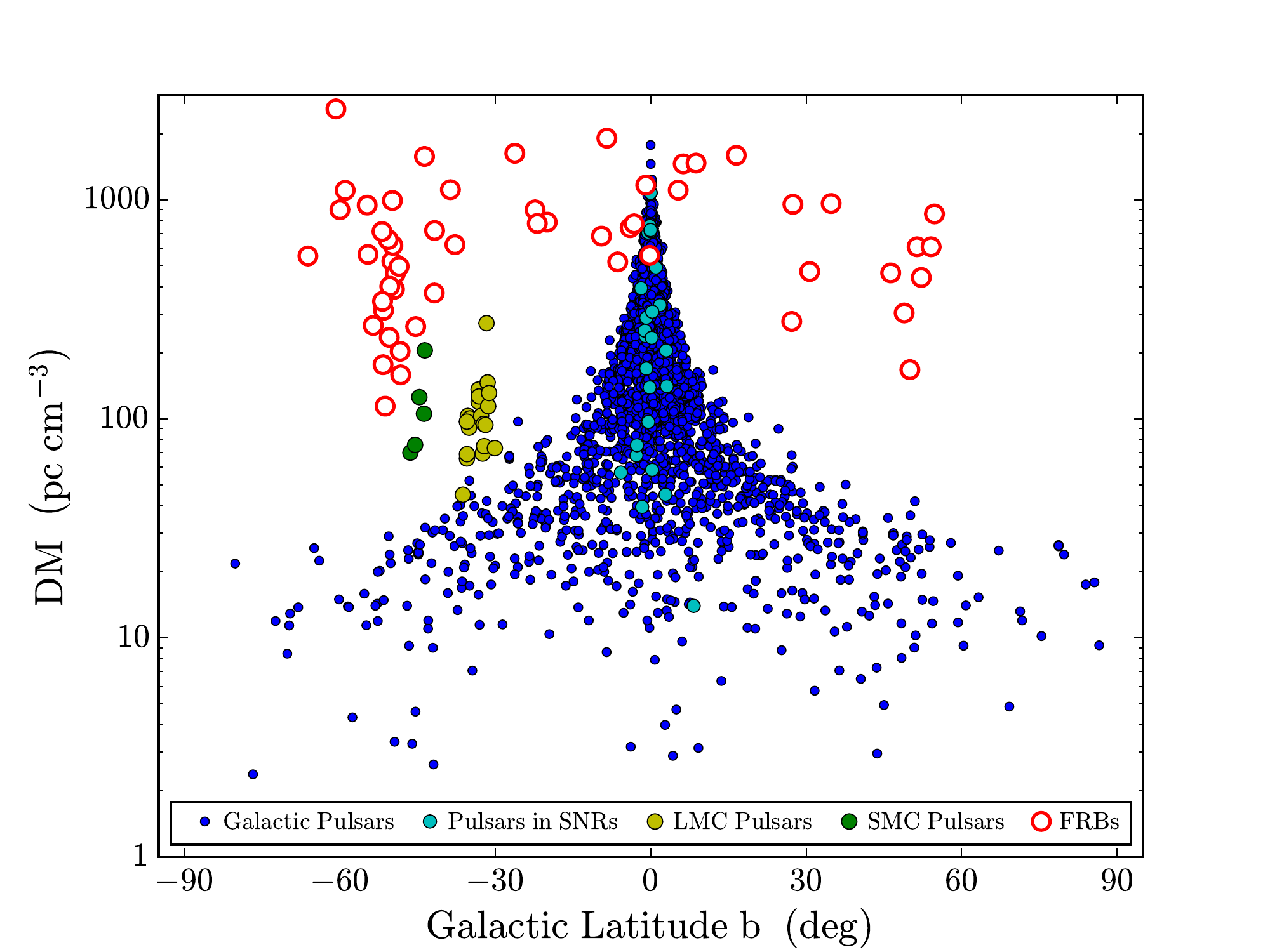}
\hspace{-2.5in}
\includegraphics[width=3.0in]{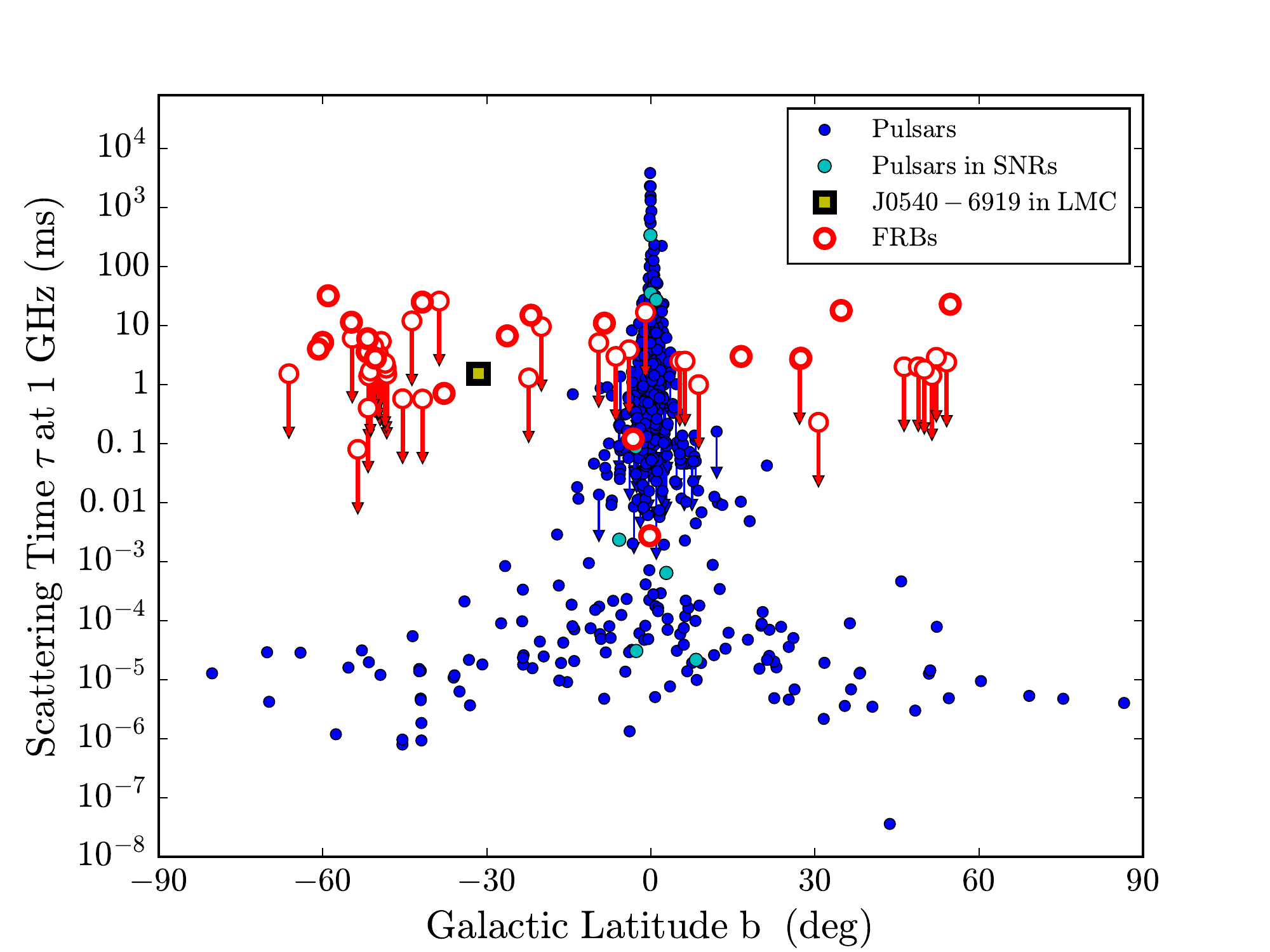}
}
\caption{
Left:
Dispersion measures plotted against Galactic latitude for pulsars and FRBs.  Different symbols are used  for Galactic pulsars (2422 objects), Galactic pulsars associated with supernova remnants (27), pulsars in the Large Magellanic Cloud (LMC, 21) and Small Magellanic Cloud  (SMC, 5), and FRBs (55). DM measurements and
pulsar associations were obtained from
\citet[][{\tt http://www.atnf.csiro.au/research/pulsar/psrcat}]{PSRCAT}.
Right:
Scattering times for pulsars and FRBs at 1~GHz plotted against Galactic latitude.
There are 421 pulsar measurements and 93 upper limits on $\taud$ compared to 18 FRB measurements and 37 upper limits.
\label{fig:dm_vs_b}
\label{fig:taud_vs_b}
}
\end{figure}

% Dispersion measures:

Figure~\ref{fig:dm_vs_b}  (left panel) shows  DMs plotted against Galactic latitude $b$
for FRBs and for pulsars  in the Milky Way and in the Magellanic Clouds (LMC, SMC).
Two conclusions can be made from the figure.
First,  the DMs of all FRBs with $\vert b \vert > 10^{\circ}$  are much larger than the outer envelope  for Galactic pulsars that approximately  follows a $\csc \vert b \vert$ dependence.
An extragalactic population of FRBs would appear just this way if the total DM includes  a large extragalactic contribution.
Second, the DMs of  FRBs cover a total range $\sim 100$ to $2600~\DMunits$  that is comparable to the range
for pulsars (3 to 1700~$\DMunits$), which is clearly due to the ISM of the Galaxy and in a few cases the ISM in
the Magellanic clouds.   The extragalactic contributions for the smallest DMs  are equal to those
expected from a dwarf galaxy, as indicated by the excesses seen in Figure~\ref{fig:dm_vs_b} for pulsars in
the  Magellanic clouds.
The largest DMs are comparable to those expected from either a long path through the IGM or  a galaxy disk,  from a galactic center like that of  the Milky Way, or from  a young supernova remnant \citep[][]{pir16}.
%\citep[][]{2016ApJ...824L..32P}.
Ionized gas in galaxies is therefore  a plausible source for some or most of the extragalactic part of DM. We discuss the relative contributions to DM from host galaxies and the intergalactic medium
(IGM) in Section~\ref{sec:distances}.

% Pulse broadening:

The right-hand panel of Figure~\ref{fig:dm_vs_b} addresses FRB scattering.
Temporal broadening of FRBs  results from  small-angle scattering by electron density variations
on scales much larger than a wavelength.
The scattered burst shape is the convolution of the emitted  burst $\F(t)$ with an asymmetric  pulse broadening function  $\PBF(t)$,
%\be
$\Fs(t) = \F(t) * \PBF(t)$.
%\ee
A one-sided exponential $\PBF(t) = \tau^{-1} \exp(-t/\tau) \Theta(t)$
is often used for modeling of measured pulses but
is a special case for thin scattering screens that only approximates realistic broadening functions.
The scattering time is a strong function of frequency, $\tau \propto \nu^{-4}$.
% resulting from thick scattering regions that may scatter according to  a
% particular wavenumber spectrum for electron density variations.

The figure shows scattering times $\tau$ scaled to  1~GHz vs. Galactic latitude for  both pulsars and FRBs.
%The data used in the figure are
%based on pulse broadening and scintillation bandwidth measurements scaled to 1~GHz.
Pulsar scattering times span more than ten orders of magnitude. The measured scattering times of FRBs, like their DMs,  are  also within the range spanned by pulsars but they  are   much larger than those of pulsars at similar Galactic latitudes in most cases.  This too is consistent with FRB  scattering occurring primarily from extragalactic gas, at least for FRBs detected so far.    However, only about
 30\% of the detected bursts show scattering.
Section~\ref{sec:distances} discusses properties of the extragalactic plasma that underly FRB scattering.

\subsection{Time-Frequency Burst Structure}

The earliest reported FRBs showed relatively simple temporal morphologies: Gaussian-like pulses modified in some cases by scattering broadening \citep[][]{lbm+07, tsb+13, sch+14} with temporal substructure in one case \citep[][]{cpk+16}.    The present understanding is that featureless bursts are in part the outcome of the limited time resolution of post-detection dedispersion used in early surveys.  Recent work enabled by coherent dedispersion of repeat pulses from FRB~121102 has revealed rich \tnu\ structure that differentiates some FRB bursts from their pulsar analogs.   Examples are shown in Figures~\ref{fig:bursts2} and \ref{fig:bursts3}.
The \tnu\ structure of FRBs is therefore substantially different from that of single pulses from pulsars, which tend to show only Galactic DISS but are otherwise continuous across a wide spectrum.

Frequency structure is best studied for the repeater FRB~121102 and is described in detail
in \citet[][]{2018arXiv181110748H}.
With adequate signal-to-noise ratio,  bursts from several FRBs show bandlimited structures of a few hundred MHz sometimes combined with narrower frequency structure, which appears to be consistent with Galactic scintillation (DISS). The broader structure is not stable between bursts from the repeater, even changing over time separations of seconds and minutes.   The broad structure appears anywhere in the 1.2 to 8~GHz frequency range used for studies of FRB~121102 though rarely in two broad receiver bands observed simultaneously \citep[][]{lab+17}.
%(Law et al. 2017).
Whether the broad structure is intrinsic to the radiation process or a post-emission propagation effect near the source
\cite[e.g.][]{cwh+17} is yet to be determined.

\subsection{Polarization}
\label{sec:polarization}

Stokes parameters are available for a relatively small subset of FRBs (Table~\ref{tab:compare}).
{\tt FRBCAT} (currently) gives polarization information on five FRBs, with four showing significant linear polarization
ranging from 8.5\% to 80\% and three showing circular polarization from 3\% to 23\%.
The repeating FRB~121102  shows 100\% polarization after removing Faraday rotation
and FRB~180301  has $\sim 30$\% linear polarization and $\sim 70$\% circular polarization.  These mixtures of linear and circular polarization are not dissimilar from those seen from  pulsars.
Polarization angles rotate across some FRBs by $\sim$tens of degrees (FRBs 110523 and 150418) while remaining constant in time for others to less than 20~degrees for FRBs~121102, 150215, and 150807.  Pulsars generally show rotation across their pulses,  often showing consistency with relativistically beamed emission from a spinning dipolar field \citep[][]{1976Natur.263..202B}.
It is unclear  if the position angles of  FRBs indicate  that the durations of FRBs are unrelated to  a similar spinning  radiation beam, that  emission comes from non-spinning objects, or that polarization is induced as a propagation effect.

% Table of FRB polarization properties
\begin{table}[h]
\tabcolsep7.5pt
\caption{Polarization of Fast Radio Bursts}
\label{tab:compare}
\begin{center}
\begin{tabular}{|   l  | llrccl  |}
\hline
\hfil FRB & \% Linear  & \% Circular & RM\quad\quad   & DM \quad\quad & $\Delta\psi$ & Reference\\
               &                  &                    & $({\rm rad \ m^{-2}})$  & $({\rm pc \ cm^{-3}})$  & (deg) & \\
\hline
%  FRB  	    L(\%)			V(\%)		         RM			         DM		$\Delta\psi$	Comment / Ref
110523 &	$44\pm3$    &		$23\pm30$   &	$-186\pm14$         &	623 & $\sim 40$         & Masui et al. 2015 \\
121102 &	$100$       &		$0$         &	$1.03\times10^5$    &	560 & $< 10$            & MJD 57747 Michilli et al. 2018 \\
       &                &                   &   $0.93\times10^5$  &       & $< 10$		        & MJD 57991 \\
140514 &	$0\pm10$    &		$21\pm7$    &		$ ...$          &	563 & $ ...$            &
	\citet[][]{pbb+15} \\
	%Petroff et al. 2015a \\
150215 &	$43\pm 5$   &	$3\pm1$         &		$1.5\pm 10.5$   &	1106& $<20$             &
	\citet[][]{pbk+17} \\
	%Petroff et al. 2017 \\
150418 &	$8.5\pm1.5$ &	$0\pm4.5$       &	$36\pm 52$          &	776 & $\sim 70$         &
	\citet[][]{2016Natur.530..453K} \\
	%Keane et al. 2016 \\
151230 &	$35\pm13$   &	$6\pm11$        &		$0$ &				960 & $...$             &
	\citet[][]{2018MNRAS.478.2046C} \\
	%Caleb et al. 2018 \\
150807 &	$80\pm1$    &		$...$       &			$12\pm0.7$  &	267 & $< 20$            &
	\citet[][]{rsb+16} \\
	%Ravi et al. 2016 \\
160102 &	$84\pm15$   &	$30\pm11$       &	$-221\pm6$          &	2596& $\lesssim 10$     &
	\citet[][]{2018MNRAS.478.2046C} \\
	%Caleb et al. 2018 \\
180301 &	$\sim 30 $  &   $\sim70$		&	$-3100$             &	520 & $\lesssim 20$     & Price et al. 2018 \\
\hline
\end{tabular}
\end{center}
\begin{tabnote}
%$^{\rm a}$ FRB pseudo luminosities based on 1~Gpc distances for all objects.
%$^{\rm b}$second table footnote.
\end{tabnote}
\end{table}

 Four objects in the catalog have quoted RM values of which three are significant but only one,  FRB~110523 \citep[][]{mls+15b},  has an RM value that is consistent with arising from propagation through a host galaxy disk. The total $\RM = -186\pm14~\RMunits$; only about 18 and 6~$\RMunits$ are from the Milky Way and IGM, respectively.
 The repeating FRB~121102 stands out by showing an  extraordinarily large RM $\sim 10^5~\RMunits$, which requires narrow frequency channels to resolve  rotation of $\psi$  with frequency.  Initial studies at $\sim 1.4$~GHz showed no linear polarization because of Faraday depolarization across the coarse frequency channels. Only higher frequency observations allowed the Faraday rotation to be resolved.   A final case is $\RM = -3100~\RMunits$ for FRB~180301
 \citep[][]{2018ATel11376....1P}.

The wide range of  RMs for FRBs is similar, perhaps coincidentally, to the range seen for Galactic pulsars, with the largest value (in magnitude) seen from the Galactic-center (GC) magnetar J1745$-$2900, $\RM \approx -0.7\times 10^5~\RMunits$.   And perhaps not so coincidentally, the RM of both the GC magnetar and FRB~121102 have decreased in magnitude  by significant amounts over periods of a few years: 5\% for the magnetar \citep[][]{2018ApJ...852L..12D} and 30\% for FRB~121102 \citep[][and ongoing work]{msh+18}.  For both objects the accompanying change in DM is very small ($< 1\%$).

\subsection{Localizations}

As pointed out by various authors \citep[e.g.][]{ebwb18,vrhs16}, sub-arcsecond localizations are required to identify  host galaxies associated with FRBs at $\sim$Gpc distances. Rapid multiwavelength follow-up to detect the analog of GRB afterglows has not been fruitful  \citep[e.g.][]{pbk+17}, and the claimed  rapidly fading radio transient associated with FRB~150418 \citep{kjb+16} was shown instead to be common AGN variability \citep[e.g.][]{wb16}. In fact, multiwavelength observations that were simultaneous with burst detections from FRB~121102 have led to upper limits on high energy and optical emission associated with the bursts \citep{sbh+17,aaa+18}.

The only reliable method so far is direct interferometric localization of the burst itself, as demonstrated for FRB~121102 \citep{clw+17,mph+17}.
But for FRBs with small extragalactic contributions to their DMs,  the number of candidate host galaxies
in the error circles with large diameters (e.g. multiple arc minutes) may be small enough for identification
of the FRB's host \citep[see, e.g.][]{mbb+18}.
%But note that with low enough extragalactic DM, there are very few candidate host galaxies encompassed within %the plausible detection volume \citep[see, e.g.][]{mbb+18}
%, Mahoney et al. ({\bf No ref yet}).

\subsection {Energetics}
\label{sec:energetics}

With peak flux densities similar to those of pulsars, FRBs originating from  $ \sim$~Gpc distances compared to $\sim$ kpc pulsar distances imply energy densities at the source and  total burst energies that are larger by factors  $\sim 10^{10}$ to $10^{14}$. For a  flux density $\Snu(t)$ in a bandwidth $\Delta\nu$, the energy density scaled to a distance
 $r = 10^{10}\,r_{10}\,{\rm cm}$ from the source is
 \be
 U_{\rm r, s} \sim  \frac{\Snu \Delta\nu}{c} \left( \frac{\dso}{r}\right)^2
 		\approx 3.2\times 10^{10}\,{\rm erg \ cm^{-3}}
		{\Snu}_{\rm, Jy} \Delta\nu_{\rm, GHz}
		\left(\frac{{\dso}_{\rm, Gpc}} {r_{10}}\right)^2.
 \ee
 The equivalent magnetic energy  $U_B = B^2 / 8\pi$ requires a field strength,
 \be
 B \sim \left( \frac{8\pi\Snu\Delta\nu}{c}\right)^{1/2} \frac{\dso}{r}
 	\approx 9\times10^5~{\rm G}\, \left({\Snu}_{\rm, Jy} \Delta\nu_{\rm, GHz}  \right)^{1/2} {\dso}_{\rm, Gpc},
 \ee
 that would be encountered, for example,  at a distance $r$ from a magnetar with a surface field
 $B = 10^{15} B_{15} $~G and radius $R = 10^6 R_6$~cm,
  \be
 r \approx  3.3\times 10^8~{\rm cm} \, R_6^{3/2} (B_{15} / {\dso}_{\rm, Gpc})^{1/2} ({\Snu}_{\rm, Jy} \Delta\nu_{\rm, GHz})^{-1/4}.
 \ee
 Expressed in terms of the velocity of light cylinder radius $\rlc = cP/2\pi$ of a spinning object with period $P$,
 \be
 \frac{r}{\rlc} = 0.07 P^{-1/2} R_6^{3/2} (B_{15} / {\dso}_{\rm, Gpc})^{1/2} ({\Snu}_{\rm, Jy} \Delta\nu_{\rm, GHz})^{-1/4}.
 \ee

 To match or exceed the  radiation energy density  with a particle energy density
 $U_{\rm p} = \gamma \melec c^2$,  electrons would have to be highly relativistic even with a  large electron density.
  For example,  a Lorentz factor $\gamma = (1 - \beta^2)^{-1/2} = 10^7$ (with $\beta = v/c$) requires
  an electron density $\nelec \approx 4\times 10^9$~cm$^{-3}$ for the same parameters as in the above equations.
 The single-particle or particle-bunch  radiation is therefore highly beamed into a solid angle $\Omega_{\rm b} \sim \gamma^{-2}$.   However, the total solid angle for an FRB is  much larger than this because bursts are incoherent sums of many individual coherent units of radiation \citep[][]{cw16}.

Isotropic emission implies a total emitted energy obtained by  integrating   over  a spherical shell of thickness $c W$.  Correcting  for the beaming solid angle gives the burst energy
 $
 E_{\rm b} \sim 4\pi \Snu W \Delta\nu \dso^2  \left(\Omega_{\rm b} /  4\pi \right)
 	\approx 1.2\times 10^{39} \, {\rm erg} \, {\Snu}_{\rm, Jy} W_{\rm ms}\Delta\nu_{\rm, GHz} {\dso}_{\rm, Gpc}^2
	\left(\Omega_{\rm b} / 4\pi \right).
 $
 % dtheta = 2\pi W / P
 % Omegab = (pi/4) dtheta^2 for dtheta << 1
 % Generally Omegab = 4\pi \sin^2(\pi W/2P)
 % Omegab / 4pi = \sin^2(\pi W/2P) \sim (\pi W / 2 P)^2.
Small beam solid angles can therefore  substantially alter burst energies.

\subsection{Are Bursts From Rotation, or Temporal Modulation?}

If a rotating beam causes observed burst widths with duty cycle $W/P\le 1$, the beam solid angle (in units of
$4\pi$) satisfies  $\Omega_{\rm b} /4\pi  =  \sin^2 (\pi W/2P) \le 1$.  Small duty cycles imply $\Omega_{\rm b} /4\pi \ll 1$, thus reducing energetic requirements for a burst.  For this to be the case,  pulse widths
$W = 1$~ms  require the spin period to exceed $P \gg 1.57\,{\rm ms} \,W_{\rm ms}$ to reduce the solid angle significantly.   The total radiated energy also  depends on the duration of radiation in the rotating beam.  To avoid seeing multiple, periodic  bursts, the duration must be less than a spin period, as for the repeating FRBs,
indicating that there is  substantial modulation of coherent radiation in the rotating frame.   This also suggests that the {\it observed} burst durations   themselves may be   from temporal modulation  rather than from a rotating beam.    In this case,  the beam solid angle cannot be constrained directly from observations.

\subsection{Fast Transients, Brightness Temperatures, and Coherent Radiation}
\label{sb:brightness}

The radiation brightness temperature  $T_{\rm b}$  is often used to characterize radio emission from astrophysical objects and it is particularly useful for distinguishing incoherent and coherent emission.
It is the effective blackbody temperature based on the Rayleigh-Jeans portion of the Planck spectrum $I_\nu = 2k\Tb/\lambda^2$, where $k$ is Boltzmann's constant and $\lambda=c/\nu$ is the wavelength. For a transient burst of duration $W$ and peak flux density $\Spk$, the specific intenisty is $I_\nu \sim \Spk / \Omegas$ where $\Omegas$ is the observed solid angle of the source. A  burst source of size $\sim cW$ at distance $d$ subtends $\Omegas \sim(cW / d)^2$, giving a brightness temperature $\Tb \sim \Spk d^2 / 2k(\nu W)^2$. A 1~Jy FRB of millisecond duration yields $\Tb = 3.4\times 10^{35}$~K, compared to $\Tb = 3.4\times 10^{23}$~K for a Galactic pulsar at a kpc distance.

Thermal sources (stars, HII regions) yield brightness temperatures equal to their physical temperatures.   Non-thermal but incoherent emission such as  synchrotron emission from active galactic nuclei (AGNs) yields
$\Tb$ as large as $\sim 10^{12}$~K indicating electron energies
$k\Tb = 86$~MeV.  AGN radio emission is limited to about this  brightness temperature
by inverse Compton scattering.

\begin{figure}[h]
\centerline{
\includegraphics[width=\textwidth]{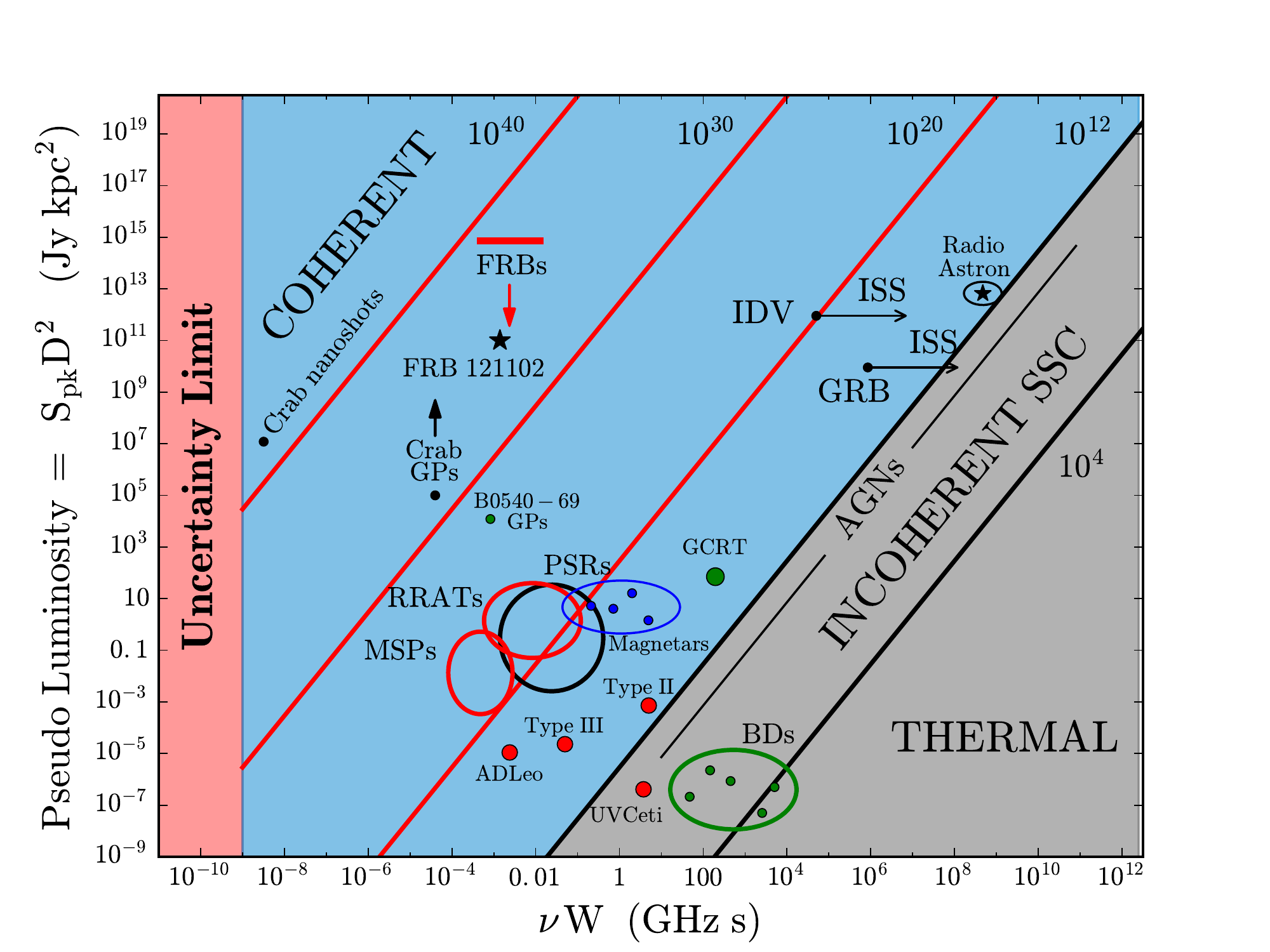}
}
\caption{Time-luminosity phase space for radio transients showing the product of peak flux $\Spk$ in  Jy and the square of the distance $D$ in kpc vs. the product of frequency $\nu$ in GHz and pulse width $W$ in s.
The ``uncertainty'' limit on the left indicates that $\nu W \gtrsim 1$ as follows from the uncertainty principle. Lines of constant brightness temperature $\Tb = S D^{2}/2k(\nu W)^{2}$ are shown, where $k$ is Boltzmann's constant.
Points are shown for the nanoshots \citep[][]{he07} and giant pulses detected from the Crab pulsar and a few millisecond pulsars, and single pulses from other pulsars.
Points are shown for  solar bursts, radio flares from stars,  brown dwarfs,  and AGNs.
The regions labeled `coherent' and `incoherent' are separated by the canonical $\sim 10^{12}$K limit for the synchrotron self-Compton process occurring in AGNs.
Arrows pointing to the right for the GRB and intra-day variable (IDV) points indicate that interstellar scintillation (ISS) implies smaller brightness temperatures than if characteristic variation times are used to estimate the brightness temperature.
Fast radio transients include rotating radio transients \citep[RRATS][]{mll+06}
(RRATs; McLaughlin et al. 2006), the Galactic center transient source, GCRT J1745-3009
\citep[][]{hlk+05} and radio emission from Galactic magnetars \citep[][]{ok14}.
\label{fig:phasespace}
}
\end{figure}

Figure~\ref{fig:phasespace} shows  the location of FRBs in the  phase space for radio transients with  a luminosity-like quantity\footnote{$\Lp$ is usefully called the pseudo luminosity in pulsar contexts to emphasize that the measured flux density is influenced by both the angular width of the beam and its direction with respect to the line of sight.}
 $\Lp = \Spk d^2$ in Jy~kpc$^2$ plotted against the dimensionless duration $\nu W$ in gigahertz-seconds; these axes allow lines of constant brightness temperature to be drawn. The region of coherent sources is designated to the left of the $\Tb = 10^{12}$~K line that represents the approximate synchrotron-self-Compton limit for AGNs.

FRBs and pulses from pulsars necessarily involve coherent radiation and  display propagation effects,  including dispersive propagation, as described above, and DISS --- the analog to optical scintillation of stars due to turbulence in Earth's atmosphere --- caused by turbulence in the interstellar electron density.    Dispersion and scintillation  are co-features  because coherent sources are typically compact, allowing radiation to have short-enough durations to show dispersive propagation as well as scintillations\footnote{Interstellar scintillation requires sources to be more compact than a critical (isoplanatic) angle in the same way that stars twinkle but planets do not, typically. Instead, pulsars and FRBs scintillate, but AGNs do not.}.
Coherent radiation mechanisms involve large numbers of particles ($N$) emitting with a distinct phase relationship, yielding collective power $\propto N^2$ rather than $\propto N$ for incoherent radiation. This can be through an antenna type mechanism with charge bunching in coordinate space or through a plasma maser involving a non-monotonic charge distribution in momentum space.

The signal processing of fast transients includes {\it dedispersion} that removes frequency-dependent delays
to improve burst detection probabilities and to
potentially restore bursts to their emitted shapes.
Two  dedispersion algorithms are used.
%as illustrated in Figure~\ref{fig:dedispersion}.
The first, {\it `post-detection' dedispersion},  is approximate and shifts intensities by the dispersion delay for the center frequency  of each channel of a digital filterbank.
%, as illustrated in the middle panel of Figure~\ref{fig:dedispersion}.
The best time resolution obtainable with this method at $\nu = 1.4$~GHz is
$\Delta t (\mu s) = 2 \sqrt{8.3\DM / \nu^3} \approx 110~\mu s$ for $\DM = 10^3~\DMunits$  \citep[][]{cm03}, which is insufficient for probing burst time structure.
The second method is {\it coherent dedispersion} \citep[][]{hr75} that applies an exact matched filter to `voltage' data proportional to sampled electric fields.  It corrects the $e^{ik(\nu)z}$ phase factor imposed by propagation and can provide sub-microsecond resolution.

\section{THE ASTRO-OPTICS OF FRBs}
\label{sec:ao}

The detectability of FRBs and their observed properties  are strongly affected by  propagation  through intervening plasmas and mass assemblies.
We  summarize propagation phenomena that affect FRB surveys and also how they can be used to probe FRB sources, their environments, and the IGM, including dark matter.

\subsection{Galactic Propagation}

Electron density variations  $\dne$  in the ionized ISM cause three important effects: angular broadening (`seeing'), pulse broadening due to angular broadening, and intensity scintillations from both refraction and diffraction.
Length scales smaller than the Fresnel scale $\sim \sqrt{\lambda d /2\pi} \sim 10^6$~km
diffract radiation  while refraction, caustics, and multiple images  from  larger scales can strengthen and
dim bursts and affect arrival times.  For reviews see \citet[][]{rickett90, 1992RSPTA.341..151N}.

An extragalactic burst viewed through the Galaxy's ISM is  broadened into an angular
diameter $\thetaG$ and temporally smeared by $\tauG \sim \dlo \thetaG^2 / 8 \ln 2 c$,
where $\dlo \sim \Lg$ is the distance of the scattering layer from the observer, approximately the Galactic scale height $\sim 1$~kpc  of free electrons.
The upper envelope of values and
the latitude dependence of $\taug$ in Figure~\ref{fig:taud_vs_b} correspond to a Galactic
seeing diameter,
$\thetaG \sim 0.3~{\rm mas} \times \sin \vert b \vert^{-3/5} \nu^{-2.2}$, where the exponents are
 for a medium with a Kolmogorov wavenumber spectrum for $\dne$.   Galactic seeing is important for  any gravitational `nanolensing' of FRBs
 \citep[e.g.][]{2017ApJ...850..159E}
because it may exceed the Einstein radius for dark-matter objects and contaminate gravitational time delays.

Intensity scintillations (DISS) are correlated over a  bandwidth
$\dnud$ related  to $\taud$ by an  `uncertainty'  relation
$2\pi\dnud\taud \approx 1$  \citep[][]{cr98, lr99}.   These  scintillations are accompanied by refractive scintillation (RISS), though typically DISS is more important.   In the strong scintillation regime where
$\dnud \ll \nu$, DISS is  100\% modulated with an exponential gain PDF, $\pdfg(g) = \exp(-g) \Theta(g)$ where $\Theta(g)$ is the Heaviside function (0 for $g  <0$, 1 otherwise).

\subsection{Propagation Model for Bursts}

The simplest model for an individual burst  relates the  emitted (pseudo) luminosity
$\Lp(\nu, t)$ to the measured flux density $S(\nu, t)$ using the distance $d$,  a propagation modulation $g(\nu, t)$,
and a  composite delay $t_{\rm d}(\nu, t)$ primarily from dispersion, scattering, and refraction,
\be
S(\nu, t) = d^{-2} g(\nu, t) \Lp(\nu, t - t_{\rm d}(\nu, t)).
\label{eq:Smodel}
\ee
Many of the observed properties of FRBs are likely a combination of the intrinsic (to the source) $\Lp(\nu, t)$ and the extrinsic
 $g(\nu, t)$ but  their relative contributions have not yet been disentangled satisfactorily.   The modulation $g$
 has both short term and long term contributions from Galactic DISS (minutes to hours) and Galactic RISS
 (hours to months), respectively.  Plasma lensing in host galaxies and gravitational microlensing will  have
 similarly long time scales.  The episodic bursting of FRB~121102 is naturally explained if bursts are heavily modulated by $g$ even if
$\Lp$ is a  process with fixed mean rate (Poissonian or otherwise).
Variability of the the total delay $t_{\rm d}$ could  potentially account for the observed aperiodicity  for a source that is intrinsically periodic.   Of course these features might also be intrinsic.

 %Additional structure can be imposed by multiple imaging that introduces factors of the form
 %$[1 + m_{12} \cos(2\pi \nu \Delta t_{12} + \phi_{12}]$ for each pair of images where $m_{12} \le 1$ and
 %$\Delta t_{12}$ and $\phi_{12}$ are the differences in arrival time and phase.

\subsection{Scattering in Host Galaxies, Intervening Galaxies, and the Intergalactic Medium}

The broadening times measured for FRBs (figure~\ref{fig:taud_vs_b}) are extragalactic and highly likely to be from scattering in host galaxies.  However there is debate about the relative roles of host-galaxy scattering and
contributions from the IGM \citep[e.g.][]{2013ApJ...776..125m, lg14, cws+16, xz16}.
Scattering near FRB sources yields a broadening time
$\tauh \sim \dsl \thetah^2 / 8\ln 2 c$, where $\dsl = \Lh$ is the scale height or source-scattering layer distance
and $\thetah$ is the range of angles over which radiation is scattered.  However, the scattering diameter seen by an observer is  much smaller (and typically unmeasureable), $\thetao = (\dsl/\dso) \thetah \ll \thetah$, where $\dso$ is the source-observer distance.

Intervening galaxies  are likely only for $z \gtrsim 1$. They will contribute to the total DM (perhaps modestly) but may broaden bursts by scattering by large amounts because of the long path lengths. The Euclidean  expression
$\taug \sim (\dsl\dlo/dso) \thetag^2 / 8 \ln 2 c$ holds in  flat $\Lambda$CDM space  where $\dsl, \dlo$, and $\dso$ are angular diameter distances; $\taug$ increases by a factor of $(1+\zg)$ while  $\thetag$ decreases by $(1+\zg)^{-2}$ (for a plasma).
The seeing disk diameter  for a face-on Milky Way galaxy will be $\thetag \sim 0.8  (1+\zg)^{-2}$~mas, giving a very large scattering time
$\taug \sim 560 \,{\rm ms}\,  (1+\zg)^{-3} (\dsl\dlo/\dso)$ for distances in Gpc that will render undetectable all but the strongest bursts.

\newcommand{\Ftilde}{\widetilde F}

In principle the IGM can contribute to the scattering of FRBs if long path lengths   compensate for the tenuous electron density.  The scattering time  per unit length depends on the square of the electron density and thus
$\taud \propto \DM^2$
with a proportionality constant dependent on the  length scales of density fluctuations and on the filling factor of larger-scale density concentrations.
\citet[][]{lg14} and \citet[][]{2013ApJ...776..125m} argue that the diffuse IGM has properties that yield negligible scattering.   This conclusion is consistent with the lack of any obvious correlation of the scattering time with the extragalactic DM \citep[][see also Section~7.3]{cws+16}.

IGM structures including Lyman-$\alpha$ clouds, damped Ly-$\alpha$ systems, and intracluster media will have larger electron densities, different filling factors, and smaller length scales than the diffuse IGM, so scattering in those regions may be significant \citep[][]{2013ApJ...776..125m} \citep[but see][]{2018MNRAS.474..318P}.
At present there is no evidence for scattering from such regions.   In fact, for a few FRBs, fine scale spectral structure is consistent with Galactic DISS and requires that
the extragalactic scattering originate from near the FRB source (within a galaxy radius)
\citep[e.g.][]{mls+15b}.

\subsection{Propagation Factors that Affect FRB Detections}

The amplitudes of bursts may be influenced strongly by lensing, scintillation, and absorption. Such effects need to be considered in analyses of both surveys and follow-up observations.

\subsubsection{Scintillation modulations and quenching}

\newcommand{\mB}{m_{\rm B}}
\newcommand{\mtheta}{m_{\theta}}

DISS typically reduces the burst amplitude but occasionally can boost it by a large amount for
large, less probable gains on the tail of the exponential distribution,
$ \pdfg(g) =  e^{- g}$.
The role of such modulations in burst detections (or missed detections) depends on the number of statistical trials in an FRB survey.  This in turn depends on the size of the burst source population and the number of bursts emitted per source.    If many trials are done,  detected bursts may have been boosted significantly by DISS (or lensing, as discussed below), with the corallary   that repeat bursts will also have low probability.

\begin{textbox}[h]
\section{Scintillation Source Size Requirements}
DISS has extraordinary resolving power because it is quenched for
sources larger than about a microarcsecond.
%Figure
%%
%source\_size\_requirement2.pdf
%%
%shows the relevant geometry.
Waves from
a source of angular size $\thetax$ have   a coherence length  $\lc \sim \lambda/\thetax$ on a Galactic scattering  screen.   The coherence length must exceed the patch size  on the screen that contributes to measured
flux densities,  or
$\thetas < \thetaiso = \lambda / \thetaG\dlo$.
For  Galactic scattering with $\dlo = 1$~kpc and  $\thetaG= 1$~mas  the isoplanatic angle  is $\thetaiso = 0.4\ \muas$.
Pulsars easily satisfy this constraint as do
FRBs, which are intrinsically  smaller in angular size by a factor  $\sim 10^{-6}$ given their millisecond durations and gigaparsec distances.
However, scattering in host or intervening galaxies reduces the coherence length and can quench scintillations.
Let  $\thetax$ and $\thetaG$ be the  scattering diameters produced by extragalactic and Galactic screens at distances $\Lx$ and $\LG$ from the source and observer, respectively,  and $\taux, \tauG$ the corresponding broadening times; then the requirement becomes (for $\nu$ in GHz, $\Lx,\LG$ in kpc, and $\dso$ in Gpc),
$$
\thetax\thetaG
	\lesssim \frac{4\ln 2}{\pi}  \frac{\lambda}{\LG}
	\sim \frac{(19.1\,\muas)^2}{ (\nu \LG)}
\quad\quad
	{\rm or}
\quad\quad
	 \taux \tauG <
 	\frac{1}{(2\pi\nu)^2}
	\frac{\dso^2}{\Lx\LG}
	\approx
 	(0.16\ {\rm ms})^2
	\left(\frac{\dso^2}{\nu^2\Lx\LG} \right)	.
	%\quad\quad {\rm for \ \nu \  in\  GHz \  \ and \  \  \LG\ in \ kpc}.
\label{eq:ssreq_theta}
$$
Host-galaxy scattering can also quench gravitational microlensing, as discussed in \S~\ref{sec:gravlens}.

 \end{textbox}

\begin{figure}[h]
\centerline{
\includegraphics[width=0.55\textwidth]{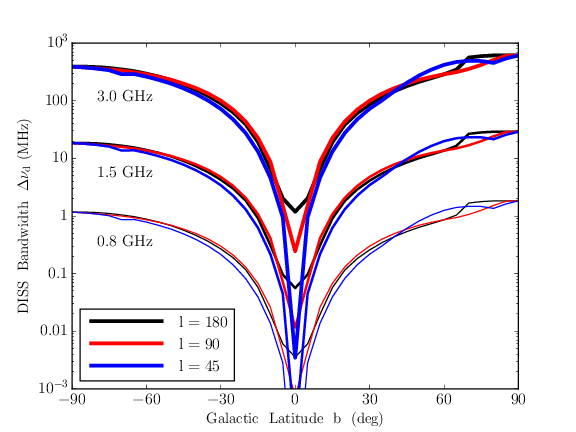}
\hspace{-2.5in}
\includegraphics[width=0.55\textwidth]{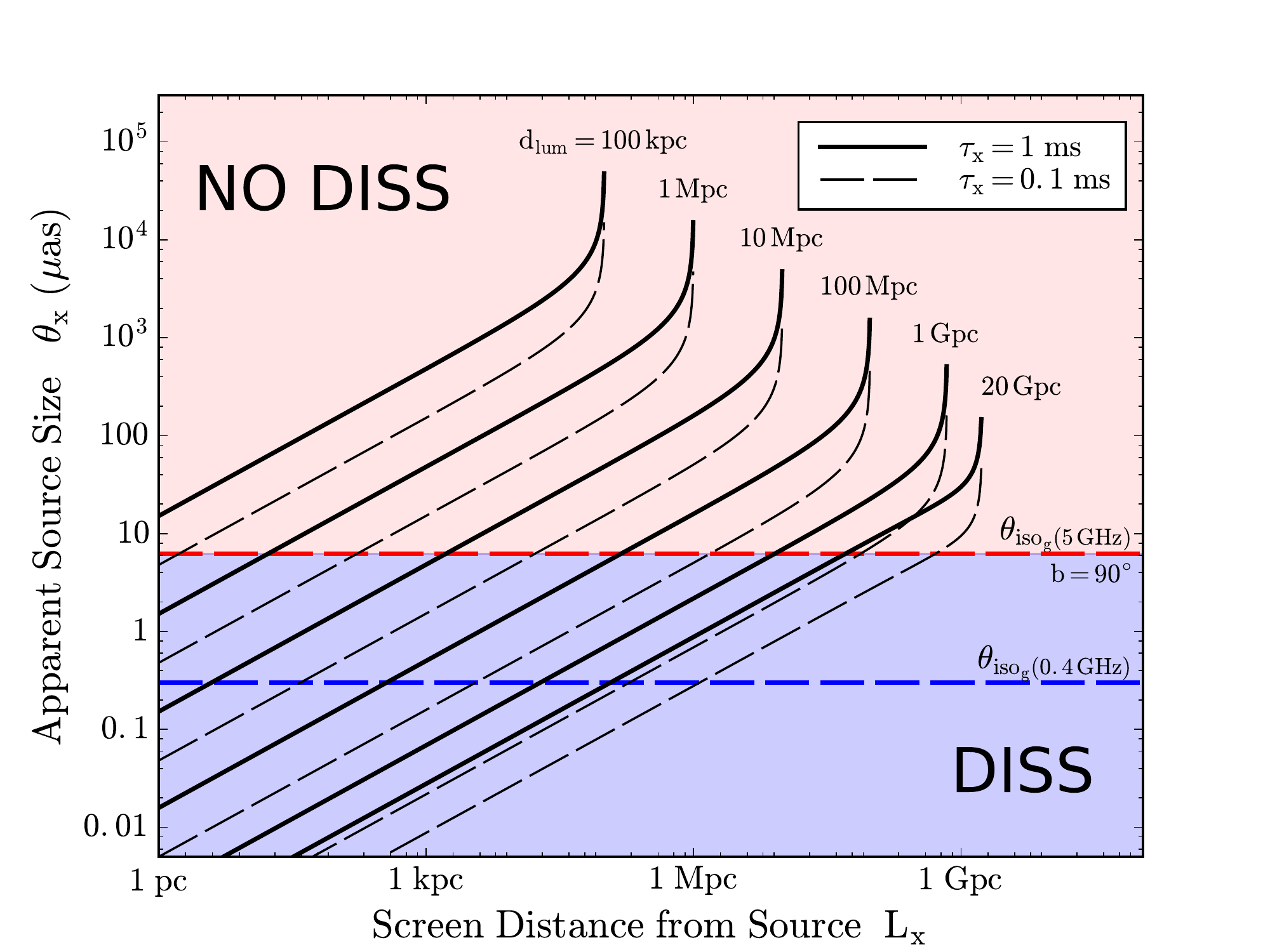}
}
\caption{
Left panel: DISS bandwidth vs. Galactic latitude for several combinations of Galactic longitude and radio frequency, as labeled.
The curves are calculated using the NE2001 electron density model.
\label{fig:diss_bw_vs_b}
Right panel:
 Apparent size of FRB source from extragalactic scattering vs. distance of an extragalactic scattering screen from the source.
The intrinsic source size is assumed negligible.
Solid lines are for an FRB scattering time of 1~ms and dashed lines for 0.1 ms.
Pairs of curves are shown for source luminosity distances of 100~kpc to 20~Gpc.
The dashed horizontal lines indicate the  critical angular size (isoplanatic angle $\theta_{\rm iso_g}$)
at the labeled frequencies  for the North Galactic pole ($b=90^{\circ}$). Apparent source sizes above this line will  have strongly attenuated diffractive interstellar scintillation (DISS) from scattering in the Milky Way, as indicated by the shading and the label `NO DISS.'  Cosmological effects are relevant only for source distances greater than a few Gpc.
\label{fig:erb_diss}
}
\end{figure}

However, DISS modulations are reduced if observations are made with bandwidths much larger than their correlation bandwidth or if the effective source size is
larger than a critical amount.
A finite bandwidth ($\Delta\nu$) reduces the RMS modulation of $g$ from unity to
$\mB \approx  2 \sqrt{\dnud / \Delta\nu}$ for a correlation bandwidth $\dnud \ll \Delta\nu$.
Figure~\ref{fig:diss_bw_vs_b} (left panel) shows the correlation bandwidth
for a few frequencies vs. Galactic latitude,
demonstrating that $\dnud$ plummets to very small values at low latitudes and that DISS is largely
quenched (for large observing bandwidths), disallowing large scintillation boosts.

Similarly, a finite source size  $\thetax$  reduces the modulation to
$\mtheta \approx   \thetaiso / \thetax$,  where $\thetaiso$ is a critical (isoplanatic) angular scale (see the sidebar titled Scintillation Source Size Requirements).
Sources of millisecond bursts are necessarily small enough to show fully modulated, Galactic DISS.  However any extragalactic scattering can make the apparent size larger than $\thetaiso$ and thus quench DISS.
In a few cases,  Galactic DISS has been identified in FRB spectra, indicating that
the extragalactic scattering also seen  must occur  in or near their host galaxies
\citep[][]{mls+15b, rsb+16, gsp+18, shb+18}.  Figure~\ref{fig:erb_diss} (right panel) shows the distance constraints
on  a scattering screen  in order for DISS to be manifest.     The occurrence
of Galactic DISS is therefore a useful probe of the host galaxies and environments of FRB sources.

\subsubsection{Reduction in survey sensitivity from free-free absorption and pulse broadening}

Dispersing and scattering electrons will also absorb FRBs.   Apart from directions through the inner Galaxy at low frequencies, free-free absorption will only be important for  any dense  gas   in host galaxies
and for low frequency observations ($\lesssim 300$~MHz).
 The emission measure is related to the DM and path length $\Lh$ of the host
 along with parameters $\zeta$ and $\epsilon$ that quantify electron density fluctuations
\citep[][]{cws+16},
\be
\EMh =
	\frac{\zeta ( 1 + \epsilon^2) \DMh^2}{\ff \Lh}
	\sim 10^4 \, \EMunits \times \frac{\zeta(1+\epsilon^2)}{\ff {\Lh}_{\rm , pc}}
	\left(\frac{ \DMh}{100\, \DMunits}\right)^2,
\ee
corresponding to an optical depth \citep[][]{2011piim.book.....D}
\be
\tauff
		%= 3.37\times 10^{-7} T_4^{-1.3} \nu^{-2.1} \EMh
		%= 3.37\times 10^{-3} T_4^{-1.3} \nu^{-2.1}
		= \frac{3.37\times 10^{-3}}{T_4^{1.3} \nu^{2.1}}
		\frac{\zeta(1+\epsilon^2)}{\ff {\Lh}_{\rm , pc}}
	\left(\frac{ \DMh}{100\, \DMunits}\right)^2.
\ee
While negligible at 1~GHz for the nominal $\DMh$,  the optical depth can exceed unity for larger
host-galaxy DMs and lower frequencies.    Free-free absorption may therefore affect detection rates
of low-frequency surveys and may provide an additional probe of source environments.

Pulse broadening, when either $\tauG$ or $\taux$ is comparable or larger than the intrinsic burst width $W$, reduces detection numbers in surveys.
It conserves fluence, so the  matched-filter output  amplitude is reduced by a factor,
\be
f_{\tau}(\nu, l, b) =  \left ( 1 +2\taud^2 / W^2 \right)^{-1/4}
		\approx \left (W / \sqrt{2} \taud\right)^{1/2} \quad {\rm for}~ W \ll \taud.
\ee
The   scattering time and   intrinsic width $W$ are implicitly frequency dependent and $\taud$  is strongly direction dependent, as implied in Figure~\ref{fig:diss_bw_vs_b}  and using $\taud \propto \dnud^{-1}$.
The scattering factor undoubtedly plays  a prominent role in low-frequency ($< 0.8$~GHz) surveys and surveys of the inner Galaxy at low latitudes.

\subsubsection{Aggregate frequency-dependent factors relevant to FRB detection}

\begin{figure}[t!]
\centerline{
\includegraphics[width=0.5\textwidth]{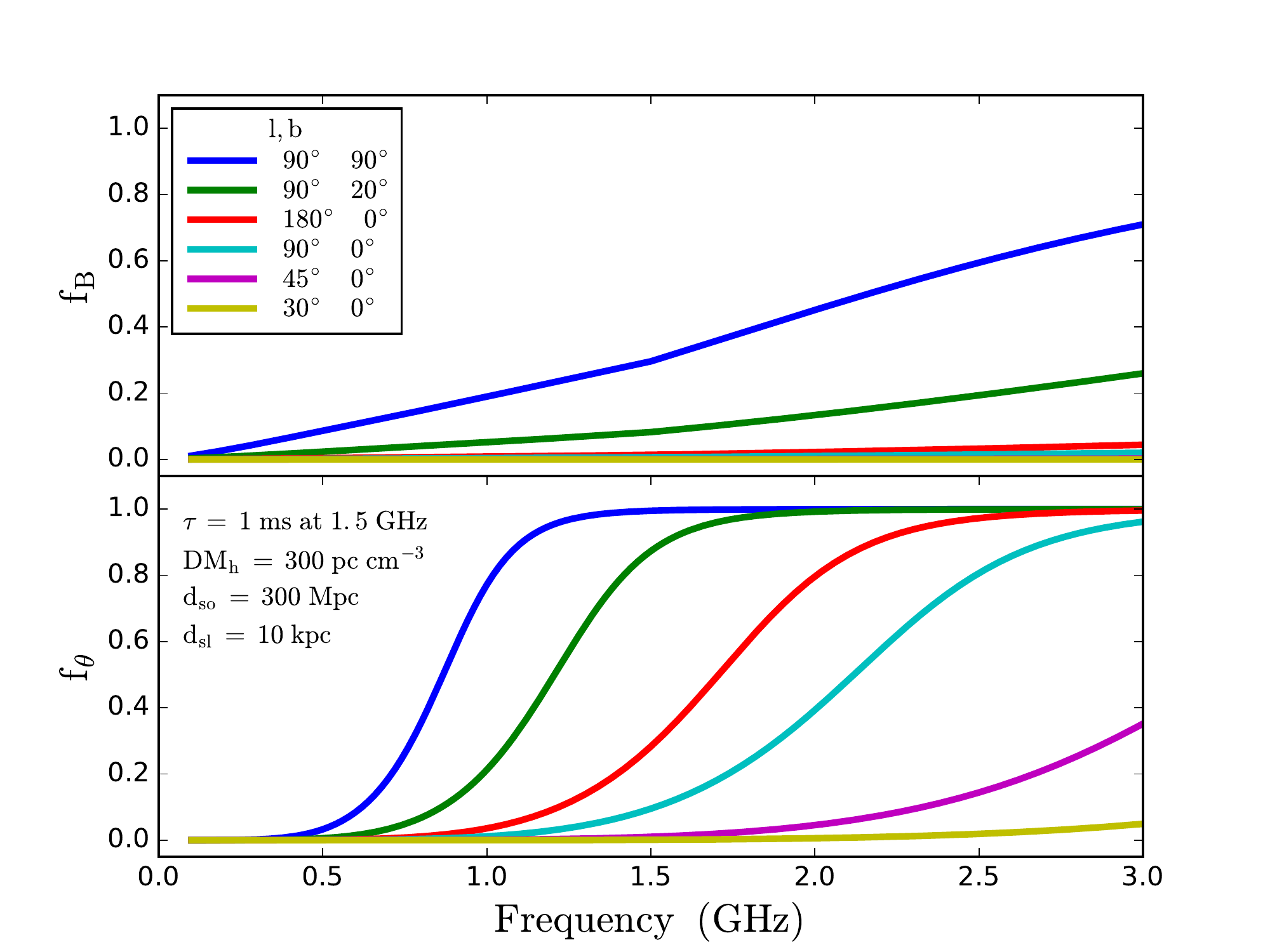}
\hspace{-2.7in}
\includegraphics[width=0.5\textwidth]{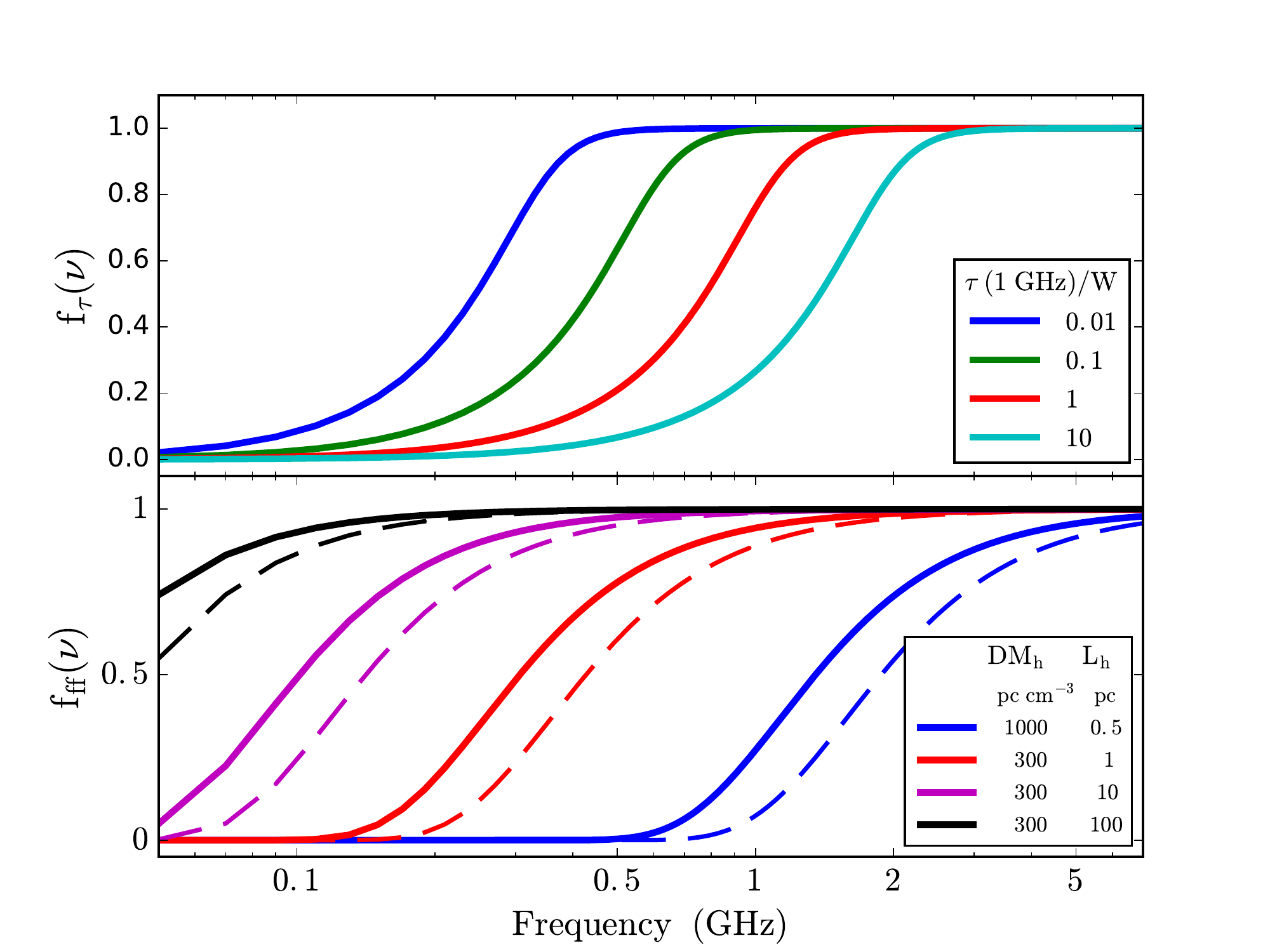}
}
\caption{
Left:
DISS reduction factors due to bandwidth averaging for different directions specified in Galactic coordinates (upper panel)  and due to quenching of Galactic DISS from source-size
quenching by scattering at 10 kpc from the FRB source (lower panel).
\label{fig:diss_reduction1}
Right, upper panel: S/N reduction factors from pulse broadening. Curves are shown for different ratios of pulse broadening time $\tau$  to intrinsic burst width $W$  at 1 GHz.
Right, lower panel: Solid curves give the factor $f_{\rm ff}$ from free-free absorption
in a host galaxy for different values of $DM_{\rm h}$ and clump thickness $L_{\rm h}$ and
with no internal density fluctuations in the clump ($\epsilon = 0$), as labeled.
Dashed lines are for full density modulations ($\epsilon = 1$).
\label{fig:snr_reduction1}
}
\end{figure}

Figure~\ref{fig:diss_reduction1}  presents the aggregate effects from propagation of FRBs through ionized gas
as a function of frequency.
The left panel shows bandwidth and source-size quenching.  The bandwidth
reduction factor in the upper frame is evaluated for six different directions and for   10\% receiver bandwidths.    The bottom frame shows the source-size reduction factor caused by extragalactic scattering.   The  pulse-broadening time  due to scattering is held fixed to
$\taud = 1$~ms at 1.5 GHz and scales as $\nu^{-4.4}$ and the apparent angular size of the source is calculated assuming a 10~kpc separation of the scattering region and source.    The conclusion is that
if Galactic DISS is important in survey detections of FRBs, by either boosting or suppressing burst amplitudes,  it will be much less important (if not negligible) at low frequencies.
Suppression of FRB amplitudes is shown in the right panel of
Figure~\ref{fig:snr_reduction1}.  The  top frame shows the suppression factor due to pulse broadening $f_{\rm \tau}(\nu)$ for four values of the ratio $\taud / W$ referenced to 1~GHz.    Suppression of the signal-to-noise ratio occurs at low frequencies even if scattering is not evident at 1~GHz.    The bottom frame shows the attentuation  from free-free absorption $\exp(-\tauff)$, that can, but need not, be important at low frequencies.

\subsubsection{Plasma and gravitational lensing}
\label{sec:plasma_lensing}

Refraction and especially  lensing from discrete structures causes  multiple imaging from the Crab pulsar
\citep[][]{glj11}, strong enhancements of pulse amplitudes from the bow shock produced by a millisecond pulsar
\citep[][]{2018Natur.557..522M},  fringes in the dynamic spectra of pulsars \citep[][]{wc87}, and events in the radio light curves of AGNs
\citep[][]{fdjh87, 2016Sci...351..354B}.
The strongly episodic burst detections from the repeating FRB~121102 may be explained most easily
from plasma lensing and the overall detection rate could also be affected.
Plasma lensing  involves  diverging or converging rays from  refraction by  electron density enhancements or deficits, respectively. These can produce strong dimming or large amplifications from caustics
along with arrival time and DM variations \citep[][]{cwh+17}. See also \citet[][]{2014arXiv1409.5766K}.

A Gaussian lens with  a dispersion measure profile $\DM$$(x) = \DMlens$$ e^{-x^2 / a^2}$ produces multiple images and caustics  for lens-observer distances larger than a
a frequency-dependent focal distance $\df(\nu)$.
The focal distance is $\sim 1$~Gpc for a modest lens  with $\DMlens = 1~\DMunits$ and
$a = 1$~AU at, say, a 1~pc distance from an FRB source at 1~Gpc from the Earth.  Amplifications
as large as $\sim 100$ are plausible.  Equivalently, caustics will be seen for
 frequencies less than a  focal frequency $\nuf \sim 1.2$~GHz for the same nominal parameters.

\begin{figure}[h]
\centerline{
\includegraphics[width=0.8\textwidth]{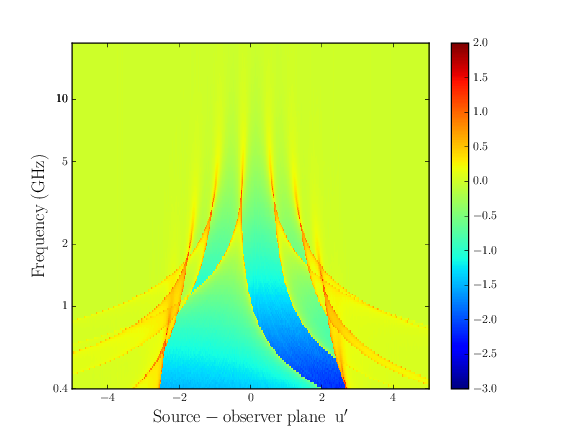}
}
\caption{
Lens gain vs. dimensionless observer position $u^{\prime}$ and radio frequency for a lens with
sinusoidal perturbations  of a Gaussian DM profile, $\DMlens(x) = [1 + 0.1 \sin(5x+\phi)]$.  The color bar represents $\log_{10} g(\nu, u^{\prime})$ and indicates that gain excursions  range from
$10^{-3}$ to $10^2$.
 \label{fig:lglens_vs_x_nu}
}
\end{figure}

The lens gain and the number of images are strong functions of frequency as well as observer location.
The dependence of the lens gain on the observer's location and frequency is shown in
Figure~\ref{fig:lglens_vs_x_nu}   for a one-dimensional Gaussian lens perturbed with 10\% oscillations.    Numerous caustics occur in this case and  the gain is strongly peaked in frequency for some observer locations while at  others, the gain $\ll 1$.   The observer's effective  location can change due to motion of the source
or lens (as well as the observer's),  serving as a possible explanation for the absence of bursts from the repeating FRB at some epochs.

\label{sec:gravlens}

\newcommand{\RE}{R_{\rm E}}
\newcommand{\DTOAg}{\Delta t_\ell}
\newcommand{\zl}{z_\ell}

Gravitational lensing can also amplify FRBs and
impose delays between multiply imaged bursts, which may also produce interference effects in dynamic spectra.
Sources with projected sizes $\dlo\thetas$ and impact parameters $b$ much smaller than the Einstein radius,
$
\RE = (4GM/c^2)^{1/2} (\dsl\dlo/\dso)^{1/2}
$,
can show double images with large amplifications  $A \sim \RE/b \gg 1$ for durations
$\Delta t \sim 2 \RE / \vperp A$ for an effective  transverse speed $\vperp$.    The  probability of
amplifications greater than $A$ is $P(>A) \sim  \pi \RE^2 L n_* /A^2$ for a population of lensing masses $M$
with a  number density $n_*$  in a region of depth $L$.

Lensing from a stellar population in a host galaxy with $\dsl\ll\dso$ has Einstein radius
$\RE \sim  4~{\rm AU}\, (M / \Msun)^{1/2} \dsl({\rm kpc})^{1/2}$,
duration $\Delta t \sim 0.1~{\rm yr}\, (M/\Msun)^{1/2} \dsl({\rm kpc})^{1/2} (Av_{100})^{-1}$ for
$\vperp = 100v_{100}\,$~km~s$^{-1}$, and
$P(>A) \sim 10^{-6} A^{-2} (M/\Msun) L_{\rm kpc} n_*({\rm pc}^3)$.
Microlensing might enhance a few bursts out of a much larger
 number of  unlensed events from many host galaxies. However, given the small $P(>A)$, microlensing is inconsistent with the episodic detections of FRB~121102 unless a particular geometry allows repeated lensing.

If it occurs for any FRBs,
gravitational lensing   is a unique  diagnostic for  dark matter comprising primordial black holes or other discrete objects, including those in binaries \citep[][]{2018A&A...614A..50W}.
Point-mass lensing yields a differential delay between a dual-imaged burst,
$\DTOAg \sim (8GM/c^3)(1+\zl) / A = 39\,\mu s\, (1+\zl)(M/\Msun) / A$ for $A\ll 1$
\citep[][]{2014ApJ...797...71Z, 2016PhRvL.117i1301M}.   This may be detectable as an oscillation in
 dynamic spectra for bursts    that are  coherently dedispersed, providing a fringe pattern of the form
$\cos 2\pi\nu\DTOAg$ with an amplitude related to the image amplification ratio
\citep[][]{2014ApJ...797...71Z, 2017ApJ...850..159E}.
Objects with masses  $\gtrsim 0.1\Msun$ are within reach.
At present, none of the spectral features seen in FRBs is clearly associated with a fringe pattern but more detailed analyses, particularly with wideband spectrometers,  are needed.
Primordial black holes (PBHs) with masses $\sim 30\Msun$  may be strongly constrained with large FRB samples if they extend to  $z\gtrsim 0.5$ because the event probability can be large enough ($\sim 0.02$) if PBHs comprise a significant fraction of dark matter (\citet[][]{2016PhRvL.117i1301M} ).
Lensing from intervening galaxies \citep[][]{lgdwz+18} has been discussed as a means for
determining  cosmological parameters, including
$H_0$ and  the spatial curvature $\Omega_k$.    Such lensing will be sustained for long times and cannot
account for the episodicity of the repeater FRB~121102.

\section{THE REPEATING FRB 121102}
\label{sec:repeater}

\newcommand{\rfrb}{FRB~121102}
\newcommand\ion[2]{#1$\;${\small\uppercase\expandafter{\romannumeral #2\relax}}}%
\newcommand{\arcsec}{\ensuremath{^{\prime\prime}}}
\newcommand{\arcmin}{\ensuremath{^{\prime}}}
\newcommand{\degree}{\ensuremath{^\circ}}
\newcommand{\hs}{\ensuremath{^{\mathrm{s}}}}
\newcommand{\hm}{\ensuremath{^{\mathrm{m}}}}
\newcommand{\hh}{\ensuremath{^{\mathrm{h}}}}

The detection of \rfrb\ \citep{sch+14} in observations acquired at the Arecibo observatory on 2012 November 2 (Figure~\ref{fig:bursts1}) laid to rest any residual concerns about site-specific interference at the Parkes observatory \citep{pkb+15} and confirmed the astrophysical nature of the FRB phenomenon. As with other FRBs, the measured DM of $557.4 \pm 2.0~\DMunits$ significantly exceeded the predicted line-of-sight maximum electron density contribution from the Milky Way ($188~\DMunits$), although the location of the source at a low Galactic latitude in the Galactic anticenter ($\ell, b = 174.95^\circ, -0.22^\circ$) added significant uncertainty, requiring extensive multiwavelength investigation to place deep limits on Galactic \ion{H}{2} regions along the line of sight \citep{ssh+16b}.

The detection of additional bursts from \rfrb\ \citep{ssh+16a} was an unexpected pay-off to routine follow-up observations (Figure~\ref{fig:bursts3}). The overlapping sky position uncertainty regions and the consistent value of DM unambiguously identified \rfrb\ as a repeating source and ruled out cataclysmic and explosive models as the source of (at least) that particular set of FRB events\footnote{It also introduced a point of confusion in the nomenclature, since prior usage had made no distinction between the burst event and its source. At present, there are a handful of examples of repeating sources, and the confusion remains unresolved except by context.}.

The repetition of the bursts, though sporadic (see \S\S\ref{subsec:periodicity}), enabled a targeted observation campaign with the Karl G. Jansky Very Large Array, where interferometric visibilities were acquired at high resolution in both time and frequency ($\Delta t = 5$~ms, $\Delta\nu = 4$~MHz over a bandwidth of $1$~GHz, limited by the maximum data rate of $\sim 300$~MB~s$^{-1}$). In 83~hours of observations over six months, nine bursts were detected, leading to a precise source localization on the sky ($\alpha, \delta = 05\hh 31\hm 58.70\hs, +33\degree 08\arcmin 52.5\arcsec$ with an uncertainty of $\sim 0.1\arcsec$; \citealt{clw+17}). The detections used complementary approaches, de-dispersing the visibilities at the known DM of the FRB and then searching for a transient source in a sequence of $5$-ms images \citep[e.g.,][]{lbb+15}, as well as phasing and summing visibilities to create time-frequency dynamic spectra over a grid of sky positions and searching those for single dispersed pulses.

\begin{figure}[h]
\centerline{\includegraphics[width=1.2\textwidth]{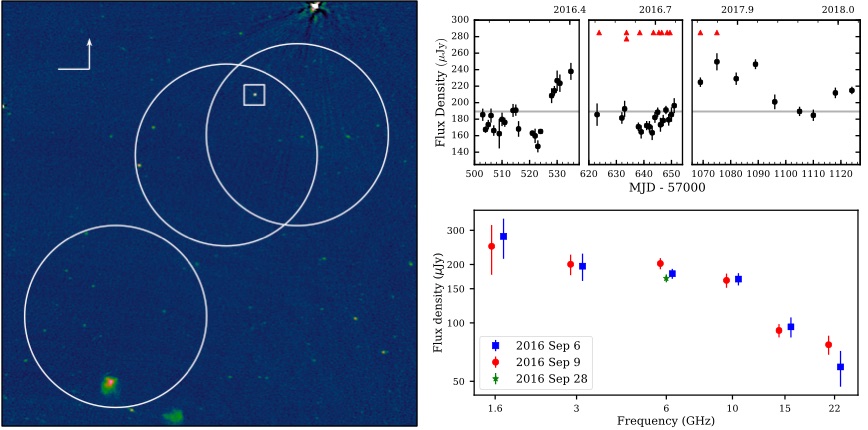}}
\caption{VLA observations of the field of \rfrb. Left: VLA image at 2--4~GHz with 2\arcsec\ resolution and an image RMS of 2~$\mu$Jy \citep{clw+17}. Arecibo detection beam positions and sizes are indicated with white circles, illustrating the positional uncertainties. The radio counterpart (PRS) is enclosed by a 20\arcsec\ square within the beam overlap area. Upper right: The light curve of the PRS, from observations reported by \citet{clw+17} and new observations, indicating the source variability. The average source flux density of $\sim 190~\mu$Jy is indicated by a gray horizontal line, and the epochs at which bursts were detected in these observations at the VLA are identified by red triangles. Lower right: The spectrum of the PRS from VLA observations spanning 1--25~GHz. The integrated flux density is plotted for each epoch of observation, and shows a spectrum inconsistent with a single power law.
\label{fig:frb121102-sky}
}
\end{figure}

\begin{figure}[ht]
\centerline{\includegraphics[width=1.2\textwidth]{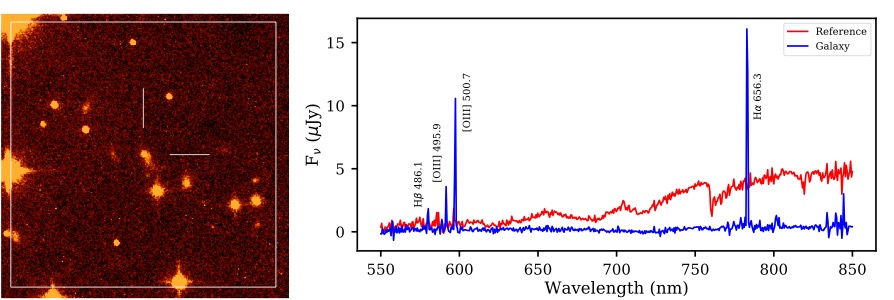}}
\caption{Left: Image from the Hubble Space Telescope WFC3 in the F110W filter (equivalent to J-band) showing the resolved irregular dwarf galaxy host of \rfrb\ \citep{bta+17}. A prominent knot of star formation dominates the optical emission. Lines indicating North and East are 3 arcseconds in length. Right: The spectra of the host galaxy and the reference object \citep{tbc+17}. Prominent emission lines are identified and labeled with their rest-frame wavelengths in nanometers, demonstrating the redshift of the galaxy.
\label{fig:localize}
}
\end{figure}

\citet{clw+17} identified a variable radio source coincident with the sky position of the bursts (Figure~\ref{fig:frb121102-sky}), with a mean flux density of $\sim 190~\mu$Jy; the nature of this persistent radio source (hereafter PRS) remains enigmatic and is discussed further below (\S\S\ref{subsec:PRS}).  Building on the initial localization, \citet{mph+17} used very long baseline interferometry with Arecibo and the European VLBI Network to detect bursts, localize them with $\sim 12$ milliarcsecond precision, and confirm their spatial coincidence with the PRS. \citet{clw+17} also identified a faint optical counterpart to the bursts, with SDSS $r^\prime$-band magnitude $m_{r^\prime} = 25.1$~AB mag. With optical imaging and spectroscopy at the 8-m Gemini North telescope, \citet{tbc+17} classified the counterpart as a low-metallicity star-forming dwarf galaxy and measured a redshift $z=0.19273(8)$, corresponding to a luminosity distance of 972\,Mpc (Figure~\ref{fig:localize}). With high-resolution optical and infra-red imaging using the Hubble Space Telescope and the Spitzer Space Telescope, \citet{bta+17} further showed that the emission is dominated by a bright knot of star formation in the irregular dwarf galaxy with a half-light diameter of 1.4~kpc compared to 5--7~kpc for the galaxy itself, and this knot coincided (within the astrometric frame-matching uncertainties) with the PRS. With its high specific star-formation rate, low metallicity, and prominent emission lines, the host galaxy resembles the preferred hosts of long duration gamma-ray bursts and superluminous supernovae \citep[e.g.,][and others]{fls+06,plt+13,vsj+15}, as discussed further below (\S\ref{sec:hosts}).

Unlike gamma-ray bursts and supernovae, though, the bursts from \rfrb\ show no afterglows, and have been observed to repeat at short enough intervals that no plausible explosive mechanism (i.e., one that destroys the central engine) could power them. For example, \citet{gsp+18} report 18 bright bursts detected by the Green Bank Telescope at 4--8~GHz within just a 30-minute span (and many more faint ones were identified by \citet{zgf+18} in the same span of data). Additionally, simultaneous coverage in X-rays \citep{sbh+17} and high-energy gamma rays \citep{aaa+18} as well as optical wavelengths reveals no other coincident electromagnetic emission with bursts from \rfrb.

However, the well-localized sky position and DM allow coherently dedispersed observations of \rfrb\ over a broad range of radio frequencies and a long period of time. Such broadband, coherently dedispersed observations have enabled the detection of bursts from \rfrb\ at frequencies as high as 8~GHz \citep{lab+17, gsp+18, shb+18}, and revealed unexpected richness of time-frequency structure in the bursts (see, e.g., Figure~\ref{fig:bursts3}).

\begin{figure}[tb]
\centerline{\includegraphics[width=1.2\textwidth]{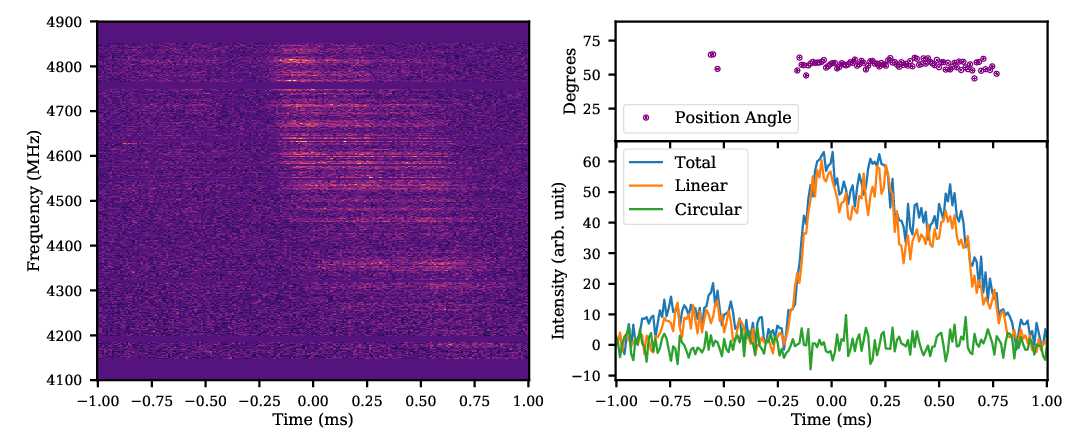}}
%\vspace{3in}
\caption{The dynamic spectrum, polarization angles, and pulse profile of a burst from \rfrb\ detected at Arecibo, adapted from \citet{msh+18}. Left: The burst dynamic spectrum, showing fine structure in time and frequency. Upper right: The polarization position angle across the burst. Lower right: Total intensity (blue), linear polarization profile (orange), and circular polarization profile (green) for the burst. The burst is almost completely linearly polarized, with a constant polarization position angle. Figure based on observations presented in \citet{msh+18}.
\label{fig:frb121102-RM}
}
\vspace{-3mm}
\end{figure}

Coherently dedispersed observations by \citet{msh+18} at Arecibo at 4.1--4.9~GHz also revealed that the bursts had nearly 100\% linear polarization (Figure~\ref{fig:frb121102-RM}) and stacking of the bursts revealed a very high and variable rotation measure of $+1.46 \times 10^5$ to $+1.33 \times 10^5$~rad~m$^{-2}$ over a seven-month period. Such a high rotation measure had previously only been observed in the environment of Sgr~A* \citep{bwf+03, mmz+07}, the supermassive black hole in our Galactic center. Further, the large changes in the value of the RM without comparable changes in the burst DM require significant changes in the line-of-sight projected magnetic field. That rules out, for example, an \ion{H}{2} region along the line of sight as the source of the high RM and implies that the source is embedded in an extreme magneto-ionic environment \citep{msh+18}. In what we suggest may not be a coincidence, such changes in RM without accompanying changes in DM have only been seen before for the Galactic center magnetar J1745$-$2900 \citep{2018ApJ...852L..12D}. The implications of the constraints from the DM, RM, and EM are discussed further below.

%%%%% Moved Fig 11 down a para to appear on a new page

\subsection{The Persistent Radio Counterpart Associated with FRB~121102}
\label{subsec:PRS}

The nature of the PRS associated with \rfrb\ remains enigmatic. The source is compact and  barely resolved with very long baseline observations, with a source size of 2--4~mas at 1.7~GHz and 0.2--0.4~mas at 5~GHz \citep{mph+17}. The radio spectrum of the source (Figure~\ref{fig:frb121102-sky}) is non-thermal and inconsistent with a single power law \citep{clw+17}, and the flux density at 2--4~GHz is variable on $\sim$day timescales, ranging between $190 \pm 50~\mu$Jy over observations from 2016 to 2018 (Figure~\ref{fig:frb121102-sky}). \citet{clw+17} show that of the 69 sources detected within a 5\arcmin\ radius, 9 (including the PRS) show significant variability as defined by a reduced $\chi^2$ metric ($\chi^2_r > 5.0$), and that the variability is uncorrelated with the detection of bursts in the uniform VLA dataset (point biserial correlation coefficient $r = 0.054$, exceeded by chance 75\% of the time).

Were it not for the knowledge that the PRS is associated with fast radio bursts, we would readily conclude that it is consistent with a weak active galactic nucleus. Indeed, given the inference of an extreme magneto-ionic environment associated with the burst source \citep{msh+18}, and its similarity to the Galactic center magnetar J1745$-$2900, the AGN model remains an attractive explanation for the PRS. An alternative model presented by, e.g., \citet{2017ApJ...841...14M} and \citet{mm18} is that the PRS is a magnetized electron-ion nebula powered by the termination shock of a relativistic magnetar wind, which implicates a very young magnetar as the source of the bursts from \rfrb.

\subsection{Burst Periodicity and Sporadicity}
\label{subsec:periodicity}

Bursts from \rfrb\ have been detected with separations as small as 22.7 seconds \citep[][Arecibo, 1.4~GHz]{ssh+16b}, at an epoch when 6 bursts were detected in a contiguous 1002~second observation. At the GBT at 4--8~GHz, \citet{gsp+18} have reported at least 18 bursts in a 1800~second scan, with bursts as close together as 11.3~seconds, although there may be even smaller separations, depending on the threshold for believable bursts \citep{zgf+18}. These detections have been searched for periodicity using Fourier techniques, Fast-Folding analyses, and brute force trials, with no significant detection of a period longer than 100~ms. Phase-coherent trials are not feasible over widely-spaced epochs, and the false alarm probability is too high for periods much shorter than 100~ms. An underlying period (due to source rotation, say) could also be masked by an accelerated binary orbit, a wide rotational phase window for the burst emission, or by plasma lensing effects that result in variable path lengths even for closely-spaced bursts.

Meanwhile, the bursts are also sporadic in nature: as shown in Figure~\ref{fig:frb121102-sky}, comparable VLA observations led to no burst detections in the first 21 epochs (1~hr each) in 2016 April--May, followed by the detection of 9 bursts over the next 16 epochs in 2016 August--September \citep{clw+17}. With GBT observations, \citet{gsp+18} report 18 bursts in 30 minutes, 21 bursts in the first hour, and no further comparably bright detections in the remaining 4 hours (although \citealt{zgf+18} report fainter bursts). Such sporadicity is consistent with time-variable focusing or lensing effects, and if such a situation holds for other FRB sources as well, there are severe implied difficulties in placing meaningful constraints on the absence of repeat bursts.

\subsection{Scattering of FRB~121102}

There is no evidence for extragalactic scattering (pulse broadening), even at the lowest frequency at which bursts have been detected (1.2~GHz).  Early detections in the 1.2-1.6~GHz band also showed no scintillation structure from either Galactic or extragalactic scattering. However, higher frequency observations
\citep[][]{gsp+18, shb+18} and coherently dedispersed 1.4~GHz measurements reveal narrow scintillation structure.
A multifrequency fit to the published data yields a frequency scaling  $\dnud \sim (7.5\pm 0.3 \, {\rm kHz}) \nu^{-4.0\pm 0.15}$ that is consistent with plasma scattering and the prediction (within a factor of two)  with the NE2001 model \citep{2002astro.ph..7156C}.
Similarly, the (deconvolved) angular diameters of the burst source and the PRS,
$\thetao \sim 2\pm1$~mas
and
$\thetao \sim 2$-4~mas at 1.7~GHz, respectively \citep[][]{mph+17} are also consistent with Galactic scattering and the relation between angular and temporal broadening.
The consistency with the NE2001 model contrasts with the YMW16 model \citep{2017ApJ...835...29Y}, which does not model scattering explicitly but instead  evaluates the scattering time $\taud$ using  empirical relation between $\taud$ and \DM\ based on Galactic pulsars.  This overpredicts the scattering by a factor of 30.

\newcommand{\RMhost}{\RM_{\rm h}}
\newcommand{\DMtotal}{{\DM_{\rm total}}}
\newcommand{\DMhost}{\DM_{\rm h}}
\newcommand{\DMRM}{{\DM_{\RM}}}
\newcommand{\lau}{l_{\rm AU}}
\newcommand{\nul}{\nu_{\,l}}

\subsection{Constraints on the Magnetoionic Circumsource Medium}

The measurements of DM and RM, the absence of detectable extragalactic scattering, and possible plasma lensing provide empirical constraints on the source environment. We attribute a nominal $\DMRM \lesssim  100~\DMunits$ to the circumsource Faraday region.
Using a nominal host-galaxy contribution $\RMhost = 10^5 \RM_5\, \RMunits$,  the parallel magnetic field is
$
\Bpar^{\rm (meas)} = 123~{\rm mG}\, \RM_5 \DMRM.
$
Equating $\Bpar$ to  that expected for a region of thickness $l$ with  plasma $\beta$ (thermal to magnetic energy) and a  geometric factor $\eta_B = \Bpar / B \le 1$   we obtain
\be
n_e &=&
	4.2\times 10^8 ~{\rm cm^{-3}}
	(\eta_B^2 T_4 / \beta)^{-1}
	\RM_5^2 \, \DMRM^{-2}
	%\left(\frac{\eta_B^2 T_4}{\beta}\right)^{-1}
\label{eq:ne1}
\nonumber \\
l &=&
	9.4\times10^{-4} ~{\rm AU}
	(\eta_B^2 T_4 / \beta)
	%\left(\frac{\eta_B^2 T_4}{\beta}\right)
	\RM_5^{-2} \DMRM^3.
\label{eq:l1}
\nonumber \\
\EM &=&  n_e^2 l =  n_e \times \DMRM = 2.2 \times 10^8 \, \EMunits ~
	(\eta_B^2 T_4 / \beta)^{-1}
	\RM_5^2 \, \DMRM^{-1}
	%\left(\frac{\eta_B^2 T_4}{\beta}\right)^{-1}.
\ee
The  free-free optical depth through the region is
\citep[][Eq. 10.22 with $\nu$ in GHz]{2011piim.book.....D}
 \be
\tau_{\rm ff}
&=&
3.37\times 10^{-7} \, T_4^{-1.3} \, \nu^{-2.1} \EM
=
 71 \, T_4^{-1.3} \, \nu^{-2.1}
 	(\eta_B^2 T_4 / \beta)^{-1}
 	\RM_5^2 \, \DMRM^{-1}
	%\left(\frac{\eta_B^2 T_4}{\beta}\right)^{-1}
\label{eq:tauff}
\ee
A small optical depth $\tauff \ll 1$ at 1~GHz requires the composite gas factor $\eta_B^2 T_4 / \beta \gg 1$, which increases
the depth $l$.
If plasma lensing occurs for a transverse lens scale $a$,  requiring the focal distance
 $\df \lesssim 1$~GHz at a frequency $\nul$ (section ~\ref{sec:plasma_lensing}),
defining the depth to be a multiple  $A$ of $a$,
 $l = Aa$,   and using $\dsl$ in pc, $\dso$ in Gpc,
\be
l &\le& 2.5\,{\rm AU}~ (\dsl\dso)^{1/2} \DMRM ^{1/2} (A/\nul)
\nonumber \\
\left(\frac{\eta_B^2 T_4}{\beta}\right)
	&\le&
	2.6\times 10^3
	\RM_5^2 \, \DMRM^{-5/2}
	(A/\nul)  (\dsl\dso)^{1/2}
\label{eq:gas_factor_upper_bound}
\nonumber \\
n_{\rm e}
	&\ge&
	8.42\times 10^4 ~{\rm cm^{-3}}\,
	\left(\dso\dsl \right)^{-1/2}
	\DMRM^{1/2}
        (A/\nul)^{-1}
\label{eq:ne_lower_bound}
\nonumber \\
\EM &\ge&8.42\times 10^4 \, \EMunits ~
	\left(\dso\dsl \right)^{-1/2}
	\DMRM^{3/2}
	 (A/\nul)^{-1}
\label{eq:EM_lower_bound}
\nonumber \\
\tau_{\rm ff}  &\ge&
 0.028 \, T_4^{-1.3} \, \nu^{-2.1}
 	\left(\dso\dsl \right)^{-1/2}
	\DMRM^{3/2}
	(A/\nul)^{-1}.
 \label{eq:tauff_lower_bound}
\ee

The situation may be more complex, of course, with distinct  Faraday and lensing regions, for example.
However, there is sufficient latitude to account for the measured Faraday rotation as well as the lensing requirements.   For a small $\DMRM = 1\, \DMunits$, the Faraday region is very thin ($l \lesssim 1$-10~AU), highly magnetized ($B \gtrsim 1$~G), and dense ($n_e \gtrsim 10^5$~cm$^{-3}$).
The optical depth then need not be large, $\tau_{\rm ff} \gtrsim 0.03$ at 1~GHz,  but will be optically thick at frequencies no lower than about 100~MHz.

\section{HOST GALAXIES AND COUNTERPARTS}
\label{sec:hosts}

The identification of host galaxies is a key step to obtain FRB distances and energetics, and to identify the progenitor population. Efforts have included mapping of the beam shape to identify plausible sky regions and candidate hosts based on multi-beam detections \citep{rsb+16}, as well as the mis-identification of a variable radio source as an ``afterglow'' \citep{kjb+16}. As pointed out by \citet{wb16}, \citet{vrm+16}, and others, radio variability is commonplace and cannot be relied on as a sole indicator for an FRB host galaxy. However, the persistent radio source (PRS) associated with \rfrb\ is variable, along with several other sources in the field (\S\S\ref{subsec:PRS}), and selecting luminous radio sources associated with galactic disks, after excluding AGN and background sources, may offer a reasonable sample for a targeted FRB search \citep{2017ApJ...846...44O}.  Rapid multiwavelength follow-up to detect the analog of GRB afterglows has not been fruitful either \citep[e.g.][]{pbk+17, tnt+18}, and the absence of high energy emission associated with \rfrb\ \citep{sbh+17} makes such routes less promising for host identification. At present, therefore, the only reliable method demonstrated is the direct interferometric localization of the burst itself. \citet{eb17} show that $\sim$1\arcsec\ localizations are required for unique host galaxy identifications, although if all FRBs are associated with PRS like \rfrb, \citet{ebwb18} show that the localization requirements are much less stringent, at the 10\arcsec\ level.

The one uniquely-identified host galaxy for \rfrb\ is an irregular low-metallicity  star-forming dwarf with a strong resemblance to the hosts of long duration GRBs and SLSNe (\S\ref{sec:repeater}), leading to a unified model for the source of the repeating bursts and the PRS as a young millisecond magnetar embedded in a nebula powered by its relativistic wind \citep{2017ApJ...841...14M, mm18}. If most FRBs come from repeating sources, \citet{2017ApJ...843...84N} find that a source association with GRBs or SLSNe and a burst emission lifetime of 30--300~years makes for a self-consistent picture. While such a model has many attractive features, the PRS does resemble a typical AGN, and it has been established that dwarf galaxies can have massive black holes \citep{rsjb11, svm+14}. The high (and time-varying) RM of the bursts, without a correspondingly large change in DM, argues for an extreme magneto-ionic environment similar to our Galactic center \citep{msh+18}. Of course, one could require both circumstances (a very young magnetar in the environment of a massive black hole) for (repeating) FRBs, a proposition difficult to rule out given a sample of one (as of February 2019).

Even in the absence of a specific host galaxy identification, the very low DM of FRB~171020 (114 $\DMunits$; \citealt{smb+18}) leads to a very restrictive upper limit on the distance to a host, and \citet{mbb+18} identify ESO~601$-$G036, a Sc galaxy at a redshift $z \sim 0.009$ as the most plausible host galaxy. However, the candidate host is not associated with a PRS, and nor is there a candidate PRS within $z \lesssim 0.06$, suggesting that it may not be a generic feature for (all classes of) FRB emission.  We discuss possible central engine models further below; further host identifications are the most urgent observational priority for FRB science.

\section{SURVEYS AND POPULATIONS}
\label{sec:surveys}

\subsection{The Fluence - DM Distribution}

Measured FRB fluences $\F$ are lower bounds in many cases (currently)  due to uncertainties in location within telescope beams  but they also show the wide range expected from a spatially distributed population having a  wide intrinsic luminosity function\footnote{Also, beamed radiation introduces the unknown angle between the beam axis and the line of sight that further increases the range of apparent luminosities}.   If the extragalactic portion of the dispersion measure  $\DMx =  \DM - \DMMWhat$ is a proxy for distance, the distribution of $\F$ vs $\DMx$ should provide some insights.

\begin{figure}[h]
\centerline{
\includegraphics[width=0.8\textwidth]{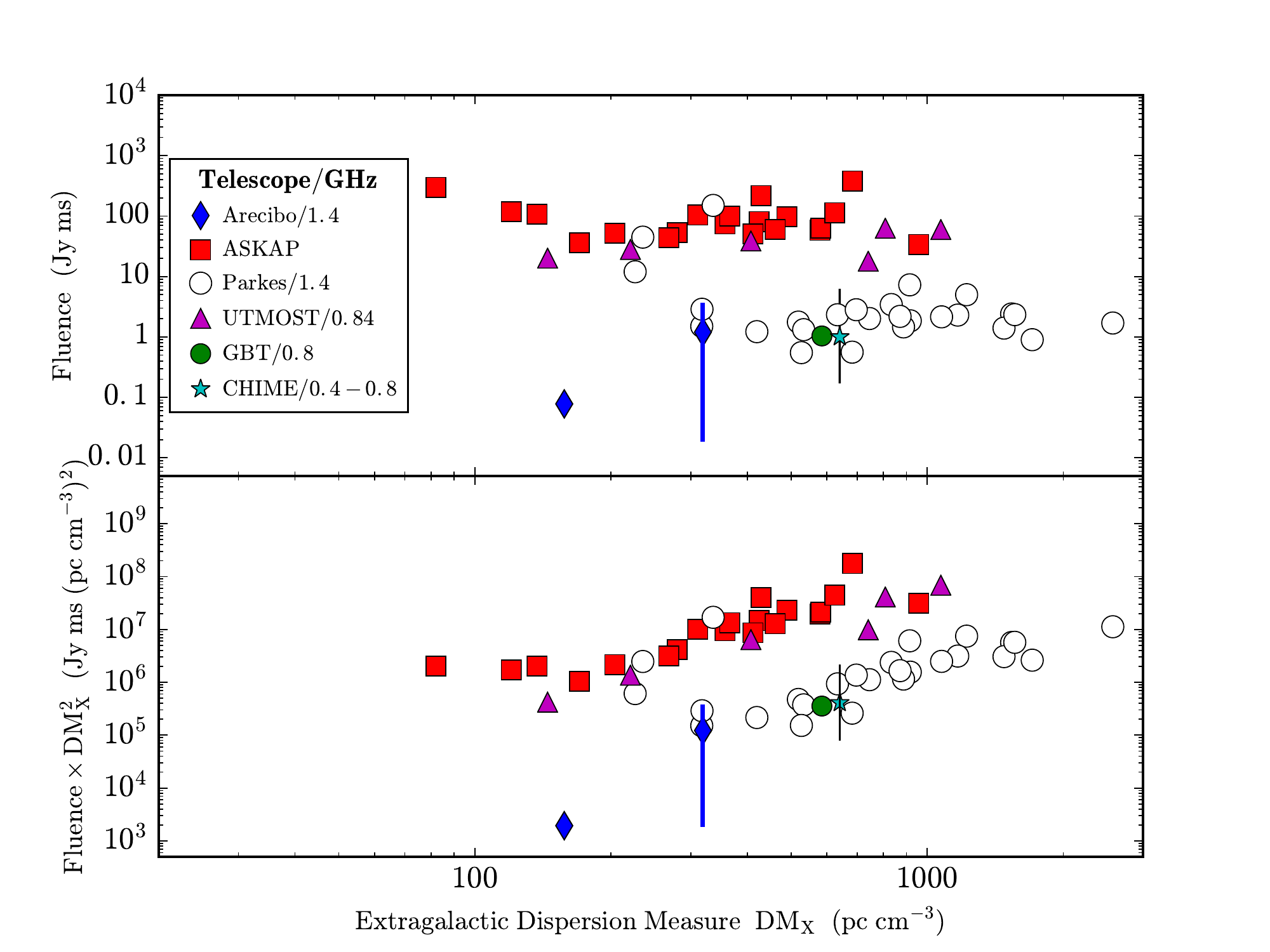}
}
\caption{
Fluence vs. estimated extragalactic dispersion measure, $\DMx$ for FRB discoveries reported in FRBCAT and Shannon et al. 2018.  The CHIME FRB~180725A has no reported fluence but the vertical line ({\em black}) represents the plausible range of values.   The blue line represents the range of values for the repeating FRB~121102.
\label{fig:F_vs_DMx}
}
\end{figure}

Figure~\ref{fig:F_vs_DMx} shows this distribution for FRBs detected in the $\sim 1.4$~GHz band from the Arecibo, ASKAP, and Parkes surveys (along with other FRBs).  Broad conclusions that can be made include the following:
\begin{itemize}
\item The ASKAP and Parkes surveys yield fluences that are largely independent of $\DMx$.
\item The mean $\langle \F \rangle$ is larger for ASKAP than Parkes by a factor $\sim 30$, which is of order the sensitivity ratio of the 64-m Parkes telescope and ASKAP's 12-m antennas  (for the fly's eye mode used by Shannon et al. 2018).
\item The scatter in fluence $\sigma_{\log \F} \lesssim 0.5$  for the two surveys.
\item The repeating FRB~121102 by itself shows a large scatter (by $10^2$) between burst amplitudes, indicating intrinsic variance that is not unlike that seen from individual  pulses of pulsars.
\end{itemize}

A simple analysis is instructive for estimating some rough numbers.
The (pseudo) luminosity $\Lp = \Spk \dso^2$ defined earlier is $\Lp \sim (\F / W) \dso^2 \propto (\F / W) \DMx^2$ if
$\DMx$ is a proxy for the source-observer distance $\dso$.
This gives
$\F \propto \Lp W / \DMx^2$.
We show $\F\DMx^2$ in the bottom panel of Figure~\ref{fig:F_vs_DMx}.
Standard candles and a constant width would imply $\F \DMx^2 =$~constant, while scatter in $\Lp$ and $W$ imply vertical scatter in the $\F$-$\DMx$ distribution.  Cosmic variance and errors in estimating $\DMx$ give horizontal scatter that combines with the variations of the true extragalactic DM.
The clear difference in average
$\F \DMx^2$ for the ASKAP and Parkes samples  demonstrates vividly that FRBs are not standard candles,
as pointed out by Shannon et al. (2018).
Only a subset of the Parkes FRBs shown in the figure come from the uniform set of surveys listed in
Table~\ref{tab:surveys_large}.  These are the events with $\DMx > 400~\DMunits$.  For this subset,
$\left\langle \log\DMx \right\rangle \sim 2.9$ compared to 2.5 for the ASKAP sample.  The corresponding fluences
have $\left\langle \log F\right\rangle \sim 0.4$ and 1.9, respectively.

The systematic rise in $\F\DMx^2$ scales roughly as  $\DMx^2$  for both the ASKAP and Parkes samples and is to be expected for   threshold-limited surveys.  However, scatter about this trend is also expected when there are significant host-galaxy contributions to $\DMx$.  The {\it apparent} flattening (beware small number statistics!) for $\DMx \le 200~\DMunits$ may result from  host-galaxy DMs.

It is also instructive to compare the {\it number} of burst detections in the ASKAP and Parkes surveys.
Let $\dminone = \left(3 / \rateone \ns \Omegas T\right)^{1/2}$ be the distance out to which only a single burst event is expected for an exposure time $T$ per sky position, where $\rateone$ is the burst rate per source and $\ns$ is the number density of sources.
The number of events occurring out to an arbitrary  distance $d$  is
$\Ne(d) = \left(d / \dminone\right)^3$.
Only a fraction of these events is detectable.   Assume that burst detection corresponds to  peak flux densities exceeding a threshold $\Smin$.

The detection number  is $\Nd(d_L) = (d_L / d_1)^3$, where $d_L$ is the maximum distance that  a burst with luminosity $\Lp$ could be detected.  Noting that both $d_L$ and $d_1$ are survey dependent,  the
ratio of  the ASKAP to  Parkes survey yields is (using values in Table~\ref{tab:surveys_large}),
\be
\frac{\Nd({\rm A})}  {\Nd({\rm P})} =
	\left[\frac{d_L({\rm A})}   {d_L({\rm P})}\right]^3
	\left[\frac{d_1({\rm P})}   {d_1({\rm A})}\right]^3
	= \left( \frac{{\Fmin}_{\rm P}} {{\Fmin}_{\rm A}} \right)^{3/2}
	\frac{(\Omegas T)_{\rm A}} {(\Omegas T)_{\rm P}}
	\approx 2.
\ee
By comparison,  the actual numbers are in the ratio $\Nd({\rm A})/ \Nd({\rm P}) \sim 1.1$.  Clearly,
the wider field and longer dwell time $T$ for the ASKAP survey more than compensate for the sensitivity difference.

\subsection{Some Population Numbers}

\newcommand{\dhat}{\widehat d}

A simple analysis of the events in Figure~\ref{fig:F_vs_DMx} illustrates some  constraints that can be made on
 the event rate density $\nsdot = \rateone\ns$.
We assume that the lowest $\DMx$ FRB in a survey is also at the lowest distance $\dhat$
with  $d_1\lesssim \dhat$, which constrains  $d_1$, the distance out to which  only
one burst occurs  in a  survey.   Using the mean IGM electron density $\nezero$ to estimate $\dhat$ and knowing the survey parameter $\Omegas T$ (Table~\ref{tab:surveys_large}),
the likelihood function for $\nsdot$ is
$
{\cal L} (\nsdot) = (\nsdot / \langle \nsdot \rangle) \exp {-\nsdot / \langle \nsdot \rangle},
$
where
\be
\langle \nsdot \rangle
	= \frac{3}{\Omegas T d_1^{3}}
	\gtrsim  \frac{3}{\Omegas T \dhat^{\,3}}
	\sim
	 \left\{
	\begin{array}{ll}
		112 \ {\rm events\  Gpc^{-3}  \ d^{-1}} & \quad\quad {\rm ASKAP\  survey} \\
		\\
		97  \ {\rm events\ Gpc^{-3} \  d^{-1}} & \quad\quad {\rm Parkes \  survey} .
	\end{array}
	\right.
\label{eq:ratedensity}
\ee

The source number density $\ns$ is not known empirically but has been estimated by \citet[][]{2017ApJ...843...84N} as $\ns \approx 10^4$~Gpc$^{-3}$ for dwarf galaxies that harbor SLSNe.   Combined with
$\nsdot \approx 100$~Gpc$^{-3}$~d$^{-1}$, we obtain the mean burst rate per source
$\rateone \approx 10^{-2}$~d$^{-1}$.   This average rate is exceeded by a large factor by the repeater FRB~121102 at some epochs but may be consistent with overall average FRB detection statistics.

\subsection{Modulated Luminosity Functions and Detection Numbers}
\label{sec:lum}

We now consider the effects of scintillations and lensing on FRB detection rates.
We assume that all individual FRB sources emit bursts according to the same   luminosity distribution
$\pdfL(\Lp)$.  The overall population luminosity function is the combination of  individual luminosity distributions  weighted by the number of sources per unit volume.
%As before, the luminosity is related to the peak flux density as $\Lp = \Spk d^2$.
Our analysis is primarily in Euclidean space and we extend to cosmological contexts as needed.   A more detailed analysis will be published elsewhere.
Cosmological effects are discussed by \citet[][]{cw16} and \citet[][]{2018MNRAS.480.4211M}.

\begin{figure}[h]
\centerline{
\includegraphics[width=0.6\textwidth]{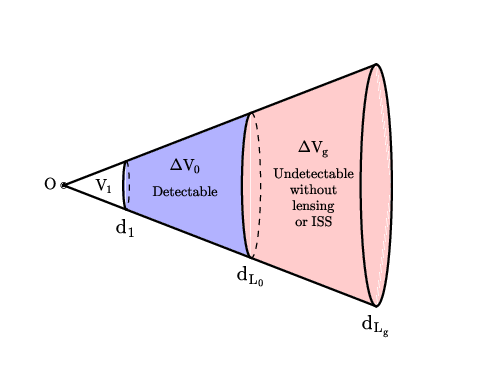}
\hspace{-2.7in}
\includegraphics[width=0.6\textwidth]{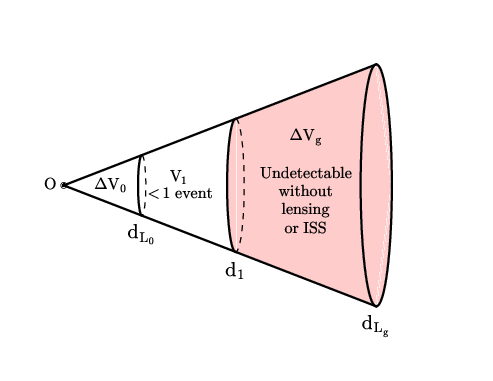}
\vspace{-6mm}
}
\caption{
Survey volumes for standard candles. Only one burst occurs within $\dminone$, on average; $\dmaxu$ is the maximum detectable distance without scintillations, and $\dmaxs$ is the maximum distance when scintillations occur.
Left:  The case where bursts are bright enough to detect without any lensing or scintillation boost, $\dmaxu > \dminone$.
Right:   Bursts are dim and sparse, so $\dminone > \dmaxu$.   Lensing or scintillations
 are required for burst detections.
\label{fig:popvol1}
}
\end{figure}

The number of detections in the absence of propagation effects is
\be
\Nd(d) = \frac{3}{\dminone^3} \int_0^d  dD\, D^2 \int_{\IT D^2}^\infty d\Lp \, \pdfL(\Lp).
\label{eq:Nd}
\ee
With scintillations or  lensing  (or absorption) characterized with a PDF $\pdfg(g)$, the modified number of detections is
\be
\Ndg(d) =
  \frac{3}{\dminone^3} \int dg\, \pdfg(g) \int_0^d  dD\, D^2 \int_{\IT D^2/g}^\infty  \!\!\!\!\!\!\!\! \!\!\!\!\!\!\!\!  d\Lp \, \pdfL(\Lp).
%= \frac{3}{\dminone^3}  \int_{0}^\infty dL \, \fL(L) \int_0^{{\rm min}(d, \sqrt{g}L/IT}  \!\!\!\!\!\!\!\! \!\!\!\!\!\!\!\! dD\, D^2.
\label{eq:Ndg}
\ee
By inspection, for a fixed gain $g(\nu, t)$,   a modulation $g > 1$ effectively lowers the threshold to $\IT/g$ or it effectively increases luminosities to $\sqrt{g} \Lp$.     When $g$ varies over an ensemble of events with unit mean (appropriate for scintillation and lensing) some nearby events become undetectable while other more distant events become detectable.

\newcommand{\dLzero}{d_{L_0}}

\begin{figure}[h]
\centerline{
\includegraphics[width=0.6\textwidth]{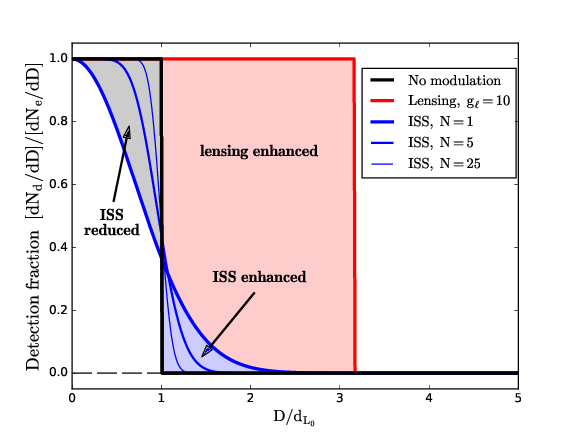}
\hspace{-2.4in}
\includegraphics[width=0.6\textwidth]{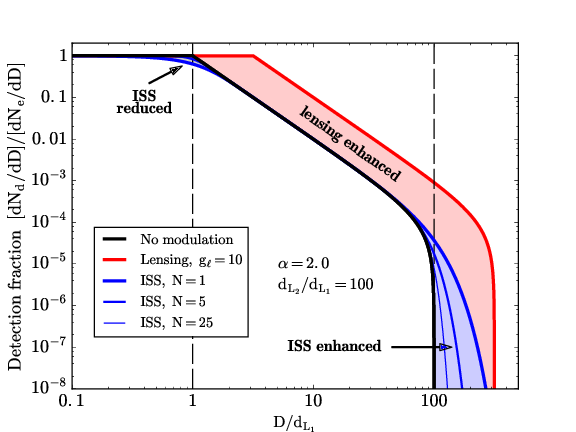}
}
\caption{
Left: Differential fraction of events detected for standard candles with and without DISS or lensing. The abscissa is the distance in units of $\dLzero$, the maximum detection distance in the absence of an extrinsic gain. ISS curves for three RMS values  $\sigma_g = N^{-1/2}$ are shown (where $N$ is the number of summed, independent `scintles') and the example lens gain is $\glens = 10$.
 \label{fig:ndet_fraction_candles}
 Right: Differential fraction of detections for events with a power-law luminosity function  with and without ISS or lensing.
The abscissa is the distance in units of $d_{L_1}$, the maximum detection distance of the weakest bursts  in the absence of an  extrinsic gain.   The corresponding distance for the strongest bursts is $d_{L_2}$.
%ISS curves for three different degrees of freedom are shown and the example lens gain is $\glens = 10$.
\label{fig:ndet_fraction_pl}
}
\end{figure}

\subsubsection{Standard candles and power-law luminosities}

A toy model comprising standard candles illustrates salient  points that also
apply to more realistic luminosity functions.   Using  $\pdfL(\Lp ) = L_0 \delta(\Lp-L_0)$, the maximum
detection distance is $\dLzero = \sqrt{L_0 / \IT}$.
Figure~\ref{fig:popvol1} shows survey volumes for two cases, one where the detection distance exceeds
$d_1$ and the other where no sources can be detected without lensing of DISS.
The differential detection number is  nonzero only for $D \le \dLzero$
\be
\frac{d\Nd}{dd} = \frac{3 D^2}{\dminone^3} \Theta(\dLzero - D),
\ee
and its ratio  to the differential number of events $d\Ne/ dd$ is
\be
\frac{d\Nd/dd}{d\Ne/dd} = \Theta(\dLzero - D).
\label{eq:dNdratio}
\ee
All events from distances smaller than $\dLzero$ are detected but none for larger distances.
Inclusion of a constant lens gain $\glens$ in a `standard lensing' model  increases the detection distance to $\sqrt{\glens} \dLzero$ and the total number of detected events from $(\dLzero / \dminone)^3$ to $\glens^{3/2} (\dLzero / \dminone)^3$.

 If the burst rate density $\nsdot = \rateone \ns$ is large enough so that $\dminone < \dLzero$,   some bursts will be detected without any lensing boost but the total number will  dramatically increase  with $\glens \gg 1$.
Episodes of lens dimming ($\glens < 1$)  reduce the number.   In the opposite case  of a sparse rate density,
detections {\it require} strong lensing.
Figure~\ref{fig:ndet_fraction_candles} shows the differential ratio from Equation~\ref{eq:dNdratio} for standard candles on the left and for a power-law case on the right, with regions indicated where detection numbers are enhanced or diminished by scintillations of lensing.

With scintillations included,  the number of detections becomes
\be
\Ndg(d) =
	\frac{
	   %\left[
		d^3 \Gamma(N, N(d/\dLzero)^2)
		+
		\dLzero^3 \gamma(N+3/2, N(d/\dLzero)^2)
	   %\right]
	   }
	{\dminone^3\Gamma(N)},
\ee
where $\Gamma(a,b)$ and $\gamma(a,b)$ are the incomplete gamma functions.
For  fully modulated DISS with $N=1$, $\pdfg(g) = e^{-g}\Theta(g)$,  and the number of detections
at distances nearer than $\dLzero$ is reduced to $\Gamma(1,1)+\gamma(5/2,1) \sim 0.57$ of the original
number because of scintillations $g<1$ but now a larger total number is detected, $\Ndg(\infty) = \Gamma(5/2) \Ne(\dLzero) \sim 1.33 \Ne(\dLzero)$ owing to scintillation boosts of sources beyond $\dLzero$.

\section{THE FRB DISTANCE SCALE}
\label{sec:distances}

 Measurement of the redshift of a securely associated galaxy is the only reliable method for determining FRB distances  and that is likely to remain the case.   Repeated bursts   from  the source of FRB~121102  were key to enabling  its sub-arcsec localization  that led to  the redshift of the dwarf host galaxy.    Absent a radio localization from the first (and perhaps only) burst from an FRB source, host galaxy associations are likely only for nearby, low-DM FRBs where  a small number of galaxies is in the positional error box.   For most bursts, which  tend to be one-offs or at least very  infrequent repeaters,   localizations need to be done at the time of discovery   using interferometric arrays.    Until such arrays operate routinely,    approximate distance estimates will be obtained from DMs. Here we summarize DM-based methods and their issues.

It is useful to consider the total DM and pulse broadening time together.
Measured values include contributions from the host, the IGM (including cosmic variance), and the Milky Way along with other contributions that can arise frome the circum-source environment,  intervening galaxies, galaxy clusters, or unmodeled HII regions in the MIlky Way,
\be
\DMfrb &=& \DMh + \DMigm + \DMg  + \DMother
\label{eq:dmfrb}
\\
\taufrb & = & \tauh + \tauigm + \taudg + \tauother
\label{eq:taufrb}
\ee

\subsection{Deconstructing Dispersion Measure}

The general approach so far has been to estimate the IGM's contribution by subtracting estimates for the host galaxy and the Milky Way while ignoring other terms,
\be
\DMigmhat = \DMfrb - \DMhhat - \DMMWhat.
\ee
Estimates for the MW term  comes from the NE2001 and YMW16 models and inclusion of a Galactic halo
contribution $\DMhalo \approx 30~\DMunits$.   Host galaxy contributions are often fixed to low constant values
such as $\sim 50~\DMunits$ (Shannon et al. 2018) or $\sim 100~\DMunits$  based on the (questionable) assumption that the host contributions arise solely from Milky Way-type ISMs averaged over inclination angles.  While MW models have systematic errors due to unmodeled HII regions (`interstellar variance'), the uncertainties in
$\DMMWhat$ are probably smaller than typical host-galaxy contributions, particularly for high-latitude FRBs.

The {\it only} empirical constraint on host-galaxy contributions comes from the repeating FRB for which Balmer-line based estimates of the emission measure translate into $\DMhhat \approx 100$-200~$\DMunits$.
The assumption of  generally small host-galaxy contributions runs counter to FRB models involving young supernovae \citep[][]{pir16},
whose expanding shock fronts imply very large DM values that can hinder detection of bursts at early times,
or models involving AGNs
\citep[e.g.][]{2017ApJ...836L..32Z}  in the centers of  galaxies.
Without other constraints, it must be allowed that a circum-source contribution could be a large fraction of the DM for  even the largest measured $\DMfrb = 2596~\DMunits$ (FRB 160102).
Consequently, the error on any given estimate for $\DMigmhat$ may be very large.

\subsection {Dispersion Measure--Redshift Relation}

Reionization at $z \gtrsim 10$ has rendered  most of the baryons in the universe to a largely invisible status
in the IGM, often referred to as the missing baryons \citep[][]{2018natur.558..406n}.   Measurements that constrain the IGM are therefore valuable for understanding the distribution and temperature-density phase structure of the ionized gas.
Future FRBs may provide much of that information once host-galaxy redshifts are measured routinely in large numbers and host-galaxy contributions are better estimated.  For now, IGM considerations have largely concerned the reciprocal problem of using $\DMigmhat$ to estimate the redshifts of FRBs.   Published analyses of the
DM-$z$ relation that conclude dominance by the IGM border on the procrustean because they attribute rather small values of the host galaxy contribution $\DMh$ in the absence of any direct measurement (other than for FRB~121102).   Assumption of small
$\DMh$  runs counter to viable models involving young NS where significant circumsource contributions to DM are expected.

% Oxford Dictionary: "Definition of Procrustean - (especially of a framework or system) enforcing uniformity or conformity without regard to natural variation or individuality."

\newcommand{\DMigmbar}{\overline{\DM}_{\rm IGM}}

Invariance of the electromagnetic phase,   $\phi = -\lambda \re \int_0^D ds\, \nelec(s)$
\citep[][]{1962clel.book.....J} implies
$$
\DM(\dso) 	= \int_0^{\dso}  \frac{d\ell\,\nelec(\ell)} {(1+z)}.
\label{eq:DMell}
$$
For a galaxy at redshift $z_{\rm g}$  with dispersion measure $\DM_{\rm g}$, an observer measures
$\DM(z_{\rm g}) = \DM_{\rm g} / (1+z_{\rm g})$
while  an arbitrary distribution of $\nelec$ gives
\be
\DM(\zs) 	= \frac{c}{H_0} \int_0^{\zs}  \frac{dz\,\nelec(z)} {(1+z)^2 E(z)} ,
\label{eq:DMzs}
\ee
where in flat $\Lambda$CDM  spacetime $E(z) = \sqrt{\Omegam(1+z)^3 + 1 - \Omegam}$.
For the IGM,  $\nelec(z) = \nezero (1+z)^3 \mu_{\rm e}/\mu_{e_0}$ where
$\mu_{\rm e}/\mu_{e_0} \sim 1$ expresses the degree of ionization of hydrogen and helium, and
the local ($z=0$) mean electron density
$\nezero = \mu_{e_0} \Omega_{\rm GM} \rho_{\rm c} / m_{\rm p}$
% \sim 2.2\times10^{-7}\,{\rm pc^{-3}}$
yields the mean DM,
\be
\DMigmbar(\zs) =
\frac{c\nezero}{H_0} \int_0^{\zs} dz\, \frac{(1+z)}{E(z)} \frac{\mu_{\rm e}}{\mu_{e_0}}.
\label{eq:DMigmbar}
\ee
We use the Planck 2015 cosmological parameters
($H_0 = 67.7~{\rm km\ s^{-1} \ Mpc^{-1}}$, $\Omega_{\rm m_0} = 0.307$, and $\Omega_{\rm b_0} = 0.0486$)
to obtain $\nezero \sim 2.2\times10^{-7}\,{\rm pc^{-3}}$ and  evaluate the  fiducial dispersion measure,
$
\DM_{\rm f} =c\nezero/ H_0 = 977~\DMunits \, (\Omega_{\rm IGM}/\Omega_{\rm b}) (\mu_{\rm e}/\mu_{e_0})$.

% Figure showing DM(z)
\begin{figure}[h]
\centerline{
\includegraphics[width=0.7\textwidth]{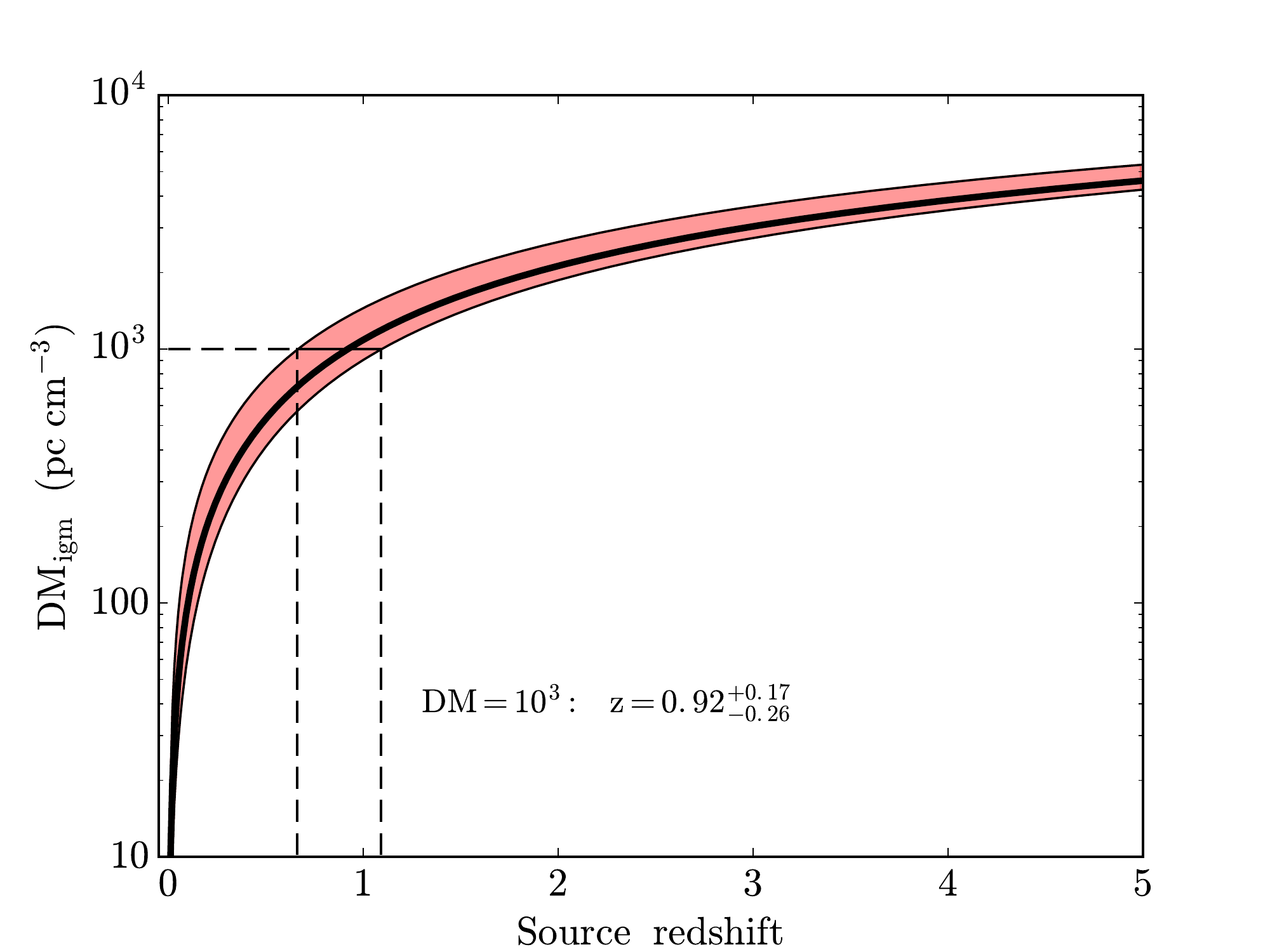}
}
\caption{IGM contribution to \DM\ vs. redshift showing the average relation from equation~\ref{eq:DMigmbar} (thick black line) and the cosmic variance in $\DMigm$
based on cosmological simulation results  characterized by
\citet[][]{iok03}, \citet{ino04}, \citet{mcq14}, and  \citet{dgbb15}.  The range of $z$ for $\DMigm = 10^3~\DMunits$ is indicated.
\label{fig:DMz}
}
\end{figure}

Cosmic variance yields significant variations about the mean from $\DMigmbar$
 that are estimated from cosmological simulations
  by several authors  and expressed in the form of a standard deviation vs. redshift, $\sigma_{\DM}(z)$.
  The fractional RMS  $\sigma_{\DM}(z) / \DMigmbar(z)$ decreases  with larger $z$ but fairly slowly.
    Consolidating the results of \citet[][]{iok03, ino04, mcq14, dgbb15, 2018ApJ...852L..11S} we show  $\DM(z)$ in Figure~\ref{fig:DMz}.        For fixed $z$ the DM distribution is asymmetric  with large positive excursions expected when the line of sight  intersects dense halos (rich galaxy clusters)  or individual galaxies for  $\zs \gg 1$, as noted by \citet[][]{dgbb15}.   Intersections with massive halos become highly probable for $z > 1$ \citep[][]{2001ApJ...548L.123V, mcq14, cw16}, so redshifts  derived from  FRBs with large DMs must be regarded with suspicion if intersections are ignored.   Future analysis can look for correlations of large FRB DMs with proximity to galaxy clusters as both the FRB sample and cluster catalog increase in size.   If DMs are IGM dominated, such a correlation should be found; conversely, the absence of a correlation is expected if FRBs are typically at $\zs < 1$ and DMs receive large host-galaxy contributions.

In addition to cosmic variance, errors in  $\DMigmhat$ due to uncertainties in $\delta\DMMWhat$ and $\delta\DMhhat$ in the Galactic and host-galaxy contributions
compound the difficulty of estimating redshifts.   The resulting $\delta\DMigmhat = \delta\DMMWhat + \delta\DMhhat$ implies [using $\zbar(DM)$ as the inverse of $\DMigmbar(z)$ and using
$\delta\DMigm^{\rm cv}$ to denote cosmic variance in the $z$-$\DM$ relation],
\be
\zhat(\DMigmhat) = \zbar(\DMigm)
	+ \frac{d\zbar}{d\DMigmbar} \left(\delta\DMigm^{\rm cv} - \delta\DMMWhat - \delta\DMhhat\right).
\ee
MW contributions are estimated using Galactic electron density models, such as NE2001
\citep[][]{2002astro.ph..7156C}
and YMW16  \citep[][]{2017ApJ...835...29Y}, which have inherent errors due to complex Galactic structure that is
not well modeled.

From Figure~\ref{fig:DMz}, $d\zbar/d\DMigmbar \approx 10^{-3}$ so each
100$~\DMunits$ of error on $\DMigmhat$  gives $\delta z = 0.1$.   What errors on $\DMigmhat$ can be expected?
Differences between the NE2001 and YMW16 models at low Galactic latitudes suggest RMS
$\delta\DMMWhat$ values exceeding 100$~\DMunits$ \citep[][]{tbc+17, shb+18}
but high latitudes have errors a factor of 5 to 10 smaller.
$\DMhhat$ for the repeating FRB likely exceeds 100$~\DMunits$  and some authors argue that host galaxy contributions will be no larger than this, based on the notion that the host contribution comes from a galaxy disk.  However, FRBs may generally be embedded in star forming regions, in galactic centers, or in a circumsource nebula that can provide much larger values.   Consequently, redshift errors may be several tenths or more for
$z\sim 1$.

\subsection{The $\taud$--Dispersion Measure Relation}

Lines of sight to FRBs span plasmas with radically different properties, including the ISM, the IGM, the host galaxy's ISM, and the circumsource medium (contributions from  the interplanetary medium and ionosphere are  minor for  FRB studies\footnote{
		We note however that any FRBs discovered in directions close to the Sun will likely be
		affected by interplanetary scintillation.}.
Turbulence will differ greatly between them just it does between intra-Galactic components.

% Figure showing tau_vs_DM for pulsars and FRBs
% Shows pulsar scattering as  a shaded region.
% tau_DM_hockey_stick.py in /Users/cordes/Research/Calcs_and_plots/NE2012

\begin{figure}[h]
\centerline{
\includegraphics[width=0.5\textwidth]{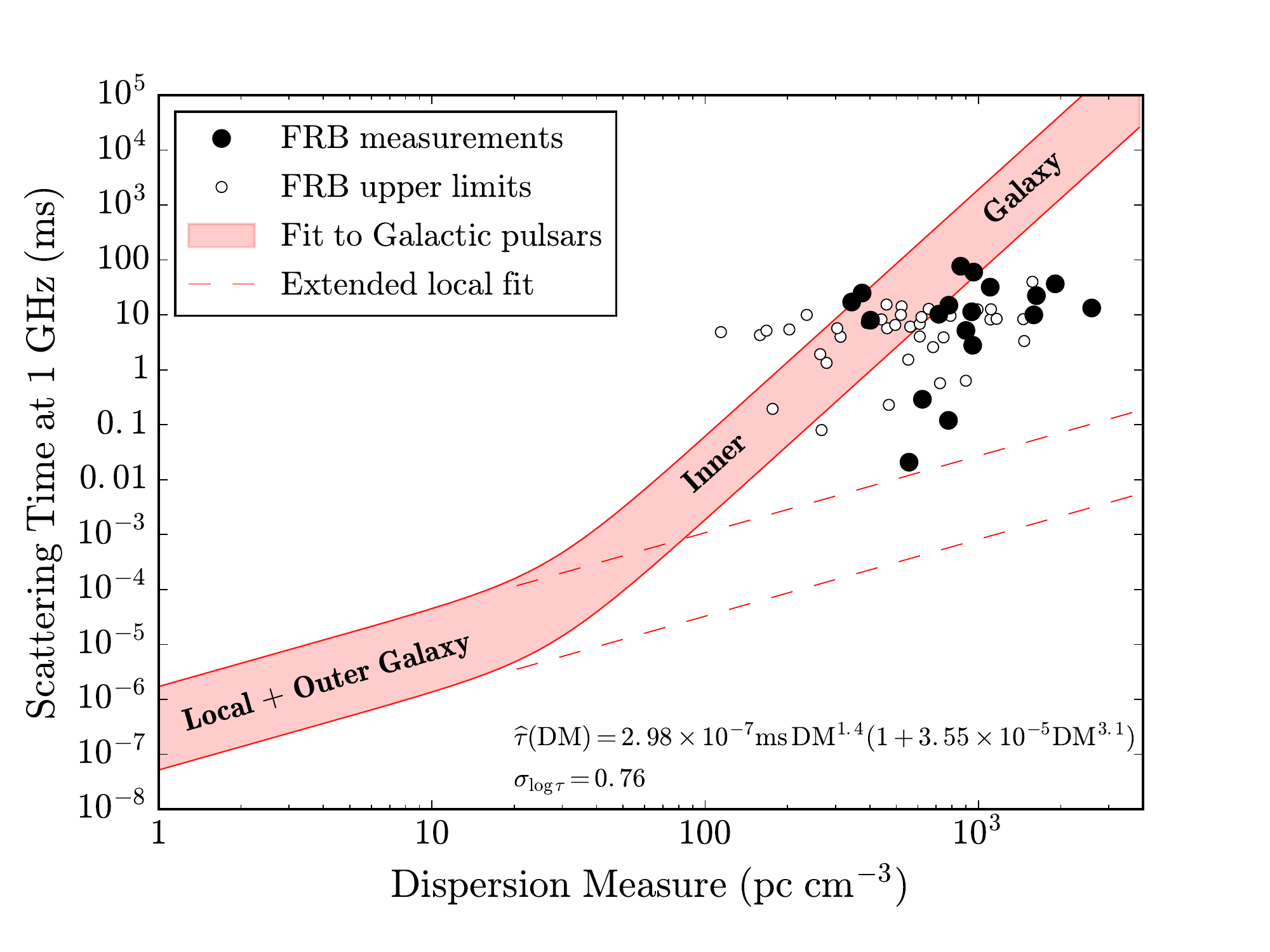}
\hspace{-2.5in}
\includegraphics[width=0.5\textwidth]{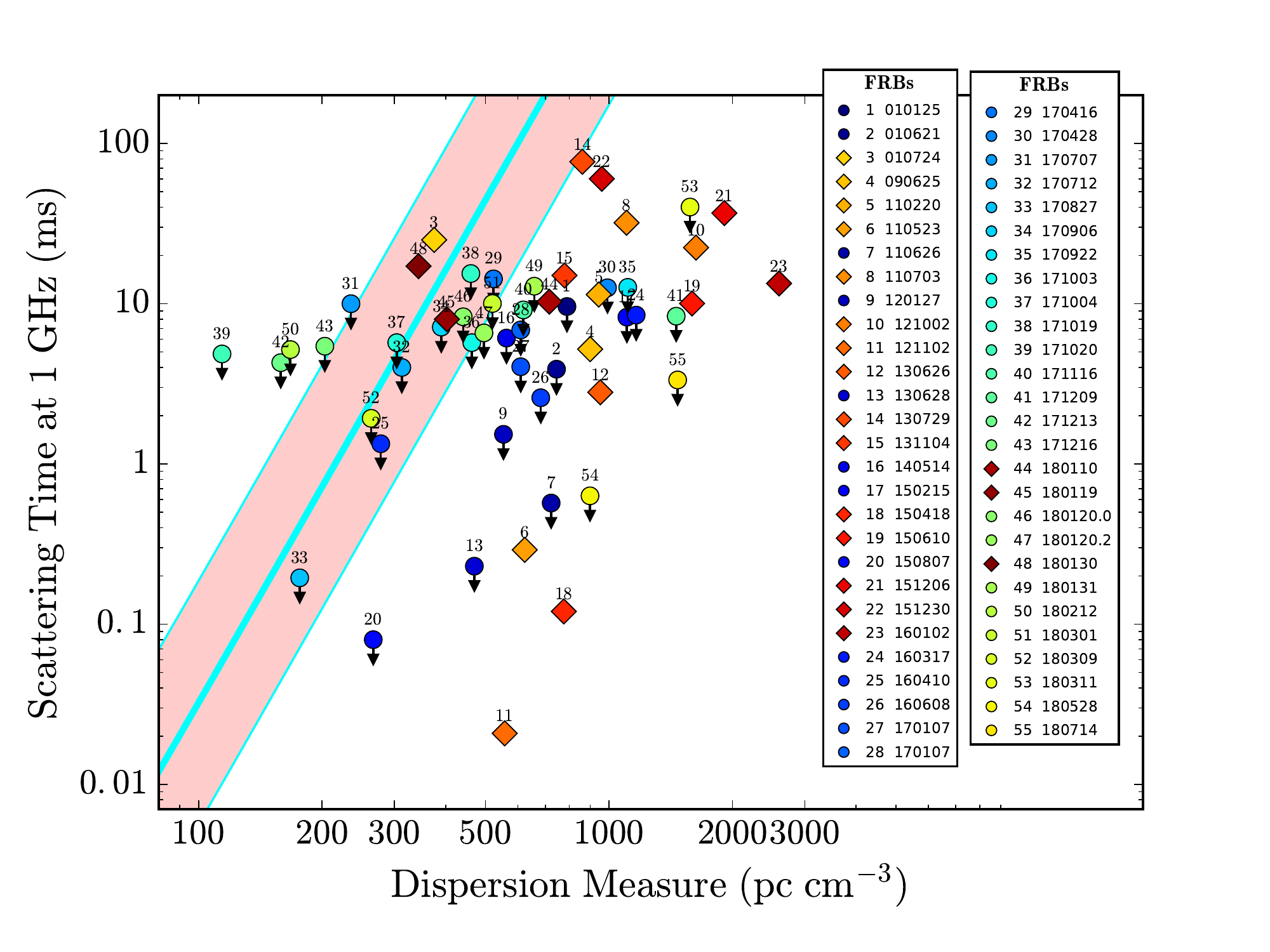}
}
\caption{
Left:
Scattering times $\taud$ vs DM for pulsars and FRBs.
Pulsar scattering is depicted  as a shaded region described by the shown equation.
FRB measurements are shown as filled circles and upper limits as open circles.
Dashed lines extend the small-DM portion of the pulsar fit.
\label{fig:tau_vs_dm_full}
Right:
Zoom-in of the FRB region of $\tau$-\DM\ space using estimates of $\DMx$, the extragalactic DM, and with the Galactic scaling law based on pulsars shifted upward by $\times3$ to account for plane wave geometry.
\label{fig:tau_vs_dm_zoom}
}
\end{figure}

To assess whether extragalactic scattering  stems from the IGM or host galaxies, we compare the $\taud$-\DM\ relation for Galactic pulsars with FRB scattering in Figure~\ref{fig:tau_vs_dm_full}.
For a fixed DM, Galactic pulsars show more than an order of magnitude variation in $\taud$.  Figure~\ref{fig:tau_vs_dm_full} shows the fit to the data of the empirical model
\citep[][]{rmd+97},
$
\taudhat(\DM) = 2.98\times10^{-7}~{\rm ms}\times \DM^{1.4} (1 + 3.55\times10^{-5}\,\DM^{3.1}),
$
with roughly 5\% errors on each parameter and a spread $\sigma_{\log\taud} = 0.76$ about the mean
 \citep[data and fit in ][update in preparation]{cws+16}.
Values for FRB broadening time measurements as well as upper limits are shown in the figure. When measurable, FRB scattering is comparable to burst widths but clearly is  biased below the pulsar band.

To interpret FRB scattering, the extreme heterogeneity of the mean scattering strength per unit DM needs to be accounted for. Galactic pulsars at large DMs sample the inner Galaxy in the spiral-arm and thin-disk components of the NE2001 electron density model. Scattering per unit length is significantly larger in those regions than in the outer Galaxy or in the thick disk component, thus causing the larger slope of the $\tau$--DM distribution in Figure~\ref{fig:tau_vs_dm_full} for $\DM \gtrsim 50~\DMunits$.  FRBs have been seen mostly at high Galactic latitudes and in the Galactic anticenter direction, which sample the more weakly scattering gas also indicated in the figure.  The measured scattering of FRBs must be extragalactic in origin, as demonstrated in Figure~4 of this review.  However, for the corresponding DMs, the scattering is weaker than it would be for lines of sight through the disk of the Milky Way.

However,  the scattering must be compared with only the extragalactic component of DM, which has contributions from the IGM and the host galaxy in a ratio that is unique to each FRB. We define the extragalactic contribution to DM as
$\DMxgal = \DMfrb - \DMg$, where the estimated Galactic contribution
$\DMMWhat = \DM_{\rm NE2001}(l,b) + \DMhalo$
 is the sum of the NE2001 model integrated to its edge and a halo contribution, taken as a uniform value
$\DMhalo = 30$~pc~cm$^{-3}$ \citep[][]{dgbb15}.
Similarly we write $\tauxg = \taufrb - \taug$, where we exclude a halo contribution  because it is likely much smaller than the Galactic disk contribution to $\taug$  that is itself small for the known FRBs.
We then redraw the $\taud$-\DM\ relation in the right-hand panel of
Figure~\ref{fig:tau_vs_dm_full}, which  shows the broadening time vs.   DM using only the extragalactic components of both quantities.

In the figure we also show the Galactic pulsar $\taud$-\DM\ relation under the assumption
 that extragalactic scattering comes from only the host galaxy.
 To compare extragalactic with Galactic scattering, we  need to  compensate for geometrical differences between
 the spherical waves from  nearby sources and plane waves from/to distant sources/observers.
The scattering time $\taud$ is thus  a  factor of three  larger for scattering in the host galaxy than implied by the Galactic $\taud$--\DM\ relation.
The figure therefore shows the Galactic $\taud$--DM band after shifting the
Galactic scaling law of Figure~\ref{fig:tau_vs_dm_full} upward by a factor of three.

\section{SOURCES, RADIATION PROCESSES, AND CENTRAL ENGINES }
\label{sec:engines}

The aggregate properties of FRBs, to the extent that they are now known, require explanations for the bursts themselves --- duration, time-frequency structure,  polarization,  energetics, and, for repeating FRBs, their low duty cycle, absence of periodicity, and rate variability --- as well as their population properties, including the sky rate vs. fluence distribution, which are linked to their spatial distribution and beaming properties.  All of these contribute to a determination and physical  understanding of the underlying sources.

\subsection{Radiation Processes and Beaming}

Emission processes for FRBs are probably most closely related to those for radio pulsars, for which there is a vast literature too large to be reviewed here.    Similarities with coherent cyclotron radiation from planets and  brown dwarfs may also be related at least by analogy.

Empirically,  bursts necessarily comprise coherent, polarized shot pulses whose short durations ($\lesssim$ns)
must be on the order of the reciprocal of the spectral width ($\sim$GHz) and  combine incoherently in large numbers,
with either a shot-rate variation or amplitude variation,   to
form the much longer burst durations $\sim$ms.   Individual shots like those seen from the Crab pulsar \citep[][]{he07, 2010A&A...524A..60J} are prototypes for FRB shot pulses \citep[][]{cw16}.   A feature of
modulated shot noise is that bursts with multipeaked structure are a natural outcome, as are spectral modulations on the reciprocal time scale
\citep[$\sim 1~\mu$s$^{-1} - 1$~ms$^{-1}$  = kHz$-$MHz;][]{cbh+04}.

\begin{figure}[h]
% So this is the brute force solution:
\includegraphics[width=\textwidth]{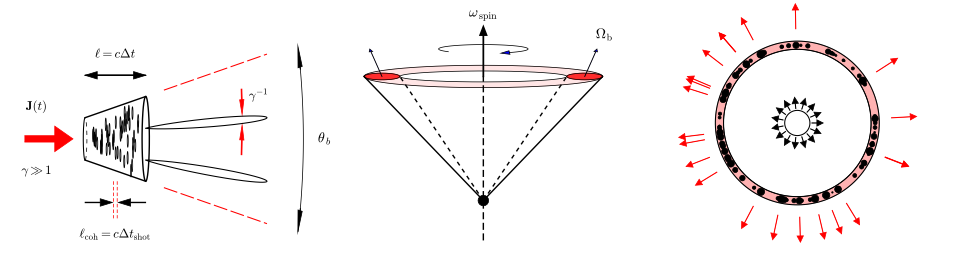}
\caption{
Possible beaming geometries for FRB sources.
Left:  A relativistic jet spanning an angle $\theta_{\rm b}$ much larger than  single particle beaming
angles $\sim \gamma^{-1}$.  Individual coherent emitters of size $\ell_{\rm coh}$ are
contained in an extended region of depth $\ell \sim c W$, where $W$ is the FRB duration.
The jet beam solid angle $\sim \theta_{\rm b}^2$.
Middle:  A rotating beam comprising a relativistic jet swept around by rotation and covering
a total solid angle $\Omega \sim 2\pi\sqrt{\Omega_{\rm b}}$.
Right:  Quasi-isotropic mission from a spherical shell containing individual coherent emitters with
a total solid angle $\Omega \sim 4\pi f_{\rm coh}$ where $f_{\rm coh}$ is the fraction of the shell
with active emitters.
\label{fig:beam_geometries}
}
\end{figure}

Relativistic beaming with large Lorentz factors $\gamma$ is certainly involved with the emission process given the burst energetics discussed in \S\ref{sec:energetics} but it is not clear if beaming plays a role in burst durations
and morphology.  Figure~\ref{fig:beam_geometries} shows three beaming configurations.    First is a non-rotating  `jetted' beam that might arise from magnetic reconnection (and would be two sided) or from a jet aligned with the spin axis of a compact object.     Its orientation might change only slowly  so burst durations  and substructure would be associated with temporal modulations of the particle flow or of the radiation coherence.      The middle frame shows a rotating pulsar-like beam that sweeps through more solid angle than that of the beam itself  and might produce polarization changes like those seen from pulsars.    Last is  quasi-isotropic radiation involving local coherent beams associated with regions where coherent emission can be established, such as by particle beams injected into shocked gas.   In this case the total solid angle is $4\pi$ multiplied by a sub-beam filling factor.    Other physical processes may also be described with these rotating vs. non-rotating paradigms.

% Figure showing beaming cases

The emitted luminosity of an individual source is determined by the radiation physics but the measured flux density  (and  luminosity $\Lp$ defined earlier) depends on beam orientations.   Individual pulsars, for comparison,  show significant variability of pulse shapes and amplitudes indicating stochasticity of the radiation process that should also be anticipated for FRBs and is seen in the repeater FRB~121102.  The  {\it population} luminosity function is  a combination of these factors with the spatial distribution of sources. For pulsars, $\Lp$  spans  at least five decades due to the combination of radiation stochasticity, beaming geometry, and spatial distribution \cite[e.g.][]{acc02} along with scintillations, so $\Lp$  will span an even greater range for the more widely distributed FRBs.

Radiation coherence makes $N$ particles radiate with $N^2$ times the single-particle emission and is responsible for the high efficiency that is needed given the large energy in radio emission alone. The underlying particle acceleration may be linear  (e.g. field-aligned electrostatic waves) or transverse (curvature and gyro-synchrotron radiation) but a coherence mechanism must also operate.    An antenna mechanism involves particle bunches $\sim \lambda$ in size (modulo beaming) with many charged particles.  Maser mechanisms require special distributions in momentum space to provide amplification.  A maser \citep[e.g.]{lm92} has the advantage of cumulative growth of radiation amplitudes over a region $\gg \lambda$, which may alleviate energy requirements that challenge  coherent curvature radiation from bunches \citep[][]{cw16, 2018MNRAS.477.2470L, 2018A&A...613A..61G}.  Nonetheless, emission is limited to the energy carried by particles, which are likely to be  strongly dissipated by radiation reaction from radio emission alone.

Polarization may provide some clues. As mentioned  in \S~\ref{sec:polarization}  for FRB~121102 and several other FRBs,  the approximate constancy of
the position angle across bursts contrasts with that often seen in pulsars, suggesting a non-rotating beam (as mentioned earlier) or a grazing beam where a magnetic axis does not make a close approach with the line of sight
\citep[e.g.][]{1976Natur.263..202B, 2012MNRAS.425..814B}.  The 100\% linear polarization of bursts from
FRB~121102 is similar to the   polarization of some pulsars but pulsars often show some circular polarization.

The spectral islands seen from several FRBs with $\sim$0.1--0.5~GHz widths  are distinct from Galactic scintillations and suggest a bandlimited emission process, such as one where the local plasma or cyclotron frequency (or their harmonics) are involved.    The electron-cyclotron maser process is well established for planetary emission, including the Earth's auroral kilometric radiation (AKR) \citep[][]{2006A&ARv..13..229T, 2011PhPl...18e6501V} and solar bursts
\citep[][]{2017JGRA..122...35C}, and produces  100\% circular polarization, contrary to
FRB emission. An $e^{\pm}$ plasma would show no net circular polarization in the simplest case but  allows linear polarization.     Measured radiation  might be, however,   a combination of emission  with normal  EM modes that are linearly polarized in the magnetosphere followed by  maser amplification without any polarization conversion.
A cyclotron frequency $\nu_{\rm c}\sim 1$~GHz corresponds to
$B\sim 360\gamma$~G that is encountered  at a radius
$r \sim \gamma^{-1/3} 10^{10}$~cm for a magnetar with a surface field of $10^{15}$~G. This exceeds or is a good fraction of the light-cylinder radius for modest $\gamma$ but would require a very large $\gamma$ to be within the light cylinder of a millisecond magnetar.    This context is similar to that for pulsars, which show polarization transfer effects and differential refraction
\citep[][]{ba86, 2010MNRAS.403..569W},
so similar complexity and diversity is to be expected from FRBs.    A distinction from pulsars is the variability of the spectral islands, suggesting `retuning'  of the emission process between bursts (if intrinsic) that might be accompanied by beam wandering \citep[e.g.][]{2017MNRAS.467L..96K}.

The absence of an observed periodicity in FRB~121102 may indicate a non-rotating object but it is easy to destroy periodicities by chaotic precession from a star with a stochastic moment of inertia tensor (e.g. crustquakes) or from  lensing that produces multiple bursts with rapidly changing delays.
The epoch dependent burst rate may have similar intrinsic or extrinsic causes.  An additional extrinsic variability mechanism is triggering by injection of asteroids into a magnetosphere \citep[][]{2016ASPC..502....1H, 2016ApJ...829...27D}.    Asteroids are difficult to inject in rapid rotators ($\lesssim 0.5$~s periods), however, because they are evaporated well before they reach the light cylinder \citep[][]{cs06}.

\subsection{Source Models}

The number of proposed  source models has long exceeded the number of detected FRBs.  Fortunately, the current rapid increase in burst numbers is not accompanied by a proportionate number of models.  In fact most
(but certainly not all)
 attention is now paid to two paradigms, those involving isolated or binary compact objects (WD, NS, and BH) and AGNs, perhaps interacting with neutron stars.    The much larger slate of models has included
technomarkers from extragalactic civilizations \citep[][]{2017ApJ...837L..23L},
superconducting cosmic strings \citep[][]{2014JCAP...11..040Y, 2017ApJ...844...65T, 2017ApJ...844..162T},
exploding  black holes \citep[][]{2014PhRvD..90l7503B},
reconnection in magnetars triggered by axion quark nuggets \citep[][]{2019PhRvD..99d3535V},
WD-NS binaries
\citep[][]{gdl+16},
NS-NS mergers
\citep[][]{2018PASJ...70...39Y},
WD-BH mergers that create reconnecting magnetic blobs
\citep[][]{2018RAA....18...61L},
collapse of supramassive NS
\citep[][]{fr14},
novae of exotic objects (quark or axion stars),  accretion or interaction of asteroids with compact objects \citep[e.g. WD, NS, BH][]{mz14}, mergers of compact objects, births of neutron stars or black holes,  as well as AGN-NS interactions and energetic activity (flares and starquakes) from magnetars.  Some of these produce GRBs
from which   associated prompt radio bursts have  long  been looked for.   However, unless beaming radically increases the prompt  radio burst rate, the GRB rate is too small by a factor $\sim 10^3$-$10^4$ to account for FRBs. Models have been suggested for intermittent pulsars and RRATs \citep[e.g.][]{lm07} that might be relevant to FRBs but the vastly different energetics may make these models less relevant.

It is not our goal to review this rich diversity, especially given page and reference list limits.
More details about the the wide range of models may be found in other reviews
\citep[][]{2016MPLA...3130013K, 2018arXiv180603628P}.
Instead  we build upon the fundamental  quantities summarized in \S\ref{sec:synopsis} on burst rates, repeatibility,  and energetics to suggest that compact objects and especially neutron stars are prime candidates for the underlying engines of many or most FRBs because they exist in sufficient numbers in the universe (a NS born roughly every second in a Hubble volume) and possess sources of free energy (rotation, magnetic) that can   account for burst energetics.   Other objects may of course also generate radio bursts but perhaps at much lower rates.

While energy reservoirs are  available, channeling it into  high brightness, coherent pulses with millisecond durations is more challenging, particularly since pulses are isolated, without obvious pre or post cursors and they certainly do not occur as an ongoing, high duty-cycle process.
This is in contrast to coherent solar bursts and radio flares, for example.

\subsection{Demographics}

Paradoxically, familiar objects in the universe are too numerous to account for the very  large all-sky FRB rate ($10^3-10^4$~d$^{-1}$), even if the beaming fraction is small.   Special objects or special circumstances are needed.
NSs are a good reference population because they can provide free energy from rotation, magnetic fields, and gravity.  In the Universe there are $\sim 10^{17}-10^{18}$ NSs in a Hubble volume (see the sidebar titled Neutron Star Populations in the Universe).   Most  pass through the pulsar channel involving birth spin rates $\sim 10-100$~ms,  electromagnetic radiation across the entire spectrum including prominent coherent radio emission, spindown, and termination of $e^{\pm}$ pair production and thus also the radio emission after 10--100~Myr.   If all NS in a Hubble volume are linked to FRBs, only about one event per NS is needed to account for the sky rate.   Clearly, only a tiny subset of NS can be involved given the $> 200$ events seen from FRB~121102.

\begin{textbox}[h]
\section{Neutron Star Populations in the Universe}

\newcommand{\Birthrate}{{\Gamma_{\rm ns}}}
\newcommand{\NGone}{N_{b}}
\newcommand{\BRminline}{\dot n_{\rm ns, M} }
\newcommand{\BRmFiducial}{\dot n_{\rm ns, M, -13}}
\newcommand{\OmstarFiducial}{\Omega_{*, 0.003}}
\newcommand{\NNS}{N_{\rm ns}}
\newcommand{\BRm}{\dot n_{\rm ns, M} }
\newcommand{\Burstrate}{\Gamma_{\rm b}}

 Extragalactic NS  formed over cosmological time are potential sources of
super-strong bursts whose  number per NS, $\NGone = \eta_{\rm b} T_{\rm b}$,
may be large or small
(where $\eta_{\rm b}$ = burst rate per NS during  a burst phase of duration $T_{\rm b}$).
The aggregate burst rate follows the NS birth rate, $\Birthrate$.  Scaling from
  the Galactic  NS formation rate per unit stellar mass,
$\BRminline \approx \BRmFiducial10^{-13}~\rm yr^{-1}\, \Msun^{-1}\, $
(e.g. one NS every 100 yr per $10^{11}~\Msun$ in stars), and
the stellar mass density $\rho_\star = \Omega_\star \rho_c$ where $\rho_c = 3H_0^2/8\pi G$
is the closure density
and $\Omega_\star \approx 0.003 \OmstarFiducial$ \citep[][]{2005RSPTA.363.2693R},
about  $\NNS \sim 10^{17}$ NS are produced  in a Hubble volume $V_H = 4\pi d_H^3/3$
for a Hubble distance  $d_H = c/H_0 = 4.3h_{0.7}^{-1}  $~Gpc and a  typical
galaxy age $T_{\rm gal} = 10$~Gyr.   A higher star formation rate at redshifts
$\sim 0.5 - 2$ increases the number by about a factor of ten
\citep[][and references therein]{2017ApJ...843...84N}.
The aggregate NS birth is then
$
\Birthrate =   \rho_c \Omega_*  \BRm V_H
	 \sim 4\times (10^4$ - $10^5) ~\rm day^{-1}\,  h_{0.7}^{-1}\, \BRmFiducial\, \OmstarFiducial
$
and the corresponding burst rate is $\Burstrate = \NGone \Birthrate$.   For burst detections out to a distance $d_{\rm max} \ll d_H$ (modulo a proper cosmological integration), $\Burstrate \approx \NGone \Birthrate (d_{\rm max} / d_H)^3$, illustrating the tradeoff between $\NGone$ and $\dmax$ in matching to the empirical FRB rate, $\ratefrb$.

\end{textbox}

FRB directions appear to be isotropic. That the  first FRB localization was to a dwarf, star forming galaxy rather than  a massive L* type galaxy suggests that FRBs do not follow star formation generally but reside in host galaxies that are themselves special.  This `sample of one' situation may change with subsequent localizations, but  the simplest provisional  conclusion is that FRBs are from special galaxies that produce appropriate  central engines.

The magnetar channel accounts for  $\sim 10$\% of NS \citep[e.g.][]{2010MNRAS.401.2675P}.  About 1\% of NS remain in binaries and become millisecond pulsars through accretion-driven spinup with radio lifetimes greater than about a gigayear.   Another $\sim 1$\% are in NS-NS binaries that ultimately merge, producing short-hard GRBs and chirped gravitational waves in the kHz band, like GW~170817 \citep[][]{2017ApJ...848L..13A}.

 If FRBs are largely one-off events per source, rendering repeaters such as FRB~121102 outliers, FRBs could be associated with NS birth events or a highly unusual crustquake, accretion event, or magnetospheric discharge that occurs only once per NS and perhaps not to every NS.  The aggregate event rate is then tied to the NS birth rate
 $\BRns$, which is  within a factor of ten of the empirical FRB rate, $\ratefrb$.   For this to be the case,  FRB events would be associated with a sizable fraction of all NS, perhaps only the magnetar channel or some kind of rare event that happens to nearly all NS.   This scenario seems implausible because spin and magnetic energies of the different NS differ by many orders of magnitude, implying that FRBs would be insensitive to this range, while the radio emission itself as extreme.   Moreover, it  seems premature to dismiss repeating FRBs as outliers because, as discussed earlier,  spectrotemporal  structure of some non-repeating FRBs is similar to that of
 FRB~121102.  One might dismiss this similarity as a feature of the radiation process  rather than of  the underlying engine, but there is currently no support for that view.   Consequently,  NS models can plausibly imply that most or all FRBs repeat, albeit at potentially different rates that have obscured the observational situation about repetitions.

\subsection{Young, Rapidly Rotating Neutron Stars}

Young, high-field NS have been a particular focus of models since the early days of FRBs and a self consistent picture is emerging in favor of these models for at least the repeating FRB~121102 and its associated PRS.
Broad features include a high magnetic field ($> 10^{13}$~G), rapid rotation (spin period $P\sim$~ms), and a young age ($\sim$ 10 - 100~yr).   The object must be old enough so that radio pulses are not free-free absorbed and young enough so that it can provide the luminosity of the PRS.
Unresolved issues include whether the objects are magnetically powered or
rotation powered and whether the coherent bursts themselves originate from the magnetosphere of the spinning object (i.e. within the light cylinder radius $ \rlc = cP/2\pi$) as giant pulses in a pulsar-like model or from
synchrotron maser activity in a distributed region well outside $\rlc$.    Other differences between models concern the mass in the supernova and pre-SN ejecta.

\subsubsection{Giant Pulse Models}

Analogs to giant pulses (GPs) from the Crab pulsar
\citep[][]{csp15, lbp16, pp16, cw16} scale burst amplitudes from the wide GP fluence distribution and from the spindown rate of the Crab pulsar.    Coherent curvature radiation may underly Crab GPs but whether it can provide $\gtrsim 10^{6}$ larger
fluences for FRBs is challenging though may be helped by local maser amplification or extrinsic lensing.    One avenue of exploration is a monitoring program to  probe the extent of the long tail of Crab GPs.
Constraints on GPs from neutron stars likely apply to other central engines, including exotic sources, because the issues in generating powerful fast bursts are generic.

\subsubsection{Magnetar models and superluminous supernovae}

\begin{figure}[h]
\includegraphics[width=0.85\textwidth]{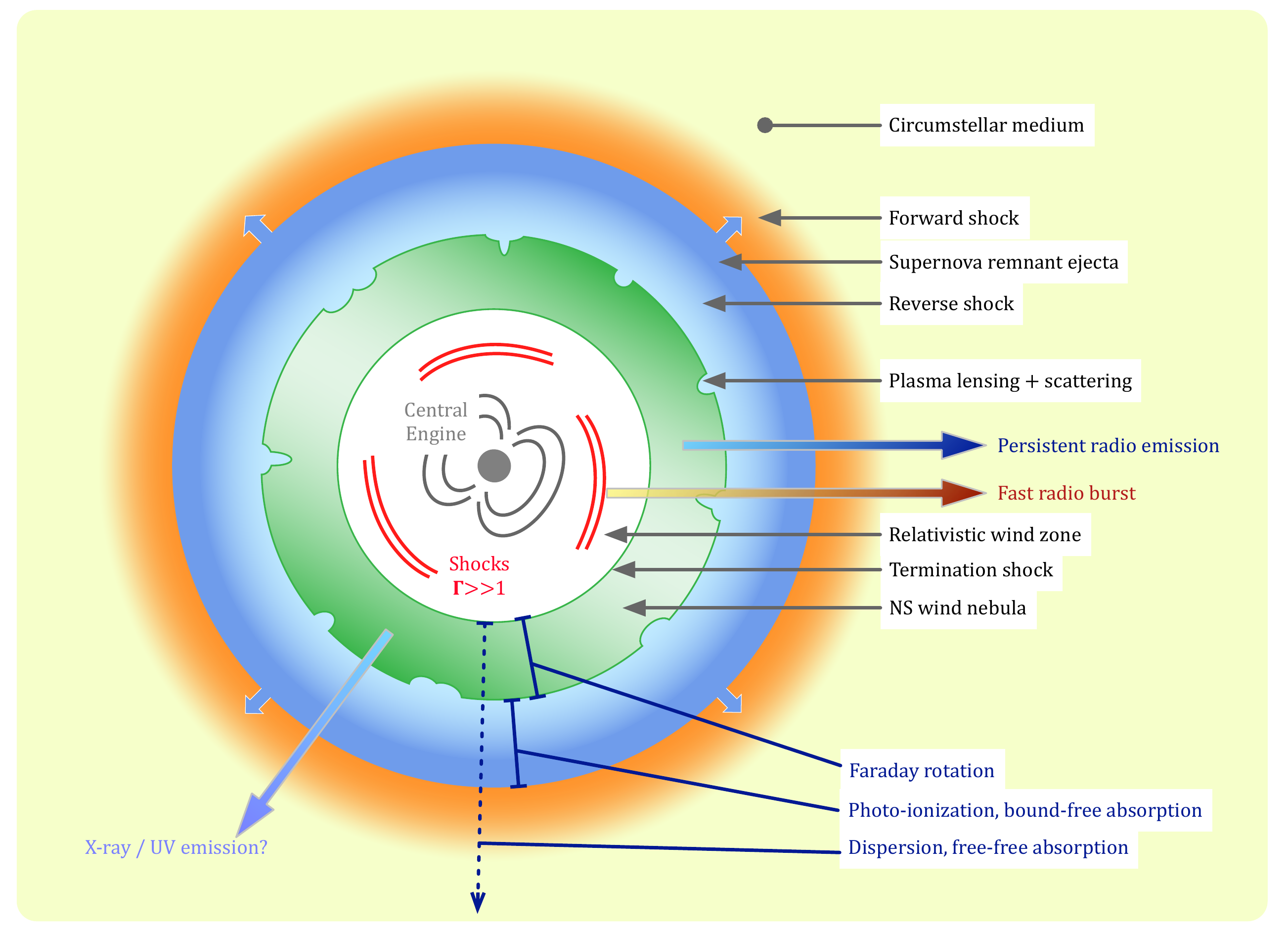}
\caption{
Schematic diagram of an FRB source engine involving a young, highly magnetized neutron star (adapted from \citealt{mm18} and B. Metzger, private communication).  Alternative models, such as compact objects orbiting AGNs, may share some (but not all ) of the same features.
\label{fig:magnetar_cartoon}
}
\end{figure}

Magnetar (and similar) models for FRBs were suggested
prior to the discovery of repeat bursts from FRB~121102 and its association with a persistent radio source in a star-forming galaxy \citep[e.g.][]{pp07, tsb+13, lyu14, kon15, pc15, kat16, mkm16, pir16,  cw16}.    Subsequent work has identified a consistent picture for FRB~121102 where the bursts and persistent source originate from the same structure, although details differ between different models
 \citep[][]{2017ApJ...842...34W, KashiyamaMurase17, mmb+18, mm18}.
 Figure~\ref{fig:magnetar_cartoon} illustrates the features of the \citet[][]{mm18} model.
 It is by no means clear that a magnetar model underlies all FRBs but the case for the repeating FRB is strong because the model can account for many features of the bursts and the PRS.  Even in the magnetar paradigm a great deal of diversity of FRB sources is expected from a range of ages, environments, and initial conditions of the sources.

\citet[][]{2017ApJ...842...34W} used the radio light curve, angular broadening from VLBI, and radio spectrum to show consistency of the persistent source with a compact ($\sim 0.1$-1~pc) region
 emitting non-self-absorbed synchrotron radiation
from gas heated by semi-relativistic shells plowing into ambient gas.  Other, highly relativistic shells produce FRBs from synchrotron maser emission at GHz frequencies determined by the local plasma and cyclotron frequencies.
Negative absorption from this process is confined to roughly a 40\% band.  The age of the source is less than a few hundred years and the  dense outer shell that confines the persistent emission provides only a small DM while providing an RM similar to the measured values.  This analysis reached  conclusions similar to those by  \citet[][]{lyu14}, although the former paper assumes an e-p plasma and the latter a pair plasma produced in magnetar flares.     \citet{2017ApJ...843L..26B} made a similar analysis but invoked specific properties of magnetars to develop a flare-driven model, also with FRBs produced by synchrotron maser emission and a similar persistent source size.

The association of FRBs and persistent source(s) with SLSNe and long GRBs
\citep[][]{2017ApJ...841...14M}
 ties together the physics of central engines and circumsource media with the demographics of SLSNe in dwarf, star-forming galaxies.  Though much of this hinges on
FRB~121102, another source similar to its PRS has been identified
\citep[][]{2018ApJ...866L..22L}, and finding such sources may be a productive avenue for finding burst sources and testing the model and even for finding bursts.
\citet{2018MNRAS.474..573O} propose that high-frequency observations with the Atacama Large Millimeter/Submillimeter Array and the Jansky Very Large Array (VLA) can detect persistent sources at earlier, optically thin epochs than at $\sim 1$-10~GHz.   A recent ALMA observation of FRB~121102 placed an upper on any persistent continuum emission that was consistent with extrapolation of the low-frequency spectrum.

Other tests for general consistency with an central engine/outburst model include the epoch dependences of DM and RM along with the flux density of the PRS.   If burst rates at present are enhanced by plasma lensing, then it too should vary.  X-rays may discriminate between models where FRBs dominate the EM budget \citep[e.g.][]{2017ApJ...842...34W}  compared to those where high-energy emission dominates, though absorption may prevent this for young objects \citep[][]{mmb+18}.

\subsection{AGNs interacting with NS}

AGNs and NS are both abundant in the universe and NS populations bound to AGNs are likely common. But rare interacting NS-AGN configurations may provide an appropriately sized population that  yields low duty cycle bursts.
\citet[][]{2017ApJ...836L..32Z, 2018ApJ...854L..21Z} presents a specific picture where AGN outbursts trigger bursts from a NS; this is also suggested by \citet[][]{cw16}.     A galaxy center is an alternative environment for providing a large RM, as demonstrated by the  (old) Galactic center magnetar J1745-2900
\citep[][]{2018ApJ...852L..12D} with  $\RM \sim 10^5~\RMunits$ that is time variable. The model implies that bursts should show periodicity at the NS orbital period and associates burst polarization with the magnetic field that interacts with the AGN's jet flow.
If a young magnetar is  required to provide bright FRBs,  it is not clear whether an external trigger or external magnetic field is really needed.

\section{FRBs AS TOOLS FOR ASTROPHYSICS AND  FUNDAMENTAL PHYSICS}
\label{sec:tools}

Radio bursts   are obvious probes of magnetized plasma of all kinds  and those originating at extragalactic distances give unique opportunities remote sensing of the extreme environments around burst sources, their host galaxies, and the IGM, as discussed at length already.    Continued monitoring of repeaters provides the means for testing models for central engines via epoch dependences of DM, RM, etc.  Numerous papers have outlined the program for using FRBs to quantify the ionized IGM (as already summarized)  and its large scale structure \citep[e.g.][]{2015PhRvL.115l1301M}. In the latter case  DMs of  $\sim 10^4$  FRBs may yield a significant clustering signal under the assumption that local (host galaxy and circumsource) DM contributions are small
\citep[][]{2014PhRvD..89j7303Z}.    Currently, only the repeating FRB has relatively good constraints on local DM (and RM) contributions, but that will change as the FRB sample continues to grow rapidly.
Probing cosmological magnetic fields using large FRB samples has been outlined by
\citet[][]{2018MNRAS.480.3907V}.

A large sample of FRBs is needed to fully ascertain the range of local DMs and contributions from galaxy clusters.  Electron density models for the Milky Way are uncertain even with thousands of DM (and other) measurements because the number of lines of sight is too small to sample all prominent HII regions.   Local DMs of FRBs may  span a similar large range  among the many host galaxies involved in a large FRB sample \cite[][]{2017ApJ...839L..25Y}.  Indeed, young-NS models imply that DM, RM, and free-free absorption  will be large in the early days of a source and that FRB detections will occur only after they have declined sufficiently \citep[e.g.][]{pc15, csp15, mm18}.    High-$z$ FRBs benefit from having their local DMs reduced by a $(1+z)^{-1}$ factor so the cosmology program may rely on identifying any large redshifts directly, which may be a challenge if only dwarf galaxies harbor FRB sources.
Large scale magnetic fields in the IGM \citep[][]{2014ApJ...797...71Z, 2018MNRAS.480.3907V} will benefit from large FRB numbers and will complement  RM measurements from AGNs already available \citep[][]{2008ApJ...676...70K}.

As previously mentioned, FRB constraints on microlensing and dark matter objects will improve greatly with both large numbers and detailed analyses of burst spectra that may be influenced by interference effects.
Finally, FRBs place limits on the photon mass \citep[][]{2016PhLB..757..548B, 2016ApJ...822L..15W,  2017PhRvD..95l3010S} but require  independent redshifts and determinations of $\DMigm$ because a non-zero photon mass contributes an arrival time delay degenerate with that of plasma dispersion.

\begin{comment}
Section Outline:

\begin{enumerate}
\item  {\bf FRBs as Tools} \quad [XX pp]
   \begin{enumerate}
   \vspace{-0.1in}
   \itemsep -2pt
      \item Magnetoionic probes (IGM, cluster gas,   hosts)
      \item Dark Matter
      \item Photon Mass
      \item Equivalence principle (best done if FRBs detected from high energies to radio)
   \end{enumerate}
\end{enumerate}
\end{comment}

\section{PROSPECTS for FUTURE WORK}
\label{sec:future}

The investigation of FRBs as a novel phenomenon has followed an explosive growth trajectory in its early phase, as measured by detections, theoretical models, publications, and citations. In the near term, increasing numbers of detections are assured. The Australian Square Kilometre Array Pathfinder (ASKAP) in its fly's-eye mode has shown remarkable success with large fields of view at low sensitivity \citep{2017ApJ...841L..12B,smb+18}, and the Canadian Hydrogen Intensity Mapping Experiment (CHIME), which views the whole sky daily as it passes overhead, has detected FRBs in its initial operations \citep{2019Natur.566..230C, 2019Natur.566..235C}. However, neither telescope is likely to produce sufficiently precise localizations in the absence of new modes of operation or outrigger telescopes.  The upgraded Molonglo telescope (UTMOST) has detected several FRBs \citep[e.g.,][]{2017MNRAS.468.3746C, 2018ATel11675....1F} but with limited localization precision in one dimension, joining blind surveys at single dish telescopes with improved instrumentation (e.g., the ALPACA phased array feed at Arecibo) to increase the FRB sample.

Reliable measurements of FRB distances and energetics require host galaxy identifications through better than arcsecond localizations. The {\em realfast} project \citep{lbb+15} should yield more such blind localizations at the VLA, as should surveys with the upgraded APERTIF at Westerbork \citep[e.g.,][]{2017ATel10693....1O}, as well as MeerKAT. New telescope projects like the 110-dish Deep Synoptic Array (DSA-110) and the Hydrogen Intensity and Real-time Analysis experiment (HIRAX) also promise a future yield of blind detections with precise localizations. A complementary approach is the targeted follow-up of blind FRB detections at higher sensitivity (e.g., with the Arecibo or FAST telescopes) in order to identify other repeating sources that can then be subjected to intensive interferometric campaigns, as done for \rfrb\ \citep{clw+17,mph+17}.

The detection of rare weak bursts in massive volumes of survey data is a difficult problem, made more challenging by the steadily worsening radio frequency interference (RFI) environment. Machine learning techniques have been proposed \cite{2018AJ....156..256C} and have already demonstrated dramatic results \citep{zgf+18}, and cross-disciplinary collaboration will continue to bear fruit.
Manifestation of the cosmologically nearby  FRB population is probably limited by the low burst rate per source, but sufficient dwell times on galaxy clusters \citep[e.g.]{2018ApJ...863..132F} may provide detections of nearby galaxies that are easier to characterize than more distant ones.
 If our Galaxy (or a neighboring one) hosts an FRB source, we might experience rare but extraordinarily bright bursts with (relatively) low pulse DM. Such bursts would be difficult to distinguish from RFI, but may be detectable with all-sky dipole antennas or as a citizen science project using mobile phone receivers \citep[e.g.][]{2014PhRvD..89j3009K,2017MNRAS.467.3920M}.

Eventually, efficient petascale computation may allow next-generation projects like the DSA-2000 (a proposed 2000-dish successor to DSA-110) and the full Square Kilometre Array \citep[e.g.][]{mkg+15} to continuously image large swathes of the sky at high enough time resolution to routinely detect and localize large samples of FRBs. Rather like the LSST event streams, the primary challenge will be the efficient allocation of foillow-up resources to extract scientific value from those detections --- no doubt a much better problem to have than the current situation.

%Disclosure
\section*{DISCLOSURE STATEMENT}
The authors are not aware of any affiliations, memberships, funding, or financial holdings that might be perceived as affecting the objectivity of this review.

% Acknowledgements
\section*{ACKNOWLEDGMENTS}
We thank our colleagues and collaborators for discussions and help, as well as access to some results in advance of publication. An incomplete list includes
Matthew Bailes,
Keith Bannister,
Cees Bassa,
Nick Battaglia,
Edo Berger,
Geoff Bower,
Patrick Boyle,
Sarah Burke-Spolaor,
Manisha Caleb,
Fernando Camilo,
Liam Connor,
Adam Deller,
Klaus Dolag,
Jean Eilek,
Ron Ekers,
Wael Farah,
Griffin Foster,
Dale Frail,
Bryan Gaensler,
Vishal Gajjar,
Avishay Gal-yam,
Tim Hankins,
Gregg Hallinan,
Jason Hessels,
Assaf Horesh,
Simon Johnston,
David Kaplan,
Vicky Kaspi,
Jonathan Katz,
Evan Keane,
Michael Kramer,
Shri Kulkarni,
Casey Law,
Avi Loeb,
Duncan Lorimer,
Ryan Lynch,
JP Macquart,
Elizabeth Mahoney,
Benito Marcote,
Maura McLaughlin,
Daniele Michilli,
Chiara Mingarelli,
Ben Margalit,
Brian Metzger,
Eran Ofek,
Stefan Oslowski,
Zsolt Paragi,
Emily Petroff,
Vikram Ravi,
Paul Scholz,
Andrew Seymour,
Ryan Shannon,
Andrew Siemion,
Lorenzo Sironi,
Laura Spitler,
Shriharsh Tendulkar,
Shen Wang,
Ira Wasserman,
Eli Waxman,
Robert Wharton,
Barak Zackay,
Bing Zhang,
Yunfan (Gerry) Zhang,
and Weiwei Zhu.
The authors acknowledge support from the National Science Foundation (AAG 1815242), and are members of the NANOGrav Physics Frontiers Center, which is supported by the National Science Foundation award number 1430284.

% What about responders who said nothing to report? Assaf, Bhaswati,
% Yashwant, Andrea Possenti, ...

% Thank Metzger, Sironi, Boyle, Shannon, Foster, Bailes, Hankins, Eilek, others who
% provided materials in advance of publication or published data in digital format,
% Kaplan, Camilo, Wasserman for discussions.

% References
%\section*{LITERATURE\ CITED}
%\bibliographystyle{ar-style2}

%\bibliography{FRBs_Cordes_Chatterjee_merged}

\end{document}